\newcommand\ignore[1]{}

\documentclass[a4paper,11pt]{article}
\pdfoutput=1
\usepackage{graphicx}
\usepackage{jheppub}
\usepackage[parfill]{parskip}    						
\usepackage{color}
\usepackage[left]{lineno}
\usepackage{amssymb}
\usepackage{amsmath}
\usepackage{amssymb}
\usepackage{graphicx}
\usepackage{tocloft}
\usepackage{float}
\usepackage{braket}
\usepackage{url}
\usepackage{subcaption}
\usepackage{upgreek}
\usepackage{outlines}
\usepackage{braket}
\usepackage{appendix}
\usepackage{xspace}
\usepackage{placeins}
\usepackage{multirow}

\newcommand{\apm}{\alpha'}

\newcommand\be{\begin{equation}}
\newcommand\ee{\end{equation}}
\newcommand\bea{\begin{eqnarray}}
\newcommand\eea{\end{eqnarray}}

\newcommand{\bdm}{\begin{displaymath}}
\newcommand{\edm}{\end{displaymath}}
\newcommand\nn{ \nonumber\\}

\numberwithin{equation}{section}
\numberwithin{figure}{section}

\def\zm{{z_{max}}}
\def\k{\kappa}

\renewcommand\t[1]{\tilde{#1}}

\DeclareMathOperator{\Disc}{Disc}

\renewcommand{\phi}{\varphi}

\renewcommand{\epsilon}{\varepsilon}

\def\pt{p_{\mathrm{T}}}
\graphicspath{{fig/,rootFigs/}}

\title{Inclusive Production Through AdS/CFT}

\author[a,b]{Richard Nally,}
\author[a,c]{Timothy Raben,}
\author[a]{Chung-I Tan}

\affiliation[a]{Department of Physics, Brown University\\
182 Hope St., Providence, RI, U.S.A.}
\affiliation[b]{Department of Physics, Stanford University\\
382 Via Pueblo Mall, Stanford, CA, U.S.A.}
\affiliation[c]{Department of Physics and Astronomy, University of Kansas\\
1251 Wescoe Hall Dr., Lawrence, KS, U.S.A.}

\emailAdd{rnally@stanford.edu}
\emailAdd{timothy.raben@ku.edu}
\emailAdd{tan@het.brown.edu}

\abstract{It has been shown that AdS/CFT calculations can reproduce certain exclusive 2$\to$2 cross sections in QCD at high energy, both for near-forward and for fixed-angle scattering. In this paper, we extend prior treatments by using AdS/CFT to calculate the inclusive single-particle production cross section in QCD at high center-of-mass energy. We find that conformal invariance in the UV restricts the cross section to have a  characteristic power-law falloff in the transverse momentum of the produced particle, with the exponent given by twice the conformal dimension of the produced particle, independent of incoming particle types.  We conclude by comparing our findings to recent LHC  experimental data from ATLAS and ALICE, and find good agreement.}


\arxivnumber{1702.05502}

\begin{document}

\maketitle

\newpage

\newpage

\section{Introduction}\label{sec:Intro}

It has been suspected for many years that large-$N_c$ QCD admits an alternate description as a string theory~\footnote{See \cite{DiVecchia:2007vd} for a detailed historical discussion.}.  Early developments were inspired by the realization that string scattering amplitudes obey Regge behavior and crossing symmetry. This conjecture was greatly spurred on with  the observation  that, in the limit of large $N_c$ with $\lambda = g^2_{YM}N_c$ fixed, the QCD perturbation series can be made to resemble the genus expansion of worldsheet string theory \cite{'tHooft:1973jz}.  With the advent of AdS/CFT correspondence~\cite{maldacenaconjecture,Maldacena:1998im,Witten:1998qj,Gubser:1998bc,Aharony:1999ti,Gubser:2002tv}, or equivalently gauge-string duality, the theoretical landscape has taken a dramatic step forward and a string realization of QCD has again become a serious goal for current studies.

 In this paper, we explore the consequences of conformal symmetry in high energy scattering experiments. In particular, we will use the AdS/CFT correspondence to examine inclusive production. Although strictly speaking QCD itself is not a CFT, it is closely related to  $\mathcal{N}=4$ super Yang-Mills, which is conformal, and the two theories are similar enough that a great deal can be learned from the conformal limit \cite{Politzer:1973fx,Gross:1973id,Braun:2003rp,Bern:2004kq}. The effects of conformal symmetry on QCD have previously been studied in inclusive scattering in both the fixed-angle \cite{Polchinski:2001tt,Brower:2002er} and in the near-forward limits~\cite{Brower:2006ea,Brower:2007qh,Brower:2007xg,Cornalba:2006xm,Cornalba:2006xk,Cornalba:2007fs,Cornalba:2008qf,Cornalba:2009ax,Costa:2012cb,Anderson:2014jia,Anderson:2016zon}.   Here, we will focus on central production at the LHC.

Inclusive processes unavoidably involve near-forward particle production.  The relevant physics  is intrinsically non-perturbative, and cannot be reduced simply to purely partonic scattering. With AdS/CFT,  one is able to address both perturbative and non-perturbative aspects of inclusive production at high energy in a unified setting. Indeed, holographic techniques based on a $t$-channel OPE \cite{Polchinski:2002jw,Brower:2010wf,Costa:2012fw,Costa:2013uia,Brower:2015hja,Nishio:2011xz,Koile:2015qsa,Ballon-Bayona:2015wra}   have been used as a complement to more traditional weak coupling methods \cite{Hentschinski:2012kr,Frankfurt:2005mc,Marquet:2006jb,Marquet:2007nf,Avsar:2007ht,Block:2014kza} to study HERA data for the deep inelastic scattering (DIS) cross section at large $s$ and small $x=Q^2/s$.

Early interest in inclusive production can be traced back to the work of Feynman~\cite{Feynman:1969ej}, Yang~\cite{Benecke:1969sh}, Wilson~\cite{Wilson:1970zzb}, and others, focusing particularly  on  the scaling properties of particle distributions. Studies of inclusive production in a CFT context began with works of Strassler~\cite{Strassler:2008bv},  Hofman and Maldecena~\cite{Hofman:2008ar} and Belitsky {\it et al.}~\cite{Belitsky:2013xxa,Belitsky:2013bja,Belitsky:2013ofa}.  Instead of focusing on the final state particle distribution, which is ill-defined in the strict CFT  limit,  the emphasis has been  on infrared safety~\cite{Sterman:1977wj,Brown:1981jv}, e.g., on energy flows, leading to  vacuum expectations
\be
\sigma_{w}(p)=\int d^4x e^{-ipx}  \langle 0| {\cal O}^\dagger(x) {\cal D}[w] {\cal O} (0) |0\rangle. \label{eq:Hofman1}
\ee
Here ${\cal O}$ serves as the source for the initial state $|{\cal O}(p)\rangle$, which carries 4-momentum $p_\mu$, and ${\cal D}[w]$ is a product of a set of local operators, measuring flows of conserved  quantities, such as energy-momentum; such an object is generically referred to as an inclusive ``event shape" distribution.

The operator product ${\cal D}[w]$ in the above expression is not time-ordered, and thus the appropriate Lorentzian correlation functions are Wightman functions.  Momentum space Wightman functions lead to 
amplitude discontinuities, so  it is necessary to deal with Landau-Cutkosky singularities~\footnote{The Landau-Cutkosky singularities  for Lorentzian correlation functions in CFTs with a gravity dual has recently been addressed in \cite{Maldacena:2015iua}.}.
The treatments in \cite{Strassler:2008bv,Hofman:2008ar,Belitsky:2013xxa,Belitsky:2013bja,Belitsky:2013ofa} have mainly focused on processes where the source  involves a single { local operator},  such as $e^+e^-\rightarrow \gamma^*\rightarrow X$, where $X$ represents all allowed final states, which are implicitly summed over.

Our discussion in this paper  will  deal primarily with scattering processes where the initial source is  {non-local}, and will be carried out  in a {\it momentum representation}.  The simplest inclusive scattering process is
\be
a + b\rightarrow X\, , \label{eq:totalXsection}
\ee 
where again X implies a sum over all possible final states.   After summing over contributions from all possible final states,  the completeness relation $\sum_X|X\rangle \langle X|=I$  leads to  
the usual optical theorem, which states that the total cross section $\sigma_{total}^{ab}(s)$ of such a process  is given by the imaginary part of the elastic amplitude in the forward limit~\footnote{Here we use canonically defined Mandelstam invariants. The elastic scattering amplitude $T(s,t)$ is parameterized by the usual center of mass energy squared $s$ and the momentum transfer squared $t$.}, 
\be
 \sigma_{total}^{ab}(s)\simeq \frac{1}{s} \, {\rm Im} \, T(s,t=0). \label{eq:Optical4}
 \ee

The next simplest inclusive process is single particle production, 
\be
a+b\rightarrow c +X\, , \label{eq:1particleInclusive}
\ee
where again X implies a sum over all possible final states,  leading to a differential production cross section, ${d \sigma_{ab\rightarrow c+X}}/d^3{\bf p}_c$.  Kinematically, single particle production can be treated  as a 2-to-2 process, with X having a variable mass $M_X^2=(p_a+p_b-p_c)^2$ often referred to as the ``missing mass"; for simplicity, we will simply call this $M^2$.   The invariant differential cross section $ {d \sigma_{ab\rightarrow c+X}}/{(d^3{\bf p}_c/E_c)}$  therefore depends on three Lorentz invariants instead of the usual two for exclusive scattering. The usual Mandelstam variables $s$, $t$, and $u$ can be used, but it is frequently more convenient to work with $s$, $t$, and $M^2$; it is easy to see that $s+t+u=m_a^2+ m_b^2+m_c^2+M^2$, so these two sets of variables encode the same information.

In a momentum space treatment,  inclusive cross sections can always be identified as \emph{discontinuities} of appropriate forward  amplitudes through the use of generalized optical theorems. The differential cross section $ {d \sigma_{ab\rightarrow c+X}}/{(d^3{\bf p}_c/E_c)}$  of the process   $ab\rightarrow c +X$  can  be identified as the discontinuity in $M^2$ of the amplitude for the six-point process $abc'\rightarrow a'b'c$; symbolically, we have 
\be
 \frac{d \sigma_{ab\rightarrow c+X}}{d^3{\bf p}_c/E_c} \simeq  \frac{1}{2i s}  {\rm Disc}_{M^2} T_{abc'\rightarrow a'b'c}\, .  \label{eq:Optical6}
\ee

The main goal of this paper is to explore the consequences of conformality on inclusive central production in proton-proton and proton-lead scattering.  We examine the use of the $t$-channel OPE~\cite{Brower:2006ea,Brower:2014wha} for high energy scattering, elucidate subtleties involved in using generalized optical theorems, and pay special attention to non-perturbative issues.  
In particular, we show that aspects of conformal invariance can be explored in a ``gluon-rich" environment~\footnote{In an AdS/CFT treatment, QCD is dominated by gluon dynamics with quark loop contributions suppressed. These contributions can become important in the ``fragmentation regions" however.} by treating  central inclusive particle production of the form   
 \be
 a+b \rightarrow X_1 + c + X_2\, , \label{eq:centralproduction}
 \ee
 where $X_1$ and $X_2$ represent   left- and right-moving ``lumps"  in the CM frame.

Our discussion can be divided into several parts. We first focus on the more formal question of how to treat CFT inclusive shape distributions as weighted discontinuities  of multiparticle momentum space amplitudes $T_{ab1'2'\cdots \,\rightarrow \,a'b'12\cdots}$, generalizing on earlier treatments.  This treatment is carried out necessarily in a Minkowski setting, with the discontinuity in the generalized missing mass, $M^2=(p_a+p_b-\sum p_i)^2$ taken in the forward limit.  This procedure applies to both events initially sourced by a single local operator, as in Eq. (\ref{eq:Hofman1}), and to scattering processes at high energy, as in Eqs. (\ref{eq:Optical4}) and (\ref{eq:Optical6}). 

By multiparticle amplitudes here we simply refer to the usual  Euclidean CFT correlations  functions, $\langle \phi(x_1)\phi(x_2)\cdots\rangle$, continued to Lorentzian signature; these lead to vacuum expectation values for time-ordered (or {\bf T} product) conformal primaries, $\langle 0|{\bf T}\{ \phi(x_1)\phi(x_2)\cdots\}|0\rangle$. We assume a standard  Hilbert space structure (e.g. a state space spanned by states associated with conformal primaries) which allow us to use completeness relations. Although our emphasis is on purely conformal characteristics, we are mainly concerned with theories that allow an IR confinement deformation so one can interpret the results in terms of canonically defined scattering amplitudes.

We will next discuss inclusive production for scattering processes and explore in particular the consequences of AdS/CFT and conformal invariance at { finite 't Hooft  coupling}, $\lambda$ large but finite. Here we review the bare necessities on how to move beyond the supergravity limit by including string corrections, and so we are effectively dealing with string amplitudes on an AdS background. Historically, the greatest obstacle to a stringy description of QCD phenomenology has been the requirement of hard partonic behavior at short distances. AdS/CFT provides a framework to resolve these phenomenological difficulties. Polchinski and Strassler \cite{Polchinski:2001tt} identified an approximation regime in which the warped geometry of the dual AdS theory provided a power-law falloff for wide-angle scattering in QCD. This argument has been extended to near-forward QCD scattering in AdS/CFT. We follow a similar approach as first described in \cite{Brower:2006ea,Herzog:2008mu}.  In \cite{Brower:2006ea} it was shown that Pomeron exchange, i.e. the leading Regge singularity with the quantum numbers of the vacuum, can be described by a Reggeized graviton~\footnote{In what follows, this will also be referred to as the BPST pomeron~\cite{Brower:2006ea}. This stands in contrast with the BFKL Pomeron~\cite{Kuraev:1977fs,Balitsky:1978ic,Kotikov:2004er}, which is based on perturbative QCD.} propagating in AdS$_5$. The unifying principle for both exclusive power-behavior at the fixed-angle limit and the Pomeron dominance for the near-forward scattering  is conformal invariance.

We next apply our analysis to single-particle 
inclusive production in the central region.  Here $X$ can be separated into left- and right-moving groups, $X_1$ and $X_2$ respectively. The event shape distribution is controlled by a matrix element,  $\langle {\cal V}_P V_{c\bar c} {\cal V}_P\rangle$,  involving two Pomeron vertex operators~\cite{Brower:2006ea}.
Just as the case of exclusive fixed-angle scattering, flat space string scatting amplitudes \cite{Detar:1971dj,Virasoro:1971zq,Gordon:1971ss,Clavelli:1978tt,Ader:1977qy} predict an exponential cutoff in the transverse momentum $\sim e^{- 4 \alpha' p_\perp^2}$.  However, we argue that a generalization of the Polchinski-Strassler regime \cite{Polchinski:2001tt,Brower:2002er} utilizes the warped AdS geometry to render the effect of confinement deformation unimportant at high $p_T$. Using this we arrive at our central result for CFT behavior at the LHC involving a partonic power-law falloff of the form   
\begin{equation}
   \frac{d \sigma_{ab\rightarrow c+X}}{d^3{\bf p}_c/E_c}
 \sim  {\rm Disc}_{M^2}\langle {\cal V}_P V_{c\bar c} {\cal V}_P\rangle
 \sim p_\perp^{-\delta} \, . \label{eq:2b}
 \end{equation} 
The exponent $\delta$ is fixed by holography and conformal invariance, given by $\delta=2 \tau$, with $\tau=\Delta-J$, where $\Delta$ is the conformal dimension,  and $J$  the spin of the produced hadron.\footnote{Amplitudes displaying a similar power-law like behavior can be described using a complimentary holographic approach where one simply considers the string zero-mode contribution. Further details can be found in~\cite{Andreev:2004sy,Andreev:2004vu} and references therein.  We focus here on the BPST approach as we believe it is more analogous to the perturbative weak coupling approach where in both cases Regge poles can be interpreted as eigenvalues of an effective Hamiltonian approach.} In the large $N_c$ limit of the AdS/CFT, the theory is dominated by gluonic interactions; the production of fermion pairs is suppressed by $1/N_c$.  The simplest bound state is then a glueball state $Tr(F^2)$ \cite{Csaki:1998qr,Ooguri:1998hq,jevickiglueball} which can be used to describe meson production via AdS/CFT \cite{Costa:2013uia,Brunner:2015oqa,Hashimoto:2007ze,Brunner:2014lya,Brunner:2015kxa} \footnote{For a brief introduction to mesons in AdS/CFT see \cite{Erdmenger:2007cm}}. In QCD, scattering processes dominated by Pomeron exchange are described via the BFKL Pomeron (reviewed in \cite{Lipatov:1996ts}).  Since the BFKL Pomeron the exchange of a Reggeized gluon ladder, the bound states lying on the trajectory are thought to be glueballs\footnote{For a recent review of the Pomeron/Glueball connection see \cite{Kisslinger:1999jk}.}.  For production via scalar glueballs, we thus have $\delta = 2\Delta = 8$.  This is analogous to the dimensional counting rule \cite{Brodsky:1973kr,Brodsky:1974vy,Matveev:1973ra}, but from a non-perturbative perspective. Finally, we test this prediction by comparing to recent ATLAC and ALICE data from the LHC. 
 
This paper is organized as follows: in Sec. \ref{sec:Mueller}, we focus on the treament of inclusive distributions as discontinuities.  Sec. \ref{sec:2-point} involves reviewing the simple, but illustrative case of 2 point functions.  Although these results can be found in the literature, we re-derive them in a consistent notation. Using this notation we then reinterpret known results about 4- and 6-point functions 
and present new analysis about generalized n-point functions in Sec. \ref{sec:Mueller-b}. Elucidating examples are left to App. \ref{sec:OPE}. In Sec. \ref{sec:Mueller-d},  by invoking AdS/CFT,  we express inclusive cross sections in terms of the discontinuities of Witten diagrams. In Sec. \ref{sec:ads-scat} we detail 2-point functions in AdS and derive analogous results to those in Sec. \ref{sec:Mueller}.  Following this we are able to posit our prediction for high energy inclusive scattering in Sec \ref{sec:incl-ads}. We turn next, in Sec. \ref{sec:CentralProduction}, to inclusive distribution in the central production. Finally, in Sec. ~\ref{sec:Test}, we test this finding  by comparing with the recent LHC  data; in Sec. \ref{sec:deviations} we discuss possible explanations for the results of the experimental fits.  We conclude with a brief discussion of our essential results in Sec. \ref{sec:discussion}.

Throughout the paper, the details of results from earlier literature are omitted from the body of the text, and are instead provided in Appendices \ref{sec:A-Mueller}-\ref{sec:FlatString}. In particular, these appendices cover the treatment of inclusive cross sections as discontinuities in QCD itself, the holographic pomeron, aspects of conformal field theory, and flat space string amplitudes, respectively. However, because much of the work here connects disparate background material we provide a bare minimum of review and examples in the main text for the paper to be relatively self contained.

\section{Inclusive Cross Sections and Discontinuities}
\label{sec:Mueller}
 In field theory, inclusive cross sections involve Minkowski space Wightman functions.
 In this section, we clarify  how these Wightman functions, in a momentum representation,  can be identified  as  ``forward discontinuities" of $n$-to-$n$ amplitudes,  e.g.,  $n=3$ for the process $a + b\rightarrow c+ X$.  We begin by reviewing the more familiar case of 2-point functions before generalizing to higher point correlators.  We conclude by  demonstrating these ideas in the context of deep inelastic scattering (DIS), where the cross section can be explicitly related to a discontinuity; we also relate the moments of the DIS distribution to a $t$-channel OPE.  
 
\subsection{2-Point Functions}
\label{sec:2-point}

The relationship between a conventional  {\it time-ordered} Green's function, $\langle 0| {\bf T}\{\phi_1\phi_2\cdots\}|0 \rangle$, 
and  a Wightman function, which is not  {\it time-ordered}, can best be understood  in a momentum representation.   Let us  illustrate this  by  first comparing   the Feynman propagator, $\langle 0 |{\bf T}\{\phi(x)\phi(0)\}| 0\rangle $,  for a free scalar, with the corresponding Wightman function, 
$\langle 0|\phi(x)\phi(0) |0\rangle $. 
   In a
momentum representation~\footnote{In this paper, we adopt the $(+,-,-,-)$ metric, so that $p^2={p^0}^2-\vec p^2$ and $p\cdot x= p^0 t - \vec p\cdot \vec x$. }$^,$ \footnote{ With interactions, in addition to the pole, $G_F(p^2)$ acquires a branch cut for $p^2>4m^2$, and can be represented in a spectral representation, again  with $G_W(p^2)$ as its discontinuity, for $p^0>0$.},   
\bea
 G_F(p^2)&=&i \int d^4x  e^{ip\cdot x} \langle 0 |T(\phi(x)\phi(0))| 0\rangle = - \frac{1}{p^2-m^2+ i\epsilon}\, , \\
G_W(p^2)&=& \int d^4x  e^{ip\cdot x} \langle 0|\phi(x)\phi(0) |0\rangle = 2 \pi \delta(p^2-m^2) \theta(p^0),  \label{eq:Wightman1}
\eea 
 $ G_F(p^2)$ is an analytic function in the invariant $p^2$, with poles at $p^0=\pm \sqrt {m^2+\vec p^2}$. 
 However, $G_W(p^2)$ defines a distribution, corresponding to  the discontinuity of $G_F(p^2)$ for $p^0>0$.

Let us turn next to CFT, using conventional CFT normalization and again in a Minkowski setting.   Consider  a generic scalar conformal primary  $\phi$ of dimension $\Delta$.  The Fourier transform for its  Feynman propagator and  the corresponding Wightman function are
\bea
G_F(p^2)&=& i \int   d^4x \frac{e^{ ip x}}{ [\vec x^2 -t^2 + i\epsilon)^2]^{\Delta}}= - d(\Delta) (- p^2)^{\Delta-2}\, ,\\
 G_W(p^2)&=&\int  d^4x \frac{e^{ ipx} }{[\vec x^2 -(t-i\epsilon)^2]^{\Delta}}= c( \Delta)  \theta(p^2) \theta(p^0)\, (p^2)^{\Delta-2},   \label{eq:Wightman2}
\eea
where  $c (\Delta) = 2d(\Delta) {\sin\pi \Delta} = (2\pi)^3   2^{2(1-\Delta)}/ \Gamma(\Delta)\Gamma(\Delta-1)$.  $G_F(p^2)$ is a real-analytic function~\footnote{Positivity  requires that $1\leq \Delta$.  For $\Delta$ approaching a positive integer $n$, coefficient $d$ diverges while $c$ remains finite, indicating the emergence of $(- p^2)^{n-2}\log (-p^2)$. We shall stay with a generic $\Delta$, away from positive integers.}  in $p^2$, with  a branch cut across $0<p^2<\infty$.  The corresponding Wightman  propagator, $G_W(p^2)$, is a distribution.  
 Although there is   no mass-gap, the relation  between time-ordered amplitudes and Wightman functions remains.  $ G_W(p^2)$    is a continuum   over $0<  p^2< \infty$, corresponding to the discontinuity of $G_F(p^2)$ across its cut. 

\subsection{Inclusive Distributions in  CFT}\label{sec:Mueller-b}
It is useful to  distinguish between two types of inclusive processes. The first type corresponds to events with a single initial local source, e.g. $\gamma^*\rightarrow c_1+c_2+\cdots+X$, which has been discussed before. The second type invovles a non-local source, as in scattering, e.g., $a+b\rightarrow c_1+c_2+\cdots+X$, which we expand on here. In the language of CFT, these inclusive cross sections can be interpreted as flow rates  for conserved quantities, such as the energy density flowing into a solid angle $d^2\Omega$ about a direction $\hat n$~\cite{Hofman:2008ar,Belitsky:2013xxa,Belitsky:2013bja,Belitsky:2013ofa}.   General inclusive flows for conserved quantities can always be expressed as weighted discontinuities~\cite{Hofman:2008ar,Belitsky:2013xxa,Belitsky:2013bja,Belitsky:2013ofa}.

Let us consider  scattering processes first.  As stated in Sec. \ref{sec:Intro}, the simplest inclusive process corresponds to Eq. (\ref{eq:totalXsection}), with the total cross section given by the imaginary part of forward amplitude, as in  Eq. (\ref{eq:Optical4}).
Consider next the inclusive production of a scalar particle, $a+ b\rightarrow c  +X$, as in Eq. (\ref{eq:1particleInclusive}).  The invariant differential cross section can be expressed as~\cite{Mueller}
\bea
 \frac{d \sigma_{ab\rightarrow c+X}}{d^3{\bf p}_c/E_c}&\propto &\sum_X (2\pi)^4 \delta^{(4)}(p_a+p_b-p_c-p_X) \Big| \langle p_c,X \Big| p_a,p_b\rangle \Big|^2\nn
 &\propto &\sum_{X} \int d^4 x e^{-ip_c\cdot x}  \langle p_{a},p_{b}| \phi_c(x)|X\rangle\langle X|  \phi_c(0) |p_a,p_b\rangle \, .   \label{eq:Optical6a}
\eea
Making use of  the completeness relation $ \sum_X |X\rangle\langle X|=I$,   
the  cross section  can also be expressed as a matrix element, 
\be
\frac{d \sigma_{ab\rightarrow c+X}}{d^3{\bf p}_c/E_c} \propto    \langle p_{a},p_{b}|   \widetilde {\cal O}_c(p_c)  |p_a,p_b\rangle \, ,
\label{eq:InclusiveProduction}
\ee
  where  $\widetilde {\cal O}_c =    \int d^4 x e^{-ip_c\cdot x}     \phi_c(x)  \phi_c(0)  $ is the Fourier transform of   product of two local operators. Here, $p_c$ is the four-momentum for the produced scalar, with $p_c^0>0$.   
Since the product $ \phi_c(x)  \phi_c(0)$ is not time-ordered, one is again dealing with a Wightman function.  

The corresponding  3-to-3 process is $a+b+ c'\rightarrow a'+b' + c$,  where the amplitude $T_{abc'\rightarrow a'b'c}$ is  given by     a {\it {\bf T}-product}  between asymptotic states~\footnote{Energy components for all external 4-vectors are positive. For simplicity, the overall delta-function due to translational invariance will be surpressed in what follows. Strictly speaking, we need to work with  amputated on-shell amplitudes, where $\phi_c$ should be replaced by  a source function, $j_c(x)=(\square -m^2_c) \phi_c(x)$. We will skip this step to avoid notational overload.},
$  \langle p_{a'},p_{b'}| {\bf T}\{\phi_{c}(x)  \phi_{c}(y)\} |p_a,p_b\rangle$, in momentum space.   One can move  from a {\bf T}-product to a Wightman function as  done earlier for  the free propagator. Because it is a matrix element between  asymptotic states, one replaces the  4-vector $p$  in  (\ref{eq:Wightman1}) by $(p_a+p_b-p_c)$, with $p^0_a+p^0_b-p^0_c>0$ and $p^0_c>0$. Therefore,  $\langle p_{a},p_{b}|   \widetilde {\cal O}_c(p_c)  |p_a,p_b\rangle$ is  the discontinuity of $T_{abc'\rightarrow a'b'c}$, in the invariant $M^2=(p_a+p_b-p_c)^2$, 
\be
 \frac{d \sigma_{ab\rightarrow c+X}}{d^3{\bf p}_c/E_c} \propto \frac{1}{2is} \,  {\rm Disc}_{M^2>0} T(p_{a'},p_{b'},p_{c}; p_a,p_b,p_{c'})\, .  \label{eq:Optical6b}
\ee

This is the process that is examined holographically in Sec. ~\ref{sec:CentralProduction} and more details can be found in Appendix \ref{sec:A-Mueller}.

Next we turn to inclusive processes involving a single local source, ${\cal O}$, for example  $e^+e^-\rightarrow \gamma^*(p)\rightarrow c_1+c_2+\cdots+X$.   The decay process can be interpreted as a CFT process as motivated by the work of Hofman and Maldacena \cite{Hofman:2008ar}. In what follows it will be useful to recast Eq (\ref{eq:Hofman1}) in the form of  a normalized distribution in a momentum representation as \be
\langle \widetilde  O_w\rangle =\frac{\sigma_{w}(p)}{\sigma_{\cal O}(p)} =\frac{\int d^4x e^{ipx}  \langle 0| {\cal O}^\dagger(x) \widetilde  O_w {\cal O} (0) |0\rangle} {\int d^4x e^{ipx}  \langle 0| {\cal O}^\dagger(x)  {\cal O} (0) |0\rangle}=   \frac{ \langle {\cal O}(p) |  \widetilde  O_w |{\cal O}(p)\rangle} { \langle {\cal O}(p )|  {\cal O}(p)\rangle} , \label{eq:Hofman2}
\ee
where $\widetilde  O_w$ is chosen to ensure infrared safety. In general, $\widetilde  O_w$ is a non-time-ordered product of a set of local operators, as in Eq. (\ref{eq:InclusiveProduction}); as discussed above, this necessitates the use of Wightman functions \cite{Belitsky:2013xxa,Belitsky:2013bja,Belitsky:2013ofa}.

We can now apply to this expression the same analysis used to argue for Eq. (\ref{eq:Optical6a}). The matrix element $\langle {\cal O}(p) |  \widetilde  O_w |{\cal O}(p)\rangle $ admits a form similar to Eq. (\ref{eq:Wightman2}), but with the momentum $p$ replaced with $p-p_c$, where $p_c$ is the momentum associated with the flow so that $p^0>p_c^0>0$. Then we can relate $\langle {\cal O}(p) |  \widetilde  O_w |{\cal O}(p)\rangle $  to a discontinuity exactly as was done for Eq. (\ref{eq:Optical6a}) earlier.

Generically, we can write the cross section for such a process as \be
 \sigma_w(p)=\sum_X (2\pi)^4 \delta^{(4)}(p-p_X)w(X) \Big| \langle X|\gamma^*(p)\rangle\Big|^2, \label{eq:completeness}
 \ee where the sum is taken over all possible $X$ and involves an integration over the phase space for each state $X$, weighted by $w(X)$. For example, the simplest inclusive single-particle production process, $e^+e^-\rightarrow \gamma^*\rightarrow c+X$, involves the measurement of a charge $Q$ by a ``calorimeter" at spatial infinity encompassing a differential solid angle $d^{(2)}\Omega$ around a direction $\hat n$. This corresponds to having $w_Q= \sum_c Q_c \delta^{(2)}(\hat p_c-\hat n_\Omega) \theta(p_c^0).$ 

 The cross section can also be re-written as
$ \sigma_Q(p,\hat n) \sim \sum_X \Big[       \langle \gamma^*(p) | X \rangle w_Q \langle X| \gamma^*(p)\rangle \Big]$.   If the factor $w_Q$ is replaced by a delta-function of four-momentum, $Q_c \delta(q_c-p_c)$, then using completeness, $\sum_{X'}  | X' \rangle  \langle X'|=I$ where the sum  over $X'$ stands for the previous sum over $X$ with a state $\phi_c$ removed.  This in turn simply leads to the discontinuity of a 4-point function in the invariant $M^2=(p_{\gamma^*}-p_c)^2$. 
One can formally introduce 
 $
 \widetilde  O_Q = \sum_c w_Q\int d^4x e^{-ip_cx}  \phi_c(x)  \phi_c(0)  
$,
  leading to  
\be
 \sigma_Q(p,\hat n) 
= \sum_c  \int {d^4p_c}\, \frac{1}{2i}\,  w_Q(p_c)\, {\rm Disc}_{M^2} \, T_{\gamma^*c'\rightarrow {\gamma'}^*c} \label{eq:event}
\ee
in the forward limit of $p_{{\gamma^*}'}=p_{\gamma^*}$ and $p_c=p_{c'}$.  The discontinuity is taken for $M^2>0$.

The same formalism can be used to study the flow of other conserved charges, such as energy and momentum, as well as higher-point correlation functions $\langle \widetilde  O_w(1) \widetilde O_w(2)\cdots \rangle 
$. For instance, the flow of energy in a direction $\hat{n}$ is given by $\langle E(\hat n)\rangle =\sigma_E(\hat n)/\sigma_{\gamma^*}(s)$, where 
\be
\sigma_E(\hat n)= \sum_{c}   \int d^4p_{c}\, \frac{1}{2i}\,  p^0_c \, \delta^2(\hat p_{c}-\hat n)\, {\rm Disc}_{M^2}\, T_{\gamma^*c'\,\rightarrow \,{\gamma'}^*c}. \label{eq:Energy}
\ee

This is related to the momentum space representation for the correlator $\langle {\cal O}(p) |  \widetilde  O_E (p_c) |{\cal O}(p)\rangle$; however, this is kinematically related to a position space three-point Wightman function of fields, $\langle \phi_1(x) \phi_2(y) \phi_3(z)\rangle$. Similarly, the two-point energy correlator $\langle {\cal O}(p) |  \widetilde  O_E(1)  \widetilde  O_E(2) |{\cal O}(p)\rangle$ is related to a position-space four-point function, and so on. It is therefore desirable  to explore directly conformal invariance for Eq. (\ref{eq:Hofman2}) in a coordinate representation, as initiated in \cite{Hofman:2008ar,Belitsky:2013xxa,Belitsky:2013bja,Belitsky:2013ofa}. 
 
A similar analysis holds for higher order correlators $\langle \widetilde  O_w(1) \widetilde O_w(2)\cdots \rangle$, where we now have 
\be
\sigma_w(\hat n_1,\hat n_2,\cdots)= \sum_{c_1,c_2,\cdots}   \int {d^4p_{c_1}}\int {d^4p_{c_2}}\cdots \, \frac{1}{2i}\,  w(p_{c_1},p_{c_2},\cdots) {\rm Disc}_{M^2} \, T_{\gamma^*c_1'c_2'\cdots\,\rightarrow \,{\gamma'}^*c_1c_2\cdots}\, .
\ee

\section{String-Gauge Duality}\label{sec:Mueller-d} 

In this section we discuss scattering via the AdS/CFT correspondence with a particular focus on scattering in the gravity theory. We first review only the essentials of scattering in AdS space needed to understand our phenomenological model and arrive to Eq. (\ref{eq:ads6pt}). Our discussion revolves around the scattering of AdS states and stringy effects beyond the super-gravity limit of $\lambda\rightarrow \infty$. A detailed dual description in terms of the $\mathcal{N}=4$ SYM theory, while interesting and informative, is not needed for our current application.  We stress that stringy effects are not only conceptually important but also phenomenologically necessary.  Due to the difficulty of full finite $\lambda$ string calculations, scattering amplitudes are most easily formulated by starting with the infinite coupling limit and then calculating $1/\sqrt \lambda$ corrections in the context of $1/N_c$ expansion: we will treat stringy effects perturbatively.

We pay special attention to two kinematic limits where the consequences of stringy corrections can be seen easily. One limit of interest is that of fixed-angle scattering, which leads to ``ultralocal" scattering in the AdS bulk and hence in the Polchinksi-Strassler regime \cite{Polchinski:2001tt}. This is briefly reviewd in Appendix \ref{sec:BPST}. A second limit of interest is scattering in the near-forward limit which is discussed below. At high energy, the most important contribution to the AdS amplitude in this limit is due to the exchange of a graviton in the $t$-channel. However, this leads to too rapid an increase for amplitudes; stringy effects can slow the increase. In \cite{Brower:2006ea,Brower:2007xg} it was shown that this leads to the introduction of a ``reggeized" AdS graviton known as the BPST pomeron; this pomeron serves as the leading contribution to the scattering in a unitarized treatment via an eikonal sum. This framework can also be extended to  multi-particle near-forward scattering~\cite{Herzog:2008mu,Brower:2012mk,Brower:2014hta}, which paves the way for the treatment of central inclusive production in Sec. \ref{sec:CentralProduction}.

\subsection{AdS Scattering}
\label{sec:ads-scat}

The AdS/CFT correspondence relates $\mathcal{N}=4$ SYM correlation functions to a dual description in terms of correlation functions of string states in a higher-dimensional via an equivalance of partition functions.\footnote{The canonical description of the AdS/CFT correspondence describes string states living in $AdS_5 \times S^5$.  Here we are only concerned with excitations in the AdS space.} From the gravity perspective, CFT states can be thought of as propagating from a four dimensional boundary theory into the gravity bulk, scattering, and returning to the boundary CFT.  In the limit of large 't Hooft coupling, this process can be described with perturbative sums of ``Witten diagrams" in analogy to weak coupling descriptions. (See Appendices \ref{sec:BPST} and \ref{sec:ConformalRegge} for further clarification.)

For most of the following calculations, it is sufficient to work with the Poincare patch of $AdS_5$, described by the metric 
\be\label{eq:OriginalMetric}
ds^{2\,Poincare}_{AdS_5} = \frac{z^2}{R^2} \Big\{ -dt^2 + d\vec x^2 + dz^2\Big\},
\ee 
where $R$ is the AdS radius.  This metric corresponds to a boundary theory with purely conformal dynamics, as can be seen by comparing the five-dimensional AdS isometry group to the four-dimensional conformal group. The radius $R$ of the bulk geometry is related to the 't Hooft coupling $\lambda \equiv g_{YM}^2 N_c$ of the boundary gauge theory by $\lambda = (R/\ell_{string})^4 $, where $\ell_{string} = \sqrt{\alpha'}$ is the string length. Therefore, the limit $\lambda\to\infty$ of strong boundary coupling corresponds to a weakly curved bulk geometry, and hence weakly coupled bulk dynamics. In these coordinates, $z\rightarrow 0$ and  $z\rightarrow \infty$ correspond to the UV and IR of the dual gauge theory, respectively.

However, we will also be interested in deforming away from a strictly conformal boundary limit, by introducing a confinement scale in the boundary theory. There are a variety of approaches to introducing a confinement deformation in AdS space \cite{Andreev:2006ct,Kapusta:2010mf,Aharony:2012jf,Witten:1998zw,Gursoy:2008za,Karch:2006pv,deTeramond:2005su,Gursoy:2007cb,Gursoy:2007er,Batell:2008zm}, but we are interested in universal features that are common to all the approaches.  Generically, a confining gauge theory has a bulk dual with metric \be ds^{2}=e^{A(z)} \Big\{ -dt^2 + d\vec x^2 + dz^2\Big\},
\ee where $A(z)$ describes both the AdS warping and the deformation away from pure AdS.  Sometimes, as in the so-called ``hard wall" models of QCD, the coordinates are restricted to lie in finite intervals. The presence of a confinement deformation introduces a new length scale $\Lambda^{-1} \gg R$; we take $\Lambda\sim \Lambda_{QCD}$.

For concreteness, in most of the rest of the discussion we will assume a hard wall deformation, where we put in a hard IR cutoff by restricting the AdS radial coordinate $z$ to lie in the interval $[0,z_{max}]$. Then the confinement scale $\Lambda$ is given by $\Lambda \sim z_{max}^{-1}$. However, we expect our main results to be essentially independent of exactly which confinement deformation is used, since they depend essentially on the conformal UV dynamics.

A  connected Green's function $\tilde G_F$ in the boundary theory can now be expressed in terms of an amplitude  in the AdS bulk via a convolution 
\be
\tilde G_F(p_1,p_2, \cdots) =\int\cdots\int  \Pi_i \{d\mu (z_i) {\cal G}_i (p_i,z_i)\}  {\cal T}_n (p_1,z_1, p_2,z_2, \cdots),\label{eq:Witten} 
\ee
where $d\mu(z) = dz \sqrt {-g} $ and $g={\rm det} \, g$.   ${\cal T}_n$ can be considered as an ``amputated Green's function", and   $ {\cal G}(p,z)$ is the bulk-to-boundary propagator, which, for a scalar of conformal dimension $\Delta$,  is given up to a normalization factor in terms of Bessel function of the second kind,
\be
{\cal G}(p,z)= z^2K_{\Delta-2}(z\sqrt{-p^2}) = z^{-\Delta} \int_0^\infty dx x^{\Delta-1} \frac{J_{\Delta-2}(x) }{x^2-z^2p^2}.\label{eq:AdS-B-Bdry-propagator}
\ee  
 We will not provide here a detailed  discussion on the Witten diagram expansion here except for several remarks, which will become relevant shortly.

Consider first the bulk-to-bulk Feynman propagator $\langle 0| {\bf T}\{\phi(x,z)\phi( x',z')\}|0\rangle$ of a scalar with conformal dimension $\Delta$. Its momentum representation, which will be designated  as $G_F(z,z',p_\mu)$, can again be expressed  in terms of Bessel functions as
\be
G_F(z,z',p^2)=(zz')^2 \int_0^\infty kdk \frac{J_{\Delta-2}(kz) J_{\Delta-2}(kz')}{k^2-p^2}. \label{eq:AdSpropagator}
\ee
Since there is no mass gap, $G_F(z,z',p^2)$ is analytic in $p^2$, with a branch cut over $0\leq p^2<\infty$. Its discontinuity over the branch cut, which corresponds to the momentum-space  representation  for the Wightman function $G_W(x,z; x',z')= \langle 0| \phi(x'z')\phi( x,z)|0\rangle$, is
\be
G_W(z,z',p^2)
=\frac{\pi}{2}(zz')^2  J_{\Delta-2}(pz) J_{\Delta-2}(pz')\theta(p^2). \label{eq:AdSWightman}
\ee 
In the limit of $z,z'\rightarrow 0$, it approaches, up to a normalization constant,  the Wightman function in Eq. (\ref{eq:Wightman2}). 

\paragraph{Confinement Deformation in the IR, Universality and Conformal Invariance:} 
 Let us return to the issue of on-shell amplitudes. For CFTs, associated with each leg of the Green's function $G_n$ is an off-shell wave-function, $e^{ip_\mu x^\mu}$, and a bulk-to-boundary propagator, $G(x',z'; z,x)|_{z'\rightarrow 0}$. In order to define on-shell amplitudes,  it is necessary to introduce a confinement deformation in the IR leading to finite a mass gap. A new dimensionful scale, $\Lambda^{-1}>> R$, enters serves as the basic length scale. Conformality holds for $z<< \Lambda^{-1}$. Conversely, confinement effect becomes important if $z\sim \Lambda^{-1}$, with $\Lambda$ expected to be of the order $ \Lambda_{QCD}$.  
	
In such a scenario, on-shell amplitudes are given by amputated Green's functions, which have a normal singularity structure as in standard flat space field theories.  After the introduction of a confinement deformation in the IR, the spectrum of the bulk theory becomes discrete, so that the propagator in Eq. (\ref{eq:AdSpropagator}) is replaced by a discrete sum, 	
\be
	G_F(z,z',p^2)\rightarrow 
	\sum_n \frac{ \phi_n(z) \phi_n(z')}{m_n^2-p^2}, \label{eq:scalarB2B}
	\ee
where the $\phi_n(z)$ are a set of orthonormal wave functions associated with an infinite set of scalar glueballs of increasing mass $m_n$\footnote{ These states also interpolate with higher spin glueball states on the same Regge trajectories, leading to the reggeized $J$-dependent propagator appearing in Eq. (\ref{eq:ReggeProp}).} More importantly, the bulk-to-boundary propagator in Eq. (\ref{eq:AdS-B-Bdry-propagator})  is also  given by a discrete sum,  
\be
{\cal G}(p,z)\rightarrow \sum_n  \frac{ c_n \phi_n(z)}{m_n^2-p^2}, 
\ee
with poles at $p^2=m_n^2$. This in turns allows us to extract on-shell amplitudes in a standard manner.

Although our discussion will turn to theories with an IR confinement deformation, there are features of the Witten diagram expansion that are model independent. As stressed in \cite{Brower:2006ea,Brower:2007xg}, it is possible to identify features which depend only on the conformal structure, such as the large $Q^2$ behavior of DIS at small-$x$. We stress here the  important fact that AdS wave functions have universal behavior~\footnote{In the hard wall model, the glueball wave function has  $\phi(z)\propto z^2 J_{\Delta-2}(m_n z)\sim z^\Delta$ as $z\rightarrow 0$. A similar explicit analytic expression can also be obtained for other deformations, such as the ``soft wall" model.} in the UV.  As $z\rightarrow 0$,
	\be
	\phi_i(z)\simeq z^{\tau_i}, \quad \tau_i= \Delta_i-J_i \, ,\label{eq:conformalWF}
	\ee
where $\tau$ is the twist and $J$ is the spin.  This behavior  is independent of the confinement deformation and depends only on the conformal properties. We shall make use of this fact when implementing the Polchinski-Strassler mechanism for large $p_\perp$ production.

It is now possible to define scattering amplitudes  as  amputated Green's functions by going on to the pole for each external state, leading to on-shell scattering amplitudes, 
	\be
	T_n(p_1,p_2, \cdots) =\int\cdots\int  \Pi_i \{d\mu (z_i) \phi_i (z_i)\}  {\cal T}_n (p_1,z_1, p_2,z_2, \cdots)\, .\label{eq:Witten2}
	\ee
For each external on-shell particle, one associates a bulk wave-function 
	$
	e^{-ipx} \phi (z)
	$.
This  can also be extended to multi-particle inclusive productions which we will turn to shortly.

\paragraph{High Energy Limit:}
In this paper, we will be primarily be interested in inclusive processes due to scattering at high energies where the source  is in general non-local. One therefore will  deal with $(2n)$-point functions for $n=2,3,\cdots.$ It is interesting to note that non-trivial dynamics already occur at the lowest level, for example the $\gamma^*p$ total cross section~\cite{Brower:2010wf,Costa:2012fw,Costa:2013uia}.  More generally, an inclusive discontinuity  can be taken through Witten diagrams in a momentum representation. This can be done  most readily for near forward scattering at high energy in the Regge limit.

\begin{figure}[ht]
	\begin{center}
		\includegraphics[scale=1]{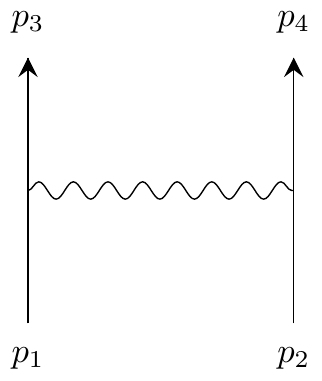}
		\caption{Schematic representation of high-energy elastic two-to-two scattering. The internal line represents the Pomeron kernel defined in Eq. (\ref{eq:Pomeron-Mellin}). } 
		\label{fig:2to2}
	\end{center}
\end{figure}

There exists a rather extensive literature on the applications of AdS/CFT to high energy near-forward scattering \cite{Brower:2007qh,Brower:2007xg,Cornalba:2006xm,Costa:2012fw,Brower:2014wha,Anderson:2014jia,Anderson:2016zon}. The factorization of AdS amplitudes has emerged as a universal feature, present in the scattering of both particles and currents. The amplitude for elastic two-to-two scattering can be represented schematically in a factorized form as 
\be
T_{2\rightarrow 2} = \Phi_{13}*\widetilde {\cal K}_P * \Phi_{24},
\label{eq:adsPomeronScheme}
\ee
where $\Phi_{13}$ and $\Phi_{24}$ are elastic vertices and the convolution, $*$, involves an integration of the vertex position over the AdS bulk, as in  Eqs. (\ref{eq:adsGraviton}-\ref{eq:bulk-integral}).  This can be seen in Fig. \ref{fig:2to2}.  The Pomeron kernel $\widetilde {\cal K}_P$, described in more detail in Appendix \ref{sec:BPST}, is defined as 
\bea
\widetilde{\cal K}_P(s,t,z,z') &=&
-  \int_{L-i\infty}^{L+i\infty} \frac{dj}{2\pi i} (
\alpha'  \widetilde s)^{j} \frac{1 + e^{-i \pi j}}{\sin\pi j}   \widetilde G_j(t,z,z') \; ,   \label{eq:Pomeron-Mellin}
\eea

where the reggeized graviton propagator $ \widetilde G_j(t,z,z')$ is defined in Eq. (\ref{eq:spectrumq3}).

Through AdS/CFT, one can identify the Pomeron with a reggeized graviton in the AdS bulk. The Pomeron kernel  $\widetilde {\cal K}_P$ can be introduced by perturbing about the super-gravity limit through a world-sheet OPE.  More formally, one can introduce a Pomeron vertex operator in AdS,  as  done in \cite{Brower:2006ea}, so that
\be
T_{2\rightarrow 2} = \langle \phi_1\phi_3| {\cal V}_P\rangle \Pi(\alpha_P) s^{\alpha_P} \langle {\cal V}_P|\phi_2\phi_4\rangle  \, , \label{eq:phase}
\ee where $\Pi(\alpha_P) $ is a complex ``signature factor" carrying information about its phase that is useful for taking the discontinuity. It is customary to normalize this signature factor as $\Pi(j)=\frac
{1 +  e^{-i \pi j } }{\sin\pi j } $  so that  ${\rm Im}\, \Pi(j)=1$.

\subsection{Inclusive Cross Sections as AdS  Discontinuities} 
\label{sec:incl-ads}

In Eq. (\ref{eq:Witten2}), we have expressed on-shell scattering amplitudes $T_n$ in the boundary theory in terms of scattering amplitudes ${\cal T}_4 (p_1,z_1, \cdots)$ in the AdS bulk.  We can now extend this treatment to inclusive cross sections.  After applying Eq. (\ref{eq:Optical4}) to Eq. (\ref{eq:Witten2}) for $n=4$, we find that the cross section for $a+b\rightarrow X$ is given in terms of a bulk amplitude as 
\be 
\sigma_{total}=(1/s) \int   \{\Pi_{i=1-4} d\mu (z_i) \phi_n (z_i)\}\, {\rm Im} \,{\cal T}_4 (p_1,z_1, \cdots). 
\ee 
Similarly, by applying Eq. (\ref{eq:Optical6b})  to Eq. (\ref{eq:Witten2}) for $n=6$, we find that the differential inclusive cross section for $a+b\rightarrow c +X$ is given by 
\be \label{eq:ads6pt}
\frac{d \sigma_{ab\rightarrow c+X}}{d^3{\bf p}_c/E_c} \simeq \frac{1}{2i s} \, \int   \{\Pi_{i=1-6} d\mu (z_i) \phi_n (z_i)\} \,{\rm Disc}_{M^2>0} \{{\cal T}_{abc'\rightarrow a'b'c}(p_i,z_i)\} \, . 
 \ee 
 Both of these discontinuities are taken in appropriate forward limits.

This key result  can also be extended to multi-particle inclusive production.  
As an explicit illustration, consider  the case of DIS.  Here one first  replaces $\Phi_{13}$ in   Eq. (\ref{eq:adsPomeronScheme})
by the appropriate product of propagators  for external currents~\cite{Polchinski:2002jw,Brower:2010wf}.  One next performs  the step  of  taking discontinuity~\cite{Brower:2010wf,Costa:2012fw,Costa:2013uia,Brower:2015hja,Nishio:2011xz,Koile:2015qsa}, leading to a factorized form for the cross section:
\be
\sigma^{total}_{\gamma^*p}\simeq  \frac{1}{s}    \Phi_{13} * [{\rm Im}\, \widetilde  {\cal K}_P] * \Phi_{24} \; . \label{eq:DIScsection}
\ee

 For the general two-to-two scattering of scalar glueballs, $1+2\rightarrow 3+4$,  one has
\be
{\rm Im} \, T_{2\rightarrow 2}(s,t=0) = \int d\mu(z)\int d\mu(z') \Phi_{13}(z) \, {\rm Im} \, \widetilde {\cal K}_P(\widetilde s,0,z,z') \, \Phi_{24}(z'),
\label{eq:adsPomeronScheme2}
\ee
where the vertex coupling
$ \Phi_{ab}(z)$
involves the normalized wave-function $\phi_a\left(z\right)$ of scalar glueball of conformal dimension $\Delta$. As indicated earlier, for a hard-wall deformation, we have $\phi_a\left(z\right)\sim z^2 J_{(\Delta-2)}(m_a z)$ and $\int d\mu(z) z^2 \phi_a(z) \phi_b(z) = \delta_{a,b}$. Similarly, the reggeized Pomeron  kernel $\widetilde{\cal K}_P$ in Eq. (\ref{eq:PomeronKernel}) can be given a more explicit form in the hard wall model \cite{Brower:2006ea,higgs1}.  We will not discuss this propagator in detail, except to note that its phase information is given by Eq. (\ref{eq:phase}). The propagator has a discontinuity in  $\widetilde s$, with its leading behavior given by 
\be
\Disc_{s} \widetilde{\cal K}_P\left(\widetilde s, 0,z,z'\right) \propto {\widetilde s}^{j_0}\, , \label{eq:disckg}
\ee
with $j_0\simeq 2-2/\sqrt \lambda$. In the  particular case of DIS, this leads to Eq. (\ref{eq:moments}), with anomalous dimensions for $j\simeq 2$ at strong coupling given by 
\be
\gamma(j) = \sqrt{2\sqrt \lambda( j-j_0)} -j  + O(\lambda^{-3/4})\, .
\ee

The results of this section rely on identifying that the analytic structure of amplitudes in AdS/CFT are analogous to that of Field Theory.  However, in the next section we turn our attention to a specific high energy process appropriate for collider scattering.  Here, collisions with large transverse momentum will be localized in a transverse space and in the Polchinski-Strassler regime; we can consider flat-space string vertices with physical momenta red shifted by the geometry.  This is more fully explored in \cite{Brower:2006ea,Herzog:2008mu} and analogous situations for specific collider physics are described in \cite{Brower:2010wf,Brower:2012kc,Brower:2012mk,Costa:2012fw,Anderson:2014jia,Brower:2014hta,Brower:2014sxa,Brower:2015hja,Anderson:2016zon}.

\section{Inclusive Single-Particle Production in the Central Region}\label{sec:CentralProduction}

We have shown that the inclusive single particle production cross section  in the boundary theory can be related to the discontinuity of the six-point amplitude in the bulk. In order to evaluate this discontinuity, we must generalize the treatment of two-to-two amplitudes given above to apply to three-to-three amplitudes. We begin by discussing the kinematics of inclusive production.

For fixed $X$, the inclusive processes $a+b\rightarrow c+X$ can be treated kinematically as a two-to-two process where we treat $X$ effectively as a particle with mass 
\be 
M^2 = \left(p_a+p_b-p_x\right)^2.
\ee 
Thus, in addition to $M^2$, we have the usual three Mandelstam invariants 
\be
s=(p_a+p_b)^2,\quad t=(p_a-p_c)^2, 
\quad  u=(p_b-p_c)^2. \label{eq:6-pt-kinematics-a}
\ee
These invariants are related by the constraint 
\be
M^2= s+t+u-m_a^2-m_b^2-m_c^2\, .\label{eq:6-pt-kinematics-b}
\ee 
Therefore, the kinematics can be parameterized by three invariants, which can be  taken to be $(s,t,M^2)$~\footnote{For central production, it  turns out to be  more convenient to use $(M^2, t, u)$ as independent variables, as explained below.}.

However, there exists an alternate parameterization that can better illuminate the simplicity of the actual process.  A universal characteristic of high energy particle production is the fact that the majority of produced particles will have small transverse momentum relative to the (longitudinal) incoming direction. In a typical hadronic collision at the LHC, the detector essentially sits at rest in the center of momentum frame of the two incoming particles, which have equal and opposite large momentum; these momenta define a longitudinal light cone (LC) direction.  To be more explicit, we choose the incoming particles $a$ and $b$ to have LC momenta 
$p_a =(p^+_a,p^-_a,\vec{p}_{\perp,a})= (m_a e^{Y/2}, m_a e^{- Y/2},0)$ and $  p_b=(p^+_b,p^-_b,\vec p_{\perp,b})=(m_b e^{-Y/2},  m_b e^{Y/2},0)$, where $Y$ is the rapidity. Then, taking $m_a=m_b=m$ for simplicity, the Mandelstam $s$ invariant is given by $ s\sim m^2 e^Y$,  and the produced particle has LC momentum given by
\be
p_c =  (m_{\perp} e^{y},m_{ \perp} e^{-y},\vec p_{ \perp })\,,  \quad m_\perp^2\equiv {m_c^2 + \vec p_\perp^2}.
\ee
Equivalently, the produced particle  has  energy $E= m_\perp \cosh y$ and longitudinal momentum  $p_L = m_{\perp}\sinh y$ so that $y=\ln [(E+p_L)/(E-p_L)]$.

Inclusive central production involves particles with fixed $\vec p_\perp$ and $y$ in the CM frame in the $s\to\infty$ limit, and therefore incoming particles have large rapidities, $-y_b \simeq y_a = Y\to\infty$. In such an event, the produced particles can be grouped in an intuitively helpful way as $a+b\rightarrow X_1 + c + X_2$, where $c$ is the centrally produced particle and $X_1$ and $X_2$ are left- and right-moving particles, respectively.  In this limit, the traditional Mandelstam variables behave as 
\begin{subequations}
\begin{alignat}{3}
 s& \simeq  \,\, M^2 \,\, \simeq m^2 e^Y&&  \rightarrow &&+\infty\\ 
 t&\simeq  -mm_\perp e^{Y/2-y}&& \rightarrow &&-\infty\\ 
 u&\simeq -mm_\perp e^{Y/2+y}&& \rightarrow &&-\infty.
 \end{alignat}
\end{subequations}
We can additionally check that the ratio 
\be
\kappa \equiv  \frac{(-t)(-u)}{M^2} \simeq m_\perp^2=m_c^2+ p_\perp^2  \label{eq:kappa}
\ee
is fixed.  These kinematic conditions can be thought of as the definition of central production.  Phenomenologically, we often prefer to use $(s, y, p_\perp^2)$ as the three independent variables describing the kinematics of central production at the LHC.  On the other hand, when we take the discontinuity in the 3-to-3 amplitude, we will see that it is more convenient to parameterize the kinematics with $(M^2,t,u)$, and to therefore treat $s$ as a dependent variable. We will return to this issue shortly.

\subsection{Inclusive Central Production  and the 3-to-3 Amplitude}

\begin{figure}[ht]
\begin{center}
\includegraphics[scale=1]{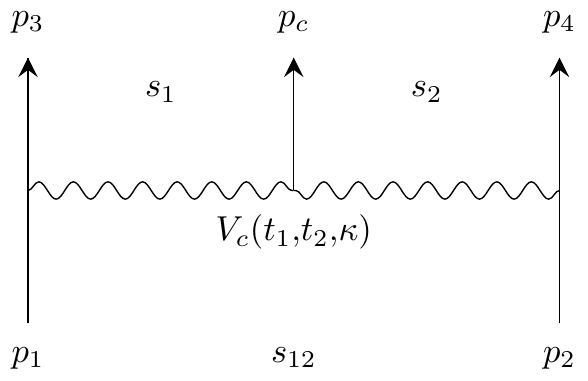}
 \caption{A schematic representation of the factorized two-to-three amplitude in Eq. (\ref{eq:adsDoublePomeronScheme}).}
 \label{fig:2to3}
 \end{center}
\end{figure}

A holographic analysis of the 2-to-3 amplitude in the double Regge limit was performed in \cite{Herzog:2008mu} by generalizing the AdS treatment of 2-to-2 scattering. 
Schematically, this 2-to-3 bulk amplitude can be represented by 

\be
T_{2\rightarrow 3}  = \Phi_{13}*  \widetilde{\cal K}_P*V_c*  \widetilde {\cal K}_P* \Phi_{24},
\label{eq:adsDoublePomeronScheme}
\ee
where we have introduced a new 3-point central production vertex, $V_c$, shown in Fig. \ref{fig:2to3}.  
In terms of the Pomeron vertex operator, $V_c$ can be expressed as $V_c = \langle {\cal V}_P| \phi_c| {\cal V}_P\rangle.$ These AdS vertex operators are closed string operators where the invariants are redshifted. In general these can be complicated expressions. However, following the analysis ~\cite{Herzog:2008mu,Brower:2012mk,Brower:2014hta,Anderson:2014jia,Anderson:2016zon}, many of the general features are shared with the much simpler flat space string theory vertex operators which we review in Appendix ~\ref{sec:FlatString}.

\begin{figure}[ht]
\centering
\includegraphics[scale=1]{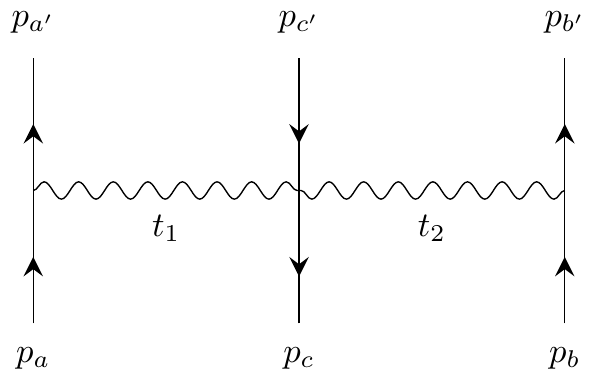}

\caption{A schematic representation of the factorized three-to-three amplitude in Eq. (\ref{eq:ads3to3PomeronScheme}).}
\label{fig:feyn6}
\end{figure}

We now move on to the six-point function, which was discussed for flat-space string scattering in \cite{Detar:1971dj,Virasoro:1971zq,Gordon:1971ss}. Following the above discussion and the logic in \cite{Herzog:2008mu,Brower:2012mk,Brower:2014hta}, we will be interested in the limit where the three-to-three amplitude takes on a factorized form, given by \be
T_{abc'\rightarrow a'b'c}= \Phi_{13}*  \widetilde{\cal K}_P*V_{c\bar c}*  \widetilde {\cal K}_P* \Phi_{24}.
\label{eq:ads3to3PomeronScheme}
\ee
Again we have had to introduce a new central vertex, $V_{c\bar c}$, shown in Fig. \ref{fig:feyn6}, which  can formally be expressed  as the matrix element involving two pomeron vertex operators
\be
V_{c\bar c}(\widetilde\kappa,\widetilde t_1,\widetilde t_2) = \langle {\cal V}_P| \phi_c \phi_{\bar c}| {\cal V}_P\rangle.
\ee

Following the flat space calculation in~\cite{Detar:1971dj,Virasoro:1971zq,Gordon:1971ss}, we can take the $M^2$ discontinuity in the amplitude to find that~\cite{Weis:1974ba}    
\be
(1/2i) {\rm Disc}_{M^2} T_{abc'\rightarrow a'b'c}= \Phi_{13}* [{\rm Im}\, \widetilde{\cal K}_P] * [{\rm Im} \,V_{c\bar c}] *  [{\rm Im}\,\widetilde {\cal K}_P] * \Phi_{24}\, . \label{eq:gptdisc}
\ee

As in two-to-three scattering, this AdS-space central vertex $V_{c\bar c}(\tilde \kappa,\tilde t_1,\tilde t_2)$  has the same functional form  as the flat space vertex, 
\be
{\cal V}_{c\bar c}= \int_0^1 \frac{d y }{y^{\alpha_{\bar a a \bar c}+1} (1-y)^{\alpha_{b\bar b c}+1}  } 
 V_c(t_1,t_2, \frac{\alpha' \kappa}{y(1-y)}), \label{eq:6pt-kappa}
\ee
 but with the arguments appropriately redshifted. (We follow here notation of~\cite{Detar:1971dj}. See Appendix \ref{sec:FlatString} for more details.) The invariant  $\kappa$ was defined in Eq. (\ref{eq:kappa}), and can also be expressed as
  \be
\kappa \simeq  \frac {(-t) (-u)}{M^2}, 
\ee
where  $t\equiv s_1<0$ and $u\equiv s_2<0$ and $M^2$  are defined in Eqs. (\ref{eq:6-pt-kinematics-a}) and  (\ref{eq:6-pt-kinematics-b}). The singularity of $T_{abc'\rightarrow a'b'c}$ in $M^2$ now appears only as a singularity of $V_{c\bar{c}}$ in $\kappa$, with discontinuity  given by
\be
{\rm Im}\,\, {\cal V}_{c\bar c}(\kappa, t_1,t_2) =\int_0^1 \frac{d y }{y^{\alpha_{\bar a a \bar c}+1} (1-y)^{\alpha_{b\bar b c}+1}  }  {\rm Im} \,V_c(t_1,t_2, \frac{\alpha' \kappa}{y(1-y)})\, . \label{eq:6pt-kappa-disc}
\ee
  At $t_1=t_2=0$, $\alpha_{\bar a a \bar c}(0)=\alpha_{b\bar b c}(0)=0$, for external tachyons, with ${\rm Im}\, {\cal V}_{c\bar c}(\kappa, 0,0)$ finite.

We can now explicitly write out the bulk six-point amplitude. Putting everything together,  Eq. (\ref{eq:ads3to3PomeronScheme}) can be expressed  as
\bea
&&T_{abc'\rightarrow a'b'c}\left(\kappa,s_1,s_2,t_1,t_2\right) \nn
& &= \frac{g_0^2}{R^4}\int_0^\zm dz_1 \sqrt{|g(z_1)|}[z_1^2\phi_a\left(z_1\right)\phi_{{a'}}\left(z_1\right)]  \int_0^\zm dz_2\sqrt{|g(z_2)|} [z_2^2 \phi_{{b'}}\left(z_2\right)\phi_b\left(z_2\right)]\nn
 &&\times \int_0^\zm dz_3\sqrt{|g(z_3)|} \, \widetilde{\cal K}_P\left(-\t{s}_1,\t{t}_1,z_1,z_3\right) \,
 I(\t{\k},\t{t}_1,\t{t}_2,z_3)\, \widetilde{\cal K}_P \left(-\t{s}_2,\t{t}_2,z_2,z_3\right), \label{eq:T6}
\eea
where  the dependence on the central vertex is collected as 
\be
I(\t{\k},\t{t}_1,\t{t}_2,z_3) = (z_3^2\phi_c\left(z_3\right)) V_{c\bar{c}}\left(\t{\k},\t{t}_1,\t{t}_2\right)  (z_3^2\phi_{c'}(z_3))\, .
\ee
In Eq. (\ref{eq:T6}), we have also introduced an explicit IR  cutoff, $z_{max}$, which should be  of the order $O(\Lambda_{QCD}^{-1})$; this amounts to implementing a hard wall confinement deformation.  It is essential that all Mandelstam invariants in this amplitude are holographic quantities, related to the flat space invariants  by the prescription in Eq. (\ref{eq:redshift}). For instance, $\tilde{s}_1<0$ and $\tilde{s}_2<0$ are given by 
$
\tilde{s}_1=\left(\tilde{p}_a-\tilde{p}_{{c}}\right)^2 = \left(\frac{z_1}{R}p_a-\frac{z_3}{R}p_{{c}}\right)^2  \sim \frac{z_1z_3}{R^2}s_1<0$ and 
$\tilde{s}_2 =\left(\tilde{p}_b-\tilde{p}_{{c}}\right)^2 = \left(\frac{z_1}{R}p_b-\frac{z_3}{R}p_{{c}}\right)^2  \sim \frac{z_1z_3}{R^2}s_2<0$.
Other  important holographic  invariants are 
\be  
\tilde{M}^2 = \left(\tilde{p}_a+\tilde{p}_b-\tilde p_c\right)^2 \sim \frac{z_1z_2}{R^2}M^2\, ,   \quad  {\rm and} \quad 
\tilde{\k} = \frac{\tilde{s}_1\tilde{s}_2}{\tilde{M^2}} \sim \frac{z_3^2}{R^2}\, \frac {s_1s_2}{M^2}= \frac{z_3^2}{R^2} \k.  \label{eq:holo33invts}
\ee
In this limit, we have $s\simeq M^2 >> |s_1|, |s_2|$.

Next we will compute the discontinuity in the missing mass $M^2$, given in Eq. (\ref{eq:gptdisc}), in the forward limit.  From Eq. (\ref{eq:gptdisc}), we see that, due to factorization, Eq. (\ref{eq:ads3to3PomeronScheme}), as schematically represented by Fig. \ref{fig:feyn6}, 
the discontinuities $\left[{\rm Im}\, \widetilde{\cal K}_P\left(-\t{s}_1,0,z_1,z_3\right)\right]$  and $\left[ {\rm Im}\, \widetilde{\cal K}_P\left(-\t{s}_2,0,z_2,z_3\right)\right]$ lead to the $z_3$ integral being entirely independent of $z_1$ and $z_2$. (See Eq. (\ref{eq:forwardPom}).) Thus, we can perform the $z_1$ and $z_2$ integrals to find an inclusive particle density $\rho$ for central production given by 
\bea
\rho(\vec p_T, y,s) &\equiv & \frac{1}{\sigma_{total}}\frac{d^3\sigma_{ab\to c+X}}{d\mathbf{p}_c^3/E} =\frac{1}{2i s\, \sigma_{total}(s)}\Disc_{M^2} T_6\left(\kappa,s_1,s_2,0,0\right) \nn
&=& \beta \, \int_0^\zm \frac{dz_3}{z_3} \,  \tilde \kappa^ {j_0}\,   [\phi_c(z_3)]^2 \left[{\rm Im} \,  {\cal V}_{c\bar c}\left(\t{\k},0,0\right)\right], \label{eq:disc3t6}
\eea
where  $\beta$ is an overall constant partially stemming from the $z_1$ and $z_2$ integrals.  This is our key result.

\subsection{Central Production at Large \texorpdfstring{$p_\perp$}{PT} and Conformal Invariance \label{sec:pwrlaw}}
It should be stressed that  Eq. (\ref{eq:disc3t6})  depends crucially on factorization in the double-Regge limit. 
In the factorization limit, the particle density is independent of both $y$ and $s$~\footnote{Saturation effects can cause dependence on these kinematics, which will be discussed briefly in Sec. \ref{sec:discussion}.}.  Conversely, the density depends on $p_\perp$ through the wavefunction $\phi_c(z)$ and the vertex ${\rm Im} \, {\cal V}_{c\bar c}(z^2\k /R^2, 0,0)$. Recall that the double Regge kinematics are such that $\kappa \simeq p_\perp^2+m_c^2$, and therefore that taking $p_\perp$ large is equivalent to working in the limit where $\kappa$ is large. We can then check that conformal dynamics emerge in this limit, as we saw above in the fixed-angle limit. 

In flat space string scattering,  the six-point central vertex  $\mathcal{V}_{c\bar c}(\kappa, 0,0)$ is an analytic function of $\kappa$, away from a branch cut along the positive real line. In the limit $\kappa\to\infty$, the discontinuity vanishes and the vertex becomes factorizable with an exponentially small imaginary part:  ${\rm Im}\, \mathcal{V}_{c\bar{c}}$ decays exponentially. From Eqs. (\ref{eq:6pt-kappa-disc}) and (\ref{eq:5pt-kappa-disc}), we have, for large $\kappa$, 
\be
{\rm Im}\,  {\cal V}_{c\bar c}\left(\k,0,0\right)\simeq  \pi (\alpha' \kappa)   \int_0^1 d y  e^{-\frac{ \alpha'\kappa}{y(1-y)}} \sim \sqrt{\alpha' \kappa}  e^{-4\alpha' \kappa}.  \label{eq:stringy}
\ee
This parallels the result for exclusive fixed-angle scattering in Eq. (\ref{eq:flatfixangle}). As emphasized in \cite{Detar:1971dj,Virasoro:1971zq,Gordon:1971ss}, this exponential suppression reflects the ``softness" of flat-space string scattering.

When the scattering occurs on an AdS background, the large $\kappa$ asymptotics are rather different. The redshifted vertex is now \be
{\cal V}_{c\bar c}\left(\widetilde \k,0,0\right)\sim  e^{-2\alpha'\k z^2/R^2}\sim e^{-2  (z^2/\sqrt \lambda)  \k,}
\ee where we have substituted $\alpha'\to\frac{1}{2}\alpha'$ to return to closed string scattering. Thus, the $z_3$ integrand picks up an exponential suppression for large $z_3$. This induces an effective cutoff $z_s$. We determine $z_s$ by demanding $2\alpha' \widetilde \k = O(1)$, so that
\be
z_s \sim \frac{R}{\sqrt{2\alpha'\k}}=  \frac{\lambda^{1/4}}{\sqrt{2\k}} \label{eq:zs}.
\ee

We can thus approximate Eq. (\ref{eq:T6}) by integrating only up to $z_3 = z_s<< z_{max}$, where the exponential factor is of order one and can be neglected. Additionally, since we are taking $\kappa\to\infty$, we can, following Eq. (\ref{eq:conformalWF}), approximate each wave-function by
$\phi(z)\simeq z^{\tau}$, where $\tau$ is the twist.  Thus Eq. (\ref{eq:disc3t6}) becomes 
\bea
\frac{1}{\sigma_{total}}\frac{d^3\sigma_{ab\to c+X}}{d\mathbf{p}_c^3/E_c} &=& \beta \int_0^{z_{s}} \frac{dz}{z} z^{2\tau_c}   ( \kappa z^2/R^2)^{j_0} e^{-(2  \kappa/\lambda^{1/2})  z^2}\nn
  &\simeq& {\beta' }\, \kappa^{-\tau_c}, 
\eea
where we have introduced a new normalization constant $\beta'$.  
 In the simplest model of bulk physics, the external particles labeled by $c$  are scalar glueballs and thus have  $\tau_c=\Delta_c  = 4$. We therefore have
 \begin{align}
\rho(p_\perp, y, s) = \frac{1}{\sigma_{total}} \frac{d^3\sigma_{ab\to X}}{d\mathbf{p}_\perp^2 dy } \sim p_\perp^{-8} \label{eq:CrossSecPrediction}.
\end{align} 
This result follows essentially from conformality, since it depends on the behavior of the external wave functions away from the confinement region; our prediction does not depend on the details of the confinement deformation chosen. It serves as a generalized scaling law  for inclusive distribution, as is the case for exclusive fixed-angle scattering~\cite{Brodsky:1973kr,Brodsky:1974vy,Matveev:1973ra}.

\section{Evidence for Conformality} 
\label{sec:Test}

We have argued that conformal symmetry is manifested in the presence of power law behavior in inclusive scattering processes. We will now test this prediction by direct comparison to experimental results. We will focus on differential cross section measurements at high $\sqrt{s}$ performed at the LHC. Many recent measurements are in the form of a double differential cross section, in which particle production is binned both in the transverse momentum $p_T$ and the pseudorapidity $\eta$; symbolically, these studies measure the cross section $\frac{1}{2\pi p_T}\frac{d^2\sigma}{d\pt d\eta}.$ Here we are interested in the region where $p_T>\Lambda_{QCD}$ where $y\approx \eta$. In principle, this is not precisely the quantity we have computed above. However, as discussed in \cite{Aad:2016mok}, these two cross sections encode essentially the same information, so we expect essentially the same dependence on the kinematic variables. More concretely, we expect that the leading order physics should be independent of $\eta$, and that  the exponent of the power law should be independent of $s$. 

Our goal is to fit conformally motivated behavior to differential cross sections.  We will use our central results, Eq.(\ref{eq:disc3t6})-(\ref{eq:zs}), to model p-p ~\cite{Aad:2016xww,Aad:2016mok} and p-pb ~\cite{1405} central production via Eq.(\ref{eq:CrossSecPrediction}).  One of our assumptions from Sec.~\ref{sec:incl-ads} going into Eq.(\ref{eq:CrossSecPrediction}) is that the incident wave functions behave as $\phi_{a,b}(z)\approx z^2 J_{(\Delta-2)}(m_{a,b}z)$. This is consistent with hard and soft wall AdS confinement schemes where the wave function scale has been shown to be  $m_{a,b}\approx 1 GeV$, or the size of a proton. ~\cite{Brower:2010wf,Brower:2012kc,Costa:2012fw,Brower:2014hta,Brower:2012mk,Brower:2014sxa,Brower:2015hja}  Although no heavy ion studies have been done, we assume a similar wave function form holds for pb as well.  As described in Sec.~\ref{sec:pwrlaw}, the simplest model of bulk physics describes the production wave function, $\rho_c$, to be that of scalar glueballs which will hadronize into the detected charged particles.

As briefly described in ~\cite{Aad:2016mok}, the central production of charged particles in pp and pb collisions is inherently non perturbative.  Described by the kinematics of Sec.~\ref{sec:CentralProduction}, the inclusive central production is described via a color-singlet exchange (Pomeron) which dominates in the Regge limit.  The only current Monte Carlo (MC) methods used to describe this data involve a combination of multi-parton interactions involving single and double diffractive dissociation (including Pomeron and gluon effects), Gribov-Regge theory, and a "semi-hard" Pomeron model.  In this kinematical region, these MC methods agree on a description of the differential cross section, but vary in describing event track multiplicities and mean transverse momentum distributions.  For the p-pb collisions there is no current MC prediction.

At large $p_T$ our result implies that the differential cross section is described by the exchange of Reggeized objects leading to power law behavior depending on conformal dimensions. However, this behavior is only expected to hold at moderately high $\pt$ above the QCD scale.  At low $\pt$, much more complicated behavior can occur~\cite{Levin}.  Some of these low-$p_T$ effects stem ultimately from saturation, which,  from a string perspective, corresponds to the emergence of eikonal physics in summing over string-loop diagrams~\footnote{Eikonalization is also responsible for saturation in the context of DIS. More discussion will be provided at the end of this section.}.  Other effects will be sensitive to confinement specifics which are partially avoided at large $p_T$ from the AdS/CFT perspective~\cite{Brower:2015hja,Brower:2014sxa}. This is borne out in the data by deviations from power-law behavior at small $p_T$ as can be seen in Figure~\ref{fig:powerlowpt}. More details are given in Appendix~\ref{sec:Validation}.

\begin{figure}
\begin{tabular}{ccc}
\multicolumn{3}{c}{\includegraphics[width=65mm]{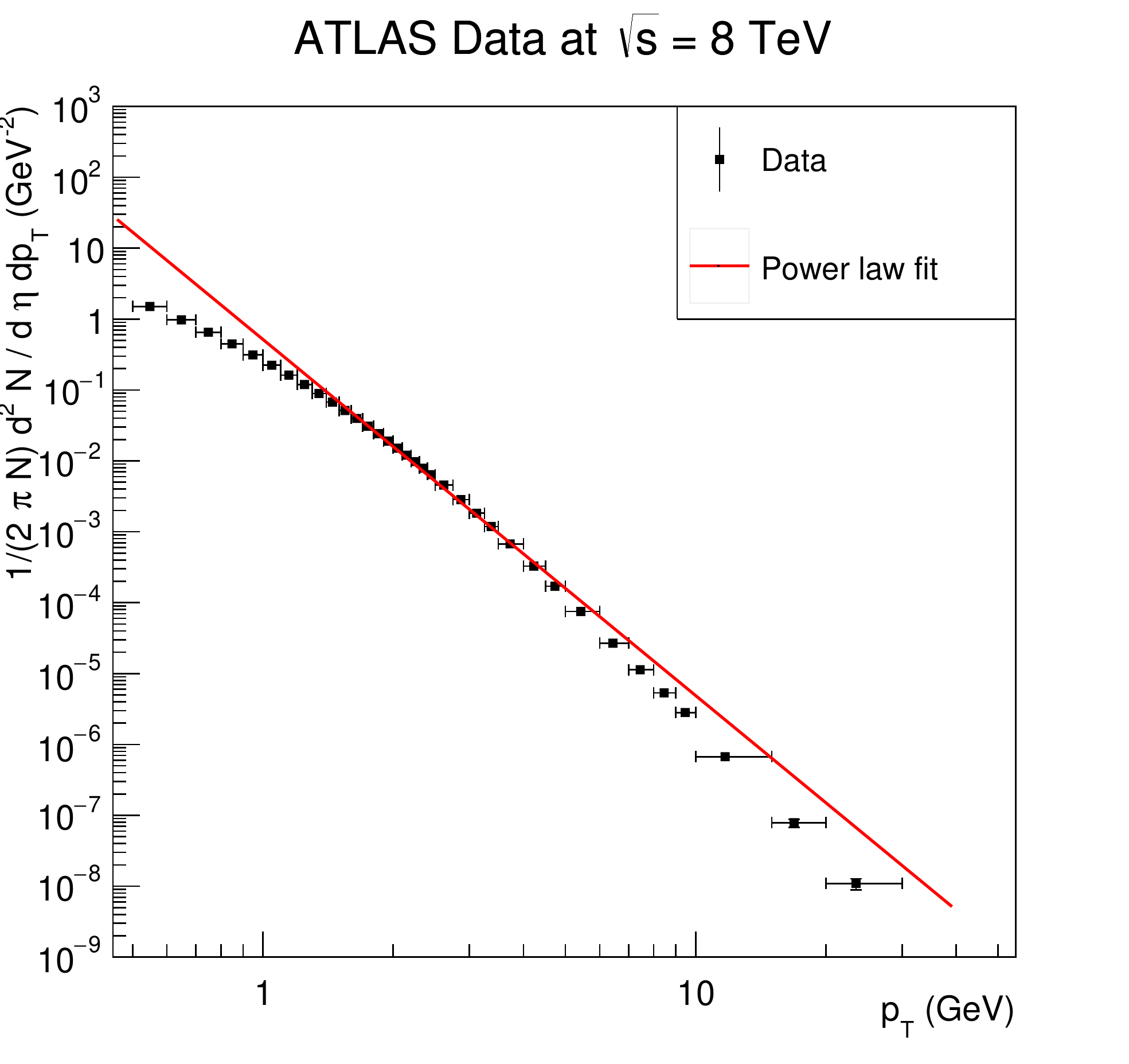} \includegraphics[width=65mm]{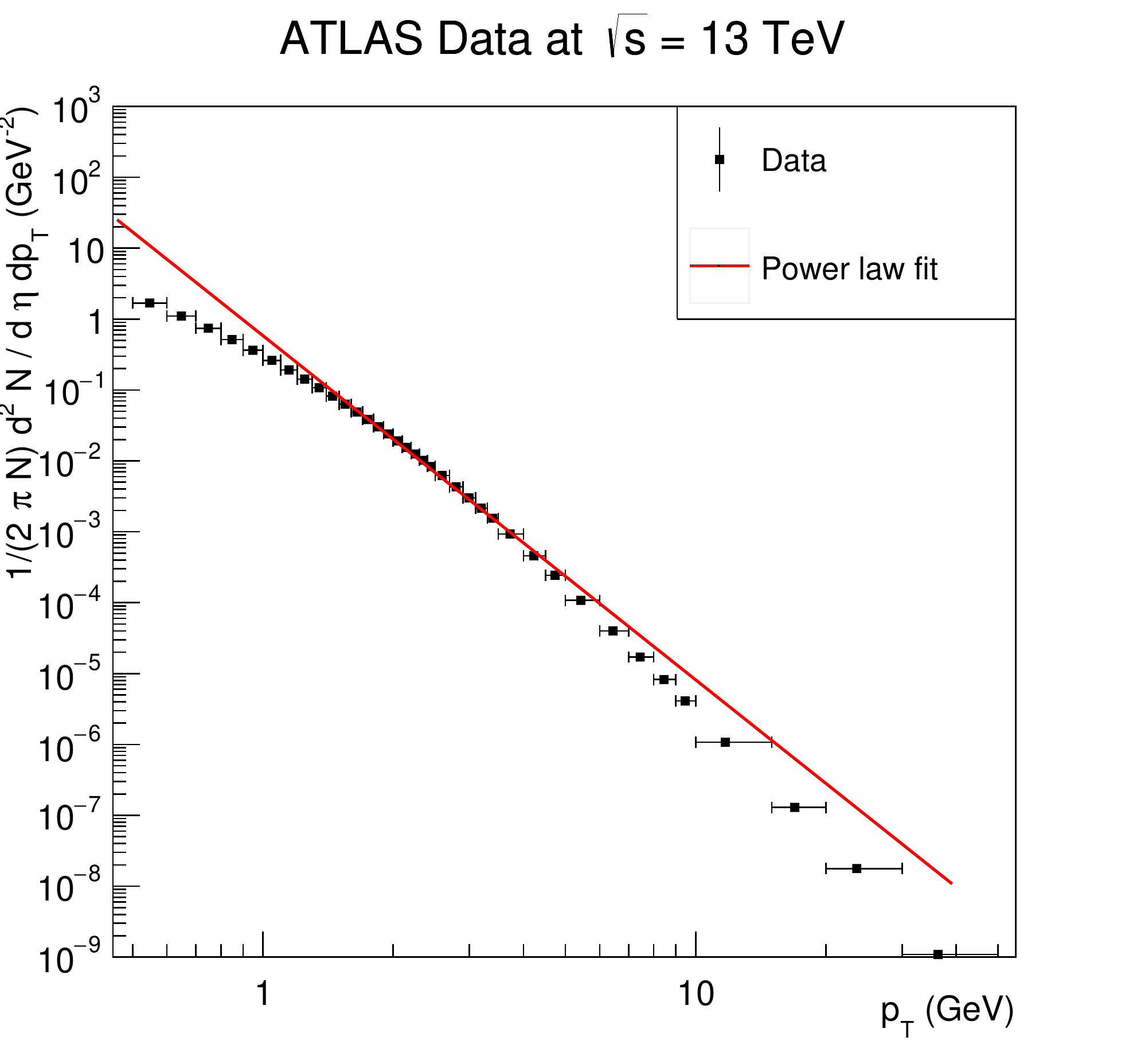}}\\
\multicolumn{3}{c}{(a)\hspace{6cm}  (b)}\\
 \includegraphics[width=50mm]{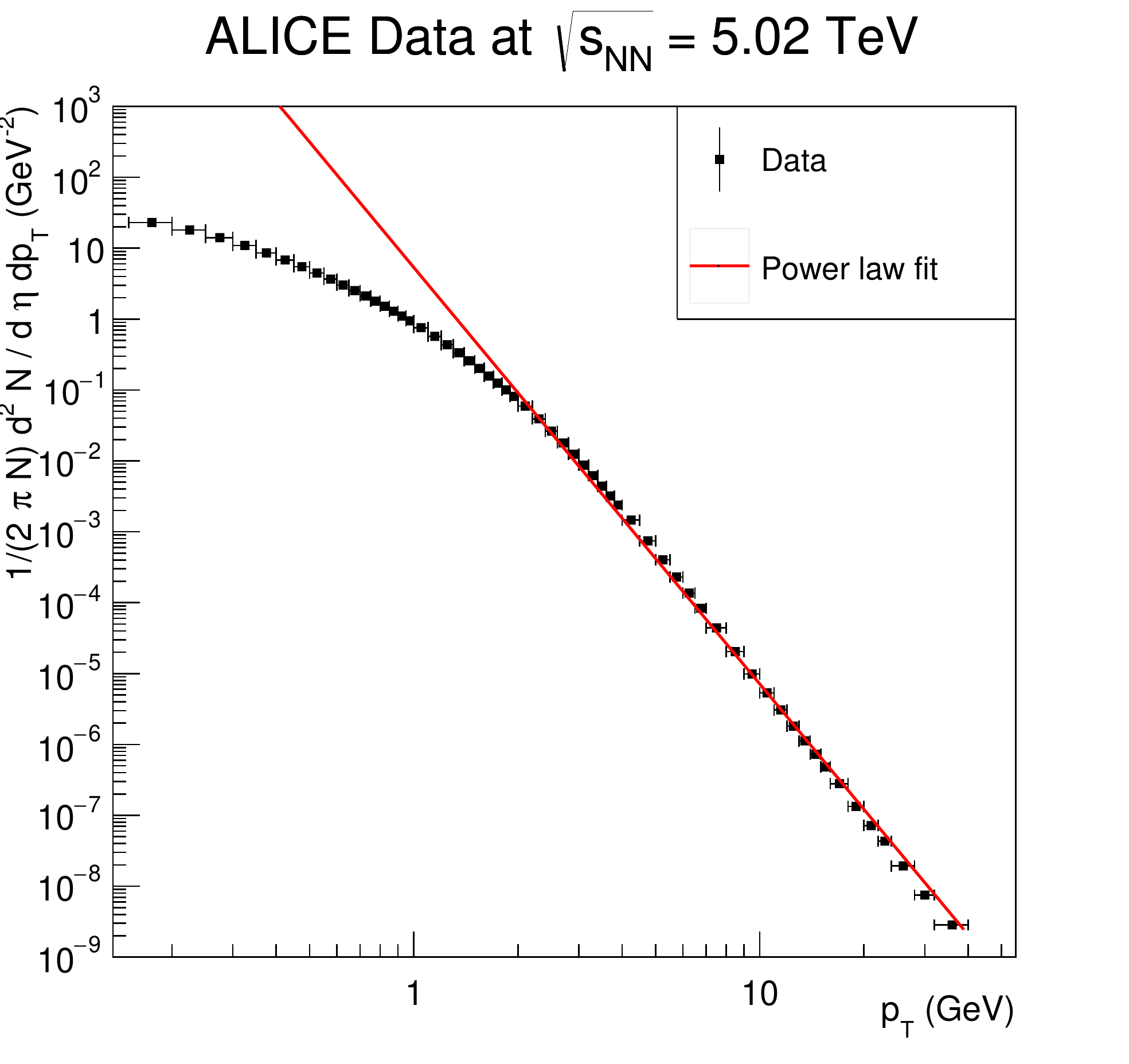} &   \includegraphics[width=50mm]{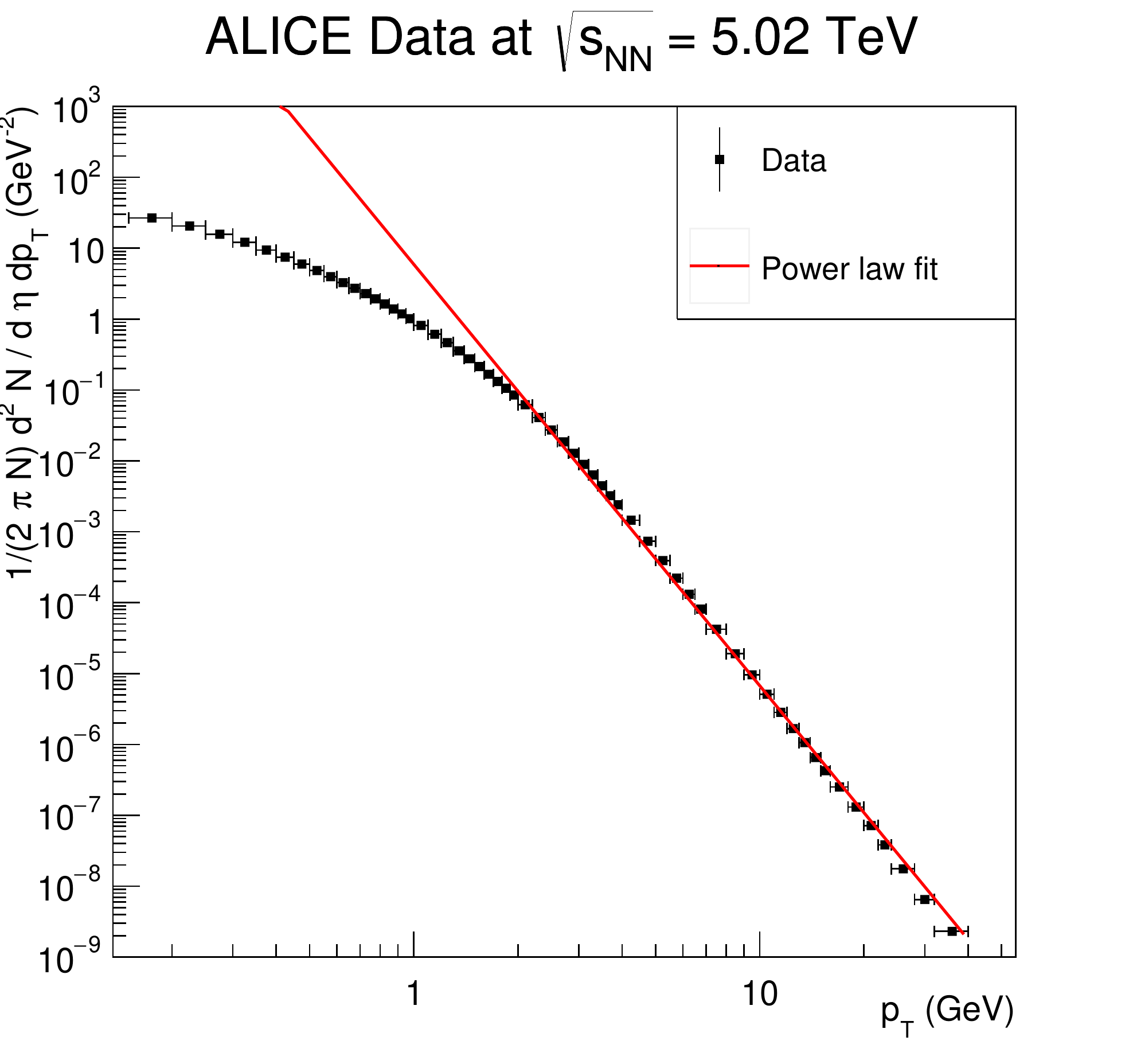} &
\includegraphics[width=50mm]{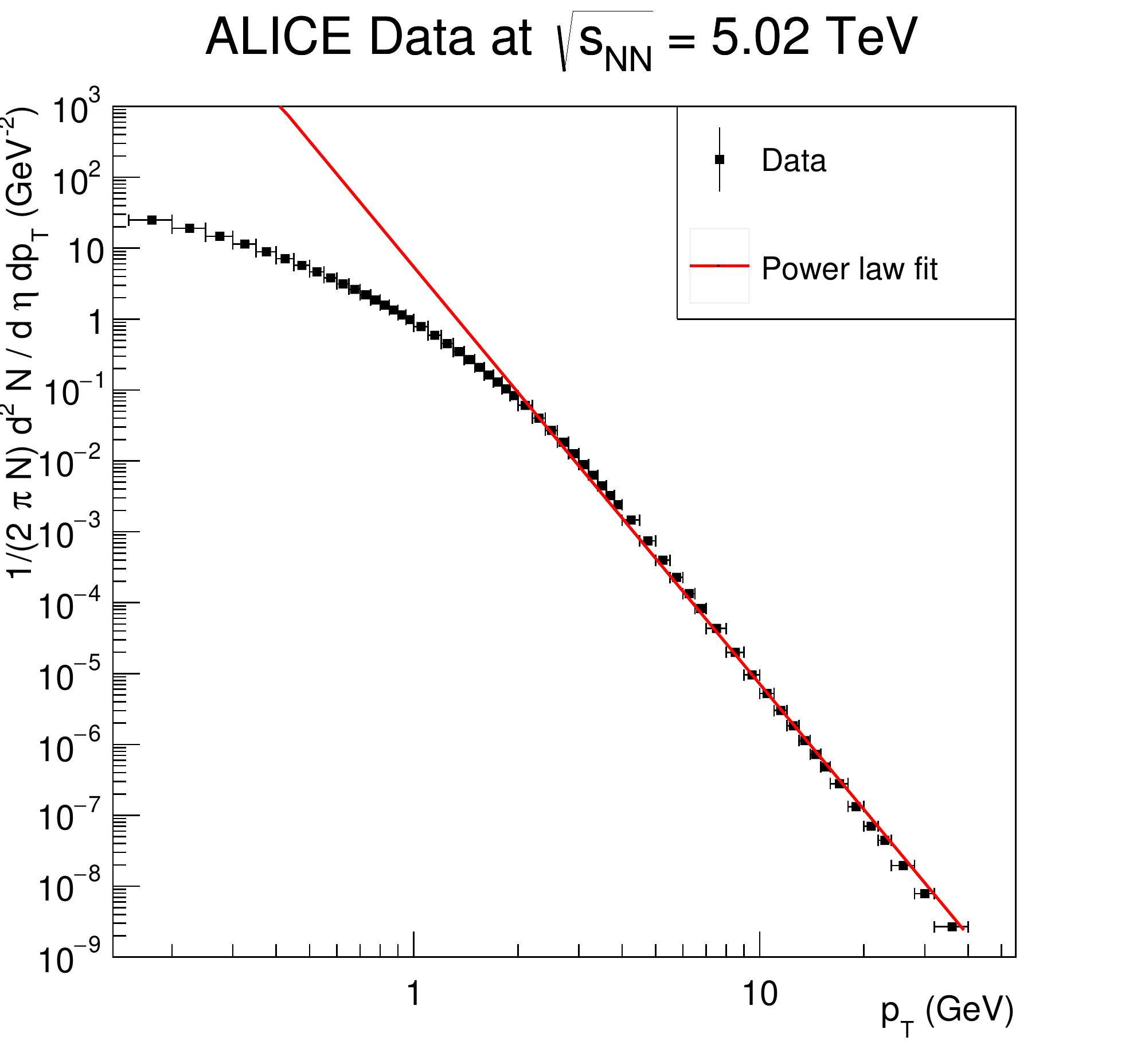} \\
(c)  & (d)  & (e)
\end{tabular}
\caption{\label{fig:powerlowpt}Pure power-law ($A/p_T^B$) fits for the ATLAS $\sqrt{s}=8$ TeV (a) and $\sqrt{s}=13$ TeV (b) data sets, as well as the ALICE $\sqrt{s}=5.02$ TeV data set at rapidity bins of $|\eta|<0.3$ (c), $-1.3<\eta<-0.8$ (d), and $-0.8<\eta<-0.3$ (e) respectively.}
\end{figure}

To avoid these complications, we will attempt to allow for such behavior by including an offset $C$, expected to be of order $\Lambda_\text{QCD}$, in our fit function.  Thus, for production mediated by  factorized Mueller diagrams, we want to fit a curve of the form 
\be\label{eq:powerfitMANY}
\frac{1}{2\pi p_T}\frac{d^2\sigma}{d\pt d\eta}  = \sum_i \frac{A_i}{\left(p_T+C\right)^{B_i}},
\ee 
where the $B_i$ are given by twice the conformal dimensions of the produced particles. More details about the reasoning leading to this fit function are given in Appendix \ref{sec:Validation}.

Theoretically, our results are most strongly suited to describe glueballs.  Because glueballs are not experimentally identifiable, we will instead focus on the production of other QCD bound states, namely mesons, via glueball decays. We will study meson production at the LHC in both proton-lead and proton-proton collisions. 
Within AdS/CFT, the  dominant contribution should be from the production  of  scalar glueballs with $\Delta=4$ (and thus $B=8$) with double-Pomeron Mueller diagram, so for simplicity we will mostly focus on a fitting function given by 
\be\label{eq:powerfit}
\frac{1}{2\pi p_T}\frac{d^2\sigma}{d\pt d\eta} = \frac{A}{(p_T+C)^B}.
\ee

We will consider here three datasets. The first comes from proton-lead collisions studied by the ALICE collaboration at $\sqrt{s_{NN}} $ = 5.02 TeV~\cite{1405}, and the last two come from proton-proton collisions analyzed by the ATLAS Collaboration at center of mass energies of $\sqrt{s}$ = 8~\cite{Aad:2016xww} and 13~\cite{Aad:2016mok} TeV. These two categories are discussed in Sections \ref{sec:ppb} and \ref{sec:pp}, respectively. Results of these studies are shown in Table \ref{tab:fits}. These results are interpreted in Section \ref{sec:interpretation}.

By comparing analysis run on the various data sets we will be able to gain some insight into the (lack) of energy dependency in this kinematic regime.  The ALICE datasets in particular have been run at various pseudorapidity, $\eta$, ranges which allows us to see that there is also essentially no variation in kinematics under changes in pseudorapidity.  The ATLAS data has been collected at psuedorapidity range covered by the end caps ($|\eta|<2.7$)~\cite{ATLASwhitesheet}, but this is still safely inside the central production limit.

\begin{table}[ht]
\begin{centering}
\begin{tabular}{|c|c|c|c|}\hline
Dataset & A/10 (GeV$^{-2}$) & B  & C/(1 GeV)  \\\hline
ALICE 5.02 TeV, $\left|\eta\right|<0.3$ \cite{1405} & 38.48 $\pm$ 8.26 & 7.23 $\pm$ 0.09 & 1.32 $\pm$ 0.04\\\hline
ALICE 5.02 TeV, $-0.8 < \eta < -0.3$  \cite{1405} & 37.60 $\pm$ 7.97 & 7.22 $\pm$ 0.08 & 1.30 $\pm$ 0.04  \\\hline
ALICE 5.02 TeV, $-1.3 < \eta < -0.8$  \cite{1405} & 43.00 $\pm$ 9.29 & 7.30 $\pm$ 0.09 & 1.31 $\pm$ 0.04  \\\hline
ATLAS 8 TeV \cite{Aad:2016xww} & 4.46 $\pm$ 2.60 & 7.03 $\pm$ 0.264 & 1.07 $\pm$ 0.123 \\\hline
ATLAS 13 TeV \cite{Aad:2016mok} & 5.77 $\pm$ 3.38 & 6.96 $\pm$ 0.265 & 1.12 $\pm$ 0.126  \\\hline
\end{tabular}
\caption{Fitted values of parameters in Eq. (\ref{eq:powerfit}) for three data sets. Both central values and statistical errors are quoted.}
\label{tab:fits}
\end{centering}
\end{table}
 
\FloatBarrier
\subsection{Proton-Lead Collisions and Pseudorapidity Dependence}
\label{sec:ppb}
The data in \cite{1405} are binned in the pseudorapidity $\eta$~\footnote{To be precise, the binning in \cite{1405} is done not in terms of the usual pseudorapidity, but instead in terms of a shifted ``center of mass" pseudorapidity. this technicality should not be important here, as it amounts to shifting the definitions of each bin by  $\delta\eta = 0.465$~\cite{1405}.}. There are three bins, corresponding to the $\left|\eta\right|<0.3$, $-0.8 < \eta < -0.3$, and $-1.3 < \eta < -0.8$ regimes, respectively. This allows us the opportunity to study the possible presence of a dependence on pseudorapidity at fixed $\sqrt{s}$. These data cover the range $0.15\text{ GeV} < \pt < 50\text{ GeV},$ and hence allow us to extend further into the high-$\pt$ regime than the above analyses. 

\begin{figure}[!htb]
\begin{centering}
\includegraphics[scale=.4]{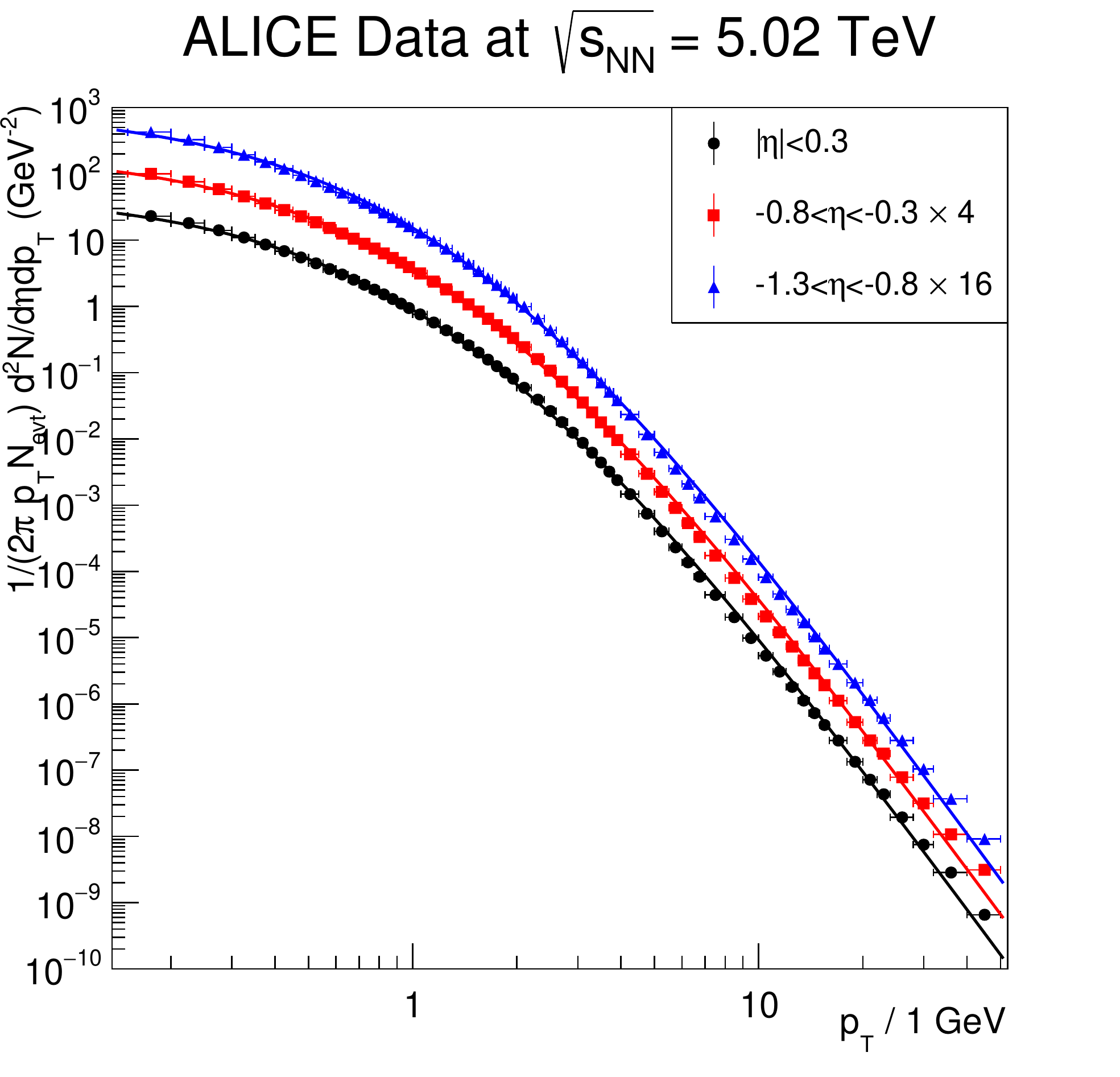}
\caption{Fit of inclusive double-differential charged hadron production cross sections obtained in proton-lead collisions at center of mass energy $\sqrt{s}$ = 5.02 by the ALICE Collaboration, presented in~\cite{1405}. Two of the datasets are rescaled by factors of four and sixteen for visual clarity.  The data are displayed alongside fits to the model in Eq. (\ref{eq:powerfit}).}
\label{fig:1405plot}
\end{centering}
\end{figure}

The results of the fits are shown in Figure \ref{fig:1405plot}. Excellent agreement between the fit model and the data is seen in all three cases. This plot is visually suggestive that the kinematic dependencies depend very slightly, if at all, on the pseudorapidity bin; this is confirmed numerically by the results in Table \ref{tab:fits}. All three fit parameters are compatible in the three bins at the one sigma level. 

\subsection{Proton-Proton Collisions and Center of Mass Energy Dependence}
\label{sec:pp}

ATLAS has also measured the inclusive double-differential single-hadron production cross section \cite{Aad:2016xww,Aad:2016mok}. Unlike the data discussed above, these data are presented in a single pseudorapidity bin, so we cannot extract any information about $\eta$ dependence.  Instead, these two datasets allow us to study the validity of our model in the energy frontier; we have worked in the limit of large center of mass energy, so this is the regime where we expect our results to be the most directly applicable. 

\begin{figure}[!tbt]
\begin{center}
\includegraphics[scale=.4]{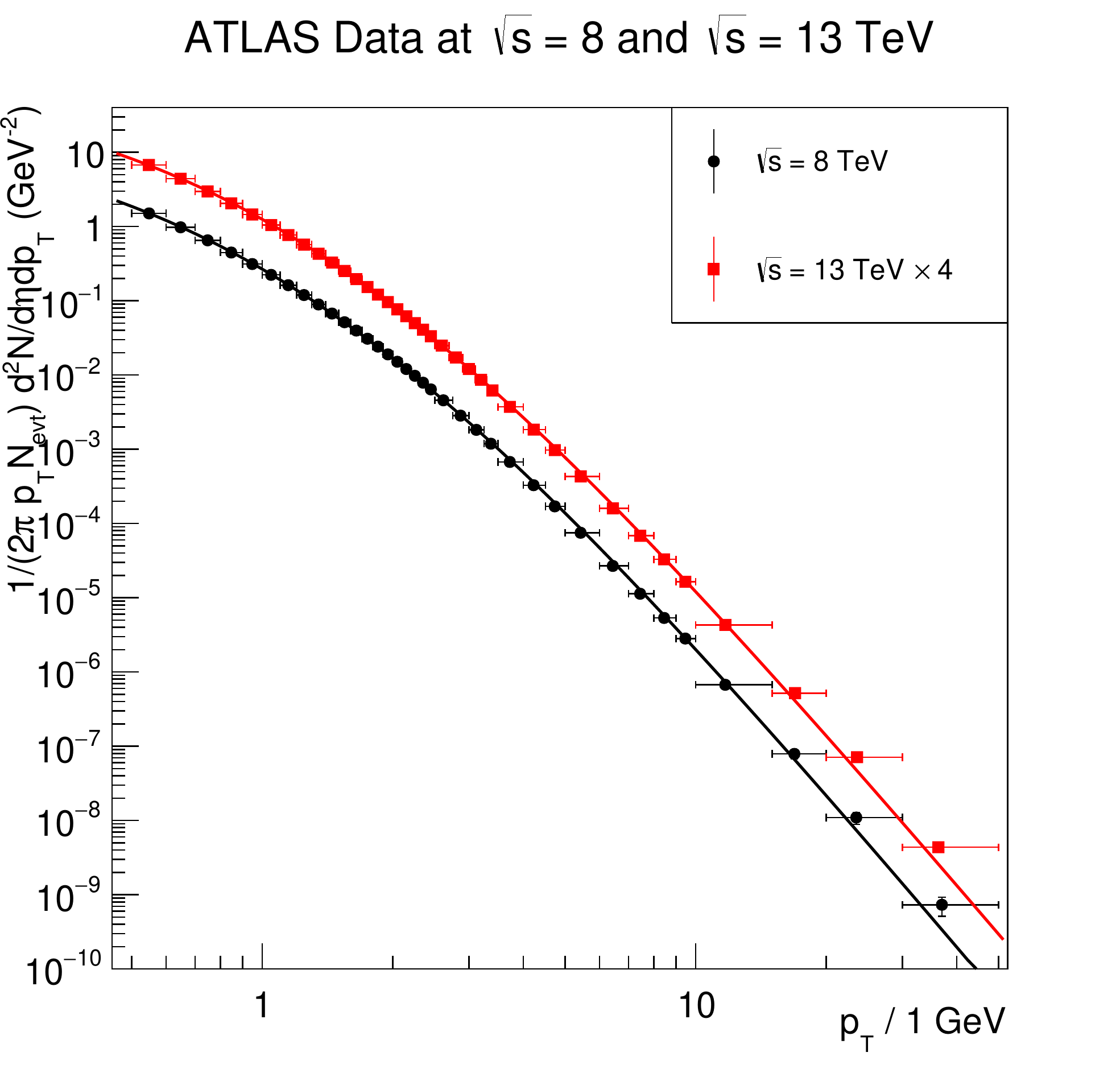}
\caption{Fit of inclusive double-differential charged hadron production cross sections obtained in proton-proton collisions at center of mass energy  $\sqrt{s}$ = 8 and $\sqrt{s}$ = 13 TeV by the ATLAS Collaboration, presented in ~\cite{Aad:2016xww} and \cite{Aad:2016mok}, respectively. The 13 TeV dataset is rescaled by a factor of four for visual clarity. The data are displayed alongside fits to the model in Eq. (\ref{eq:powerfit})}
\label{fig:atlasfigs}
\end{center}
\end{figure}

The results of the fit are shown in Figure \ref{fig:atlasfigs}. As before, the model is seen to correspond closely to data. Within one sigma, the results are seen to match between the two ATLAS datasets, although given the smaller number of data points the uncertainties are of course larger than in the above analysis.

\FloatBarrier
\subsection{Interpretation}
\label{sec:interpretation}
Overall, the preceding results, summarized in Table \ref{tab:fits}, match up rather well with our predictions. The fits are compatible at the two-$\sigma$ level with the power law exponent being independent of both the pseudorapidity and the center of mass measurement. This agrees with the results of Section 4. There are two important caveats, however. First, the overall normalization of the distributions varies sharply between the two types of measurements, with the proton-lead collisions seeming to have a cross section enhanced by an order of magnitude relative to the proton-proton collisons. 
 That the overall normalizations vary so strongly is not altogether surprising. The holographic argument presented here does not offer an easy way to compute this prefactor, so we have no real prediction for it. Certainly we expect higher-order corrections, which are unaccounted for in our tree-level calculation, to importantly influence the normalization. Moreover, from considerations of the mechanisms for proton-lead and proton-proton scattering, it is clear that the difference between these two can have a physical interpretation, rather than being interpreted as an artifact of our calculation.

Let us turn to our predicted value $B=8$ for   scaling dimension.  In \cite{Tannenbaum:2009fg}, it was found that this value is consistent with low energy data. In a perturbative treatment for inclusive production, one generally expects a $p_T$-dependence of the type
 \be
\frac{E d\sigma}{d p^3} \simeq  F(p_T/\sqrt s)\, p_T^{-n}\, ,\label{eq:scaling}
\ee
with $n=4$ for naive scaling.    It is also interesting to point out that a picture based on ``constituent-quark-interchange"~\cite{Blankenbecler:1972cd}  also leads to an effective value of $n\simeq 8$. However, our expectation of  $n=8$ follows from  the  assumption that gluon dynamics dominates  in central production and particle distribution follows that for production of scalar glueballs. It would be  interesting to explore  how the ``constituent-quark-interchange"  approach could be made compatible with our  dual picture of strong-coupling AdS-Pomeron  for central production in a gluon-dominated setting.

It is equally important to point out that  the fitted values for the scaling dimension, although comparable,  are not strictly compatible with the predicted value of $B=8$.
In the context of our paper, that the experimental data do not appear strictly consistent with the interpretation of production mediated by $\Delta_c=4$ glueballs could be  significant. In general, the best we can hope for  from AdS/QCD is an understanding of event kinematics, so the value of the power law exponent is of central importance to our results. We therefore turn now to a discussion of this small but possibly significant discrepancy.

\paragraph{Deviations from Conformality:}
\label{sec:deviations}
The five values for the exponent $B$ are all consistent with $B\simeq 7$, which seems to correspond to a process with $\Delta \simeq 3.5$ instead of our expected $\Delta = 4$.  Given the small numerical uncertainties on our fits, it is extremely unlikely that this is a fluctuation, and we must reconcile this result with our expectations. We will outline below some possible explanations for this effect. Although we cannot conclusively claim that any or all of these suggestions completely explain the fit results, they are within the realm of possibility, and would provide conceptually appealing physical interpretations.

Note that, strictly speaking, our CFT prediction yields a power $B=2\tau$, where $\tau$ is the twist, $\tau=\Delta-J$, $J$ being the spin. For scalar glueball, with $J=0$, we thus have $\tau=4$. A more appealing version has the additional power law terms originating in the production  of  object with twist $\tau\neq 4$. The dominant scaling behavior is due to the production  of scalar glueballs, with $\tau=\Delta=4$.  However, if there is a significant production  via tensor glueballs, $\tau=4-2=2$, thus leading to a term with power $2\tau=4$.  If we allow production to be mediated by both types of glueballs, we would naturally find a cross section of the form

\be
\label{eq:twoterm}
\frac{1}{2\pi p_T}\frac{d^2\sigma}{dp_Td\eta} \sim A (p_T + C)^{-B} + D (p_T + C)^{-E},
\ee
where we expect $B\sim8$ and $E\sim4$. For $A\gg D$, at small $p_T$ the $\delta=4$ term will dominate (See App. \ref{sec:Validation} for other small $p_T$ information) and at large $p_T$ the $\delta=2$ term dominates. In the crossover region of intermediate momentum, the two terms can compete, causing a lowering of the effective power law exponent to be lowered, as alluded to above.

\begin{figure}
\begin{center}
\includegraphics[width = 4.50in]{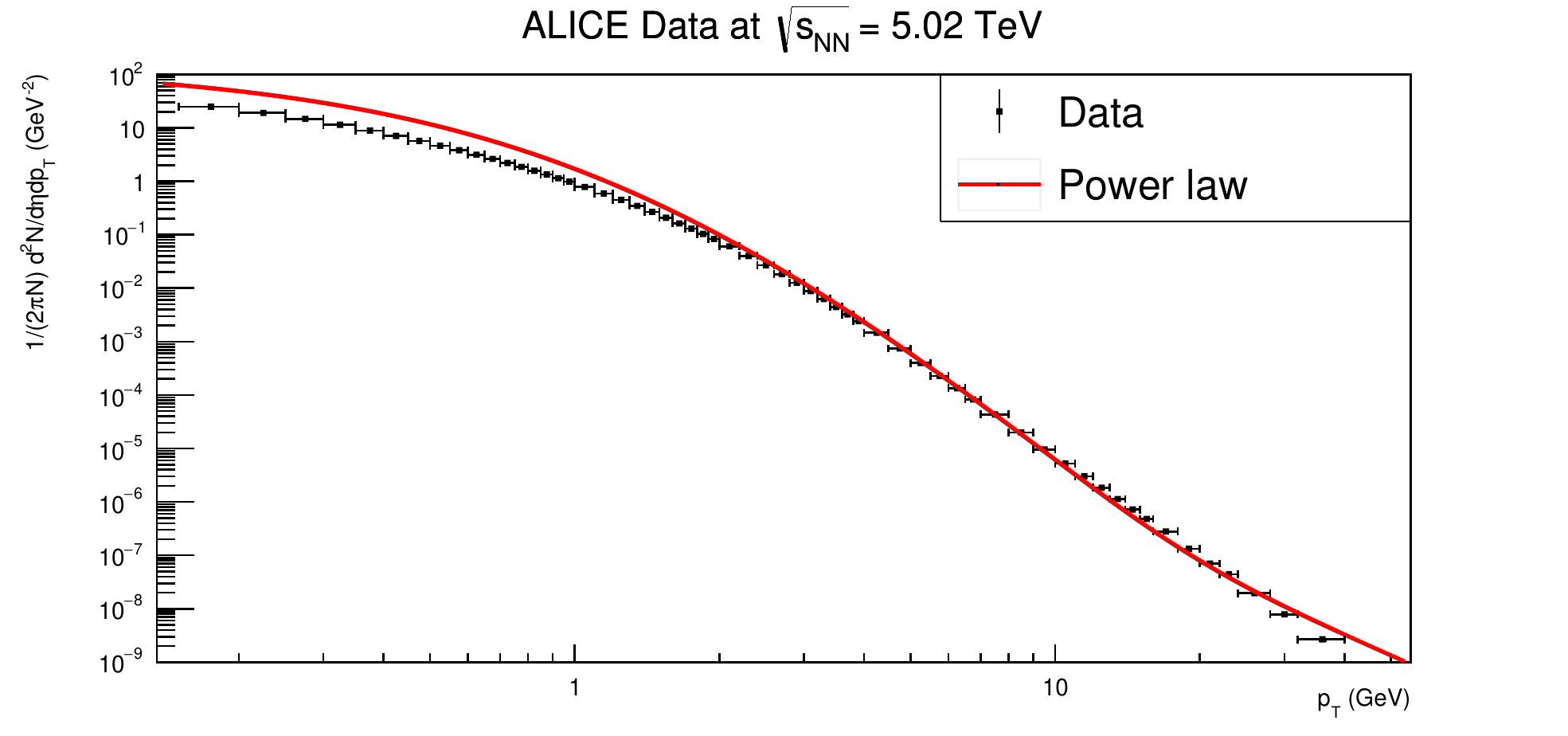}
\caption{Two term power law fit to the first ALICE data set.  A 3 GeV cutoff has been used.}
\label{fig:twoterm}
\end{center}
\end{figure}

Because of the competing effects of these two terms, it is difficult to fit a function of this form directly to data. However, it is possible to make some simplifying assumptions to demonstrate that it is at least a plausible model. If we expand the cross section in Eq. (\ref{eq:twoterm}) about intermediate momentum, fix by hand the values $B=8$ and $E=4$, and import the value of the offset $C$ from Table \ref{tab:fits}, we can float the normalizations $A$ and $D$ to compare this model to data. Such a fit is shown in Fig. \ref{fig:twoterm} for the  ALICE data set with $-0.8\leq \eta \leq -0.3$; to mitigate low-$p_T$ effects, we have discarded data with $p_T<3$ GeV. We do not claim that this is a legitimate fit to data per se; instead we aim to show that such a two-term fit is not an unreasonable form for the cross section.

Along similar lines, one could imagine quark-antiquark ($q\bar{q}$) mixing becoming significant. The calculation in Sec. 4 occurs at large $N_c$, where $q\bar{q}$ mixing is suppressed. However, in real-world QCD we have $N_c=3$, so to obtain phenomenologically viable results it would be beneficial to consider the effects of glueballs mixing with $q\bar{q}$. One could imagine performing this calculation in a top-down Sakai-Sugimoto picture~\cite{Sakai:2004cn}. From a power-counting argument, we expect scalar $q\bar{q}$ states to lead to wavefunctions with $\tau_c=2$, and thus would contribute identically to a tensor glueball. It is unclear how these two scenarios might be distinguished either phenomenologically or experimentally.

As a last incarnation of this argument, we could have considered the effects of mixed Pomeron-Reggeon exchange; it was argued in \cite{Anderson:2014jia,Anderson:2016zon} that these contributions could remain important at LHC energies. Such Mueller diagrams could effect the $z$-cutoff in Eq. (\ref{eq:disc3t6}), which would clearly have effected the final result. For instance, one might expect a correction of the order $ (\sqrt  s/p_T)^{-a}\,  p_T^{-8}\sim s^{-1/2}\, p_T^{-8+2a}$, with $a\simeq 1-2$, which will move the fit closer to the LHC data.    In worldsheet terms, these diagrams would involve additional twist-two operators contributing to the $t$-channel OPEs. This could in general lead to an additional $\eta$-dependence of the final result, which a more refined treatment could become sensitive to.

As another possible line of reasoning, we can consider the effects of finite coupling. The earlier discussion mostly focused on the strong coupling limit of $\lambda\to\infty$. However, other attempts to fit holographic calculations to data have demonstrated that finite-$\lambda$ effects can be important \cite{Brower:2015hja,Brower:2014sxa,Costa:2013uia,Brower:2010wf}. In Appendix \ref{sec:BPST}, we argue that the Reggeization of the graviton depends crucially on finite-$\lambda$ effects, i.e., from stringy physics beyond the supergravity limit. Thus, we expect finite-$\lambda$ physics to effect the glueball wavefunctions that must be convoluted  with the scattering kernel. Related to this is the possibility of nontrivial anomalous dimensions. The central argument of this paper involved a holographic prediction for the kinematics of $\mathcal{N}=4$ SYM. In this theory, superconformality protects the conformal dimensions of scalar glueballs. However, real-world QCD has no such protection, and hence we might expect QCD glueballs to pick up nonvanishing anomalous dimensions. Such an effect could easily account for the observed deviation from $\Delta=4$ production.

\FloatBarrier

\paragraph{Eikonalization:}  
Another  possibility for lowering  the effective exponent is due to  corrections coming from string-loops, although  it is not immediately clear how such effect would emerge.  See Appendix \ref{sec:BPST} for more details.  When the eikonal, Eq. (\ref{eq:eikonal}), becomes large, $\chi(s,\vec b, z, z')=O(1)$, multiple Pomeron exchange becomes important, leading to ``saturation". 
Indeed, such effect should be important for inclusive production with $p_T=O(\Lambda_{QCD})$. Since this region depends crucially on how  confinement deformation is implemented, our single-Pomeron analysis can be modified significantly~\footnote{For a perspective possibly different from ours, see \cite{Kharzeev:2006zm}. A universal $e^{-c p_T}$ behavior for the region $p_T< \Lambda_{QCD}$ was advocated in \cite{Bylinkin:2014qea}. See also \cite{Kharzeev:2006aj,Castorina:2007eb} and App. \ref{sec:Validation}. Since the data in this region is spare, a more conventional  behavior such as $e^{-c' p^2_T}$  cannot be ruled out.}. 
 However, for production at large $p_T$, our current treatment should be reliable.  Further study in this direction will be pursued. 

\paragraph{Naive Scaling:}

 In a perturbative treatment for inclusive production, in the absence of dimensionful scales,  the function $F$ in (\ref{eq:scaling}) would be dimensionless, leading to $n=4$; this is known as  ``naive scaling".  However, our non-perturbative result in Eq. (\ref{eq:CrossSecPrediction}), differs significantly  from the naive scaling expectation, and the corresponding function $F$ in (\ref{eq:scaling}) depends also on confinement scale $\Lambda_{QCD}$; this dependence enters  through the ``string cutoff" $z_s$ in Eq. (\ref{eq:zs}), as well as through the total cross ection $\sigma_{total}$.  We note  that LHC data has also been examined in~\cite{Bylinkin:2014qea,lhcfit}, against such  naive expectation of $p_T^{-4}$.   Clearly, this is not evident  at LHC energies.  This perturbative scaling law was also mentioned peripherally  in \cite{brodsky}. Assuming the parameter $B$ is energy dependent, it was speculated in \cite{Bylinkin:2014qea,lhcfit}  that one would reach $B\simeq 4$ at $s\, \sim 10^3$ TeV, far beyond the LHC range.  Our study, on the other hand, is based on the belief that there are no unexpected new scales involved other than $\Lambda_{QCD}$, and therefore that our AdS/CFT based analysis should be applicable at LHC energy.

\section{Summary and Discussion}\label{sec:discussion}

We have explored the consequences of conformal invariance in inclusive QCD production at high energy by means of the AdS/CFT correspondence. As mentioned in Sec. \ref{sec:Intro}, although QCD is not strictly a CFT, it is nevertheless possible to address in certain kinematic limits where effects of confinement deformation are not expected to be important. In this treatment, we have focused on inclusive central production at large $p_\perp$ where we demonstrate that particle density obeys a power law fall-off that depends only on conformal dimension of the produced particle,
 \begin{equation}
 \frac{1}{\sigma_{total} } \frac{d \sigma_{ab\rightarrow c+X}}{d^3{\bf p}_c/E_c}
 \propto  {\rm Disc}_{M^2}\langle {\cal V}_P V_{c\bar c} {\cal V}_P\rangle.
 \sim p_\perp^{-2\tau_c}. \label{eq:2f}
 \end{equation}    
The analysis is carried out   in a momentum-space setting.  With inclusive cross sections as discontinuities, it is  important to  include stringy effects, e.g., taking  the discontinuity for the matrix element of the central vertex, $V_{c\bar c}$,   between two Pomeron vertex operators. As is the case of exclusive fixed-angle scattering, this power fall-off occurs due to the geometry of warped AdS space,    via a generalized  Polchinski-Strassler mechanism~\cite{Polchinski:2001tt,Brower:2002er}. The form of the power law is fixed by conformal invariance. This prediction appears to be well-supported by recent LHC data.
   
In the first part of this paper, we concentrated on more formal aspects of inclusive cross sections as discontinuities. We first focused on general CFT and useed DIS at small-$x$ as an illustration on how to invoke a $t$-channel OPE. We next discussed AdS/CFT via Witten diagrams, and additionally  introduced a confinement deformation in the IR. Lastly, we discuss gauge-string duality beyond the strict supergravity limit,  which leads to the inclusion of stringy effects and, in turn, the AdS-Pomeron. 

In the second part of the paper, we turned to the calculation of inclusive distribution for central production, with a particular focus  on the kinematic limit of large $p_\perp$ production. We discussed the generalized optical theorem  for   $3\to3$  amplitude and   computed the curved-space string theory prediction for the inclusive cross section, which lead to the conformal behavior in Eq. (\ref{eq:2f}). Finally, we test this finding by examining the recent LHC data, coming from both proton-lead and proton-proton collisions analyzed by the ALICE and ATLAS collaborations.

 We end by mentioning  some possible future directions for inclusive  study of conformal invariance. On a more theoretical side, a better understanding on   the  $x$-space and $p$-space connection would be desirable. For a CFT with gravity dual,  this can be done most easily  through a perturbative Witten diagram approach. Another possible avenue of attack  is through the use of Mellin representation, as discussed by Mack~\cite{Mack:2009mi}.  Equally interesting is to extend the study to multi-particle production~\cite{Hofman:2008ar,Belitsky:2013ofa,Jen:1971ex,Hasslacher:1972ne}.  Other phenomenological applications include inclusive production in other kinematical regions where the consequences of conformality can appear ~\footnote{Single-particle inclusive cross section in other kinematic regions has been addressed in ~\cite{Tannenbaum:2009fg,Blankenbecler:1972cd,brodsky} from a perturbative dimensional counting perspective.}, such as the triple-Regge limit,  explore  heavy quark production in the central region\footnote{For heavy quark production, this can be treated with the perturbative BFKL approach.  See~\cite{Chachamis:2009ks} and references therein.}, and tetra quark production \footnote{Although quark contributions are 1/N suppressed holographically- the $\Delta=4$ tetra quark contribution~\cite{Ebert:2012glz,Brodsky:2016yod,Brodsky:2015wza}, which has already been investigated holographically~\cite{Sonnenschein:2016ibx}, could compete with a scalar glueball.} Also interesting  would be the study of two-point correlations, such as $\gamma^*\rightarrow c_1+c_2 +X$ or $a+b\rightarrow X_1+c_1+c_2 +X_2$. 
  Study in some of these issues are currently underway.

 \section*{Acknowledgments}
The work of T.R. and C.-I T. are supported in part in part by the Department of Energy under contact DE-Sc0010010-Task-A. T.R. is also supported by the University of Kansas Foundation Professor grant.  R.N. is funded by the Stanford University Physics Department, a Stanford University Enhancing Diversity in Graduate Eduction (EDGE) grant, and by NSF Fellowship number DGE-1656518.

\newpage

\bibliographystyle{JHEP}
\bibliography{Inclusive}

\newpage

\appendix

\section{Inclusive Cross Sections and Applications} 
\label{sec:A-Mueller}
Inclusive cross sections as discontinuities also follow from unitarity. Here we give more
detail first on the single particle inclusive amplitude and also provide examples of the
power of taking discontinuities to calculate cross sections. The issue of analytic structure
is necessarily more involved in the case of CFT, which can be simplified in strong coupling
via the use of Witten Diagrams in momentum-space representation.
\subsection{Single Particle Inclusive}
\begin{figure}[ht]
\begin{center}
\includegraphics[scale=1]{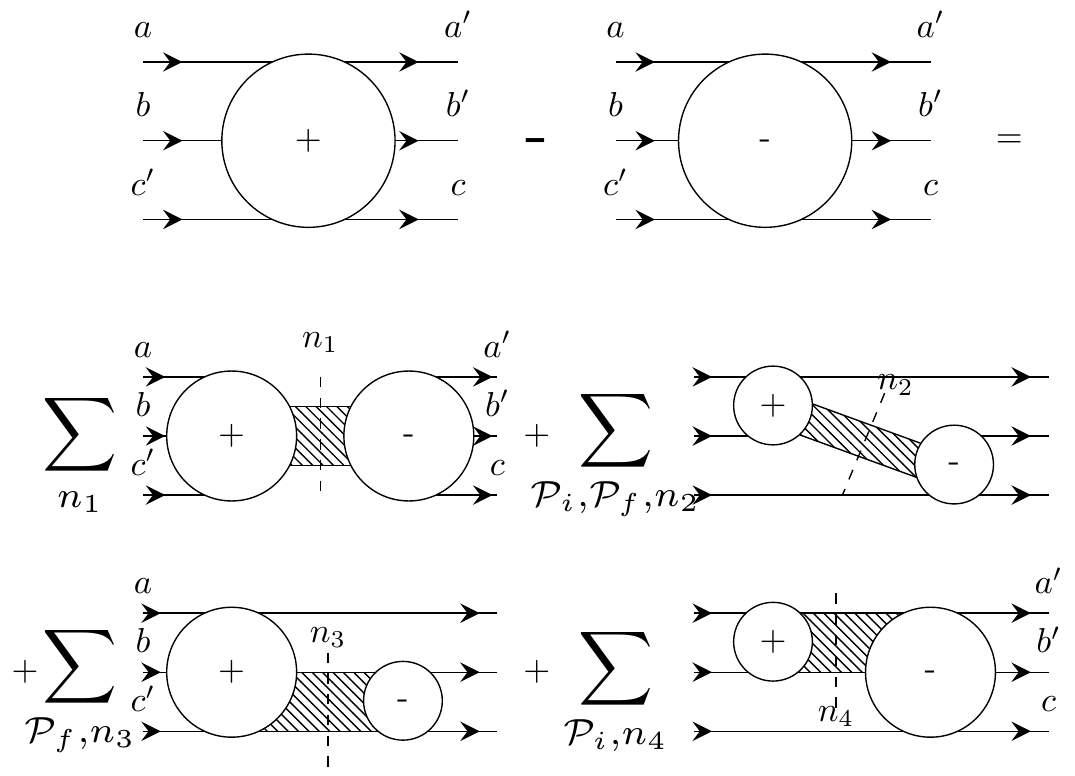}
\caption{ A schematic representation of unitarity equation  for  connected three-to-three scattering amplitudes from \cite{Tan:1972kr}.}

\label{fig:33unitarity}
\end{center}
\end{figure}

The discontinuity in Eq. (\ref{eq:Optical6b}) is taken in the forward limit, where $p_{a'}=p_a$, $p_{b'}=p_b$, and $p_{c'}=p_c$.   This corresponds to a generalized optical theorem~\cite{Mueller,Stapp:1971hh,Tan:1972kr}, and is also known  as the Mueller formula.   Just as the familiar optical theorem in Eq. (\ref{eq:Optical4}) follows from unitarity for the 2-to-2 elastic amplitude,  this  $M^2$-discontinuity   enters as a particular term in the 3-to-3 unitarity relation, schematically represented in Fig.  \ref{fig:33unitarity}.

Each term on the right-hand-side of the  equation can be identified as the discontinuity in an appropriate invariant~\cite{Stapp:1971hh}.  There are four types of discontinuity diagrams, with ${\cal P}_i$ and ${\cal P}_f$ summing over all possible permutations of initial and final states, while the $n_i$-sum over all allowed states.  The  missing-mass discontinuity enters in the second group, i.e., that indicated by $n_2$ sum in the unitary equation. 

The discontinuity in Eq. (\ref{eq:Optical6b}),
\be
 \frac{d \sigma_{ab\rightarrow c+X}}{d^3{\bf p}_c/E_c} \propto \frac{1}{2is} \,  {\rm Disc}_{M^2>0} T(p_{a'},p_{b'},p_{c}; p_a,p_b,p_{c'})\, ,  \nonumber 
\ee
is taken in the forward limit,
where $p_{a'}=p_a$, $p_{b'}=p_b$, and $p_{c'}=p_c$.   This identification, as explained in Sec. \ref{sec:2-point},  is in exact correspondence  to that for the free propagator. The discontinuity in $M^2$ enters as a term in the 3-to-3 unitarity relation, as represented schematically in Fig.  \ref{fig:33unitarity}.  In this figure,
  shaded bands represent allowed intermediate states and all amplitudes, indicated by circles,  involved are connected.  We denote amplitudes in the physical region by ``$+$"   and complex conjugation by ``$-$".  As also explained in Sec. \ref{sec:2-point}, each term  on the right-hand-side of the unitarity equation can be identified as the discontinuity in an appropriate invariant~\cite{Stapp:1971hh}.

From the perspective of the process $a+b+c'\rightarrow a'+b'+c$, $M^2$ is a ``cross-channel"  invariant, as opposed to ``direct-channel" invariants, such as $s_{ab}=(p_a+p_b)^2$, $s_{abc'}=(p_a+p_b+p_{c'})^2$, etc. Because of the Steinman rule, there are no double-discontinuities in overlapping invariants in the physical region~\cite{Stapp:1971hh,Tan:1972kr}. 
This discontinuity in $M^2$, Eq. (\ref{eq:Optical6b}), yields a sum over all allowed multi-particle states $X$, multiplied by  a delta function factor, $\delta((p_a+p_b-q_c)^2-M_X^2)$.  Each state $X$  contributes a term which is the product of an on-shell amplitude  for $a+b\rightarrow c+X$ with its conjugate, 
$
  T^*_{a'b'\rightarrow c'X} T_{ab\rightarrow cX}.
$
The total discontinuity involves a sum over each allowed state $X$; for each $X$, the sum involves an integral over the appropriate multi-particle phase space.

\subsection{DIS, OPE  and Anomalous Dimensions} \label{sec:OPE}

As an explicit illustration, consider the inclusive scattering $\gamma^*+{\rm proton}   \rightarrow X$ of a virtual photon with momentum $q$ off of a proton of momentum $p$ in the limit of  $Q^2=-q^2\rightarrow \infty$ with $x=Q^2/s$ fixed. That is, one is dealing with the photon-proton total cross section, $\sigma^{total}_{\gamma^* p}$, as a function of $Q^2$ and $x$.
  This cross section can be expressed as a product of  photon polarization vectors and the hadronic tensor,
$
W^{\mu\nu}(p,q)$, defined as the Fourier transform  of   the current  commutator, $
  \langle p |[J^\mu(x) ,J^\nu (0)] |p\rangle
$.    
It can be expressed  in terms of  two scalar structure functions, 
\be
W^{\mu\nu} =     F_1(x,Q^2) \Big(g_{\mu\nu}-\frac{q_\mu q_\nu}{q^2}\Big) + F_2(x,{Q^2})\Big(p_\mu+\frac{q_\mu}{2x}\Big)\Big(p_\nu+\frac{q_\nu}{2x}\Big).\label{eq:DISstructure}
\ee

For virtual Compton scattering, $q+p\rightarrow q'+p'$,  the amplitude $T^{\mu\nu} (p,q; p', q')$ is  given by the Fourier transform of the {\bf T}-product
$
 \langle p'| {\bf T}\{J^\mu(x) J^\nu (0)\} |p \rangle
$. 
In the  forward scattering limit, $p=p'$ and $q=q'$, $T^{\mu\nu}$ has a Lorentz covariant expansion similar to that of $W^{\mu\nu}$, with new form factors $\widetilde F_\alpha(x,Q^2)$ replacing $F_\alpha(x,Q^2)$.     The hadronic tensor is related to the forward amplitude by the Optical Theorem, which implies that $W^{\mu\nu}(p,q)=\frac{1}{2i} \,  {\rm Disc}_{s>0} T^{\mu\nu} (p,q; p, q)$.    Treating the $F_\alpha(x,Q^2)$ as real-analytic functions  of $x$ with a branch cut over $[0,1]$, one has~\footnote{ As explained  in \cite{Polchinski:2002jw}, in a large-$N_c$ treatment for QCD, the  discontinuity consists of an infinite sequence of delta-functions, coming from Regge recurrences for the proton.  These discontinuities can also directly be related to $\sigma_T$ and $\sigma_L$ for transverse and  longitudinal off-shell photons.} 
\be
F_\alpha(x,Q^2) = 2\,{\rm Im}\, \widetilde F_\alpha(x,Q^2)\, .
\ee

DIS in QCD is strictly speaking not conformal. However, it is possible to explore conformal dynamics if one assumes a fixed coupling and focuses on the kinematic region of small-$x$. 
DIS structure functions are strongly peaked phenomenologically at $x\rightarrow 0$, which can be used to infer the dominance of gluon dynamics, consistent with the large $N_c$ expectation~\cite{Polchinski:2002jw,Brower:2010wf,Brower:2015hja}.  This singular small-x behavior   allows a direct measurement of the anomalous dimensions, $\gamma_n$,  for twist-two operators, $\mathcal{O}_n$, since these operators dominate in the $t$-channel OPE of two currents,  $J^\mu(x)J^\nu(0)=\sum_n |x|^n c^{\mu\nu}_n{\cal O}_n(0)$.

\begin{figure}[ht]
\begin{center}
\includegraphics[width = 3.0in]{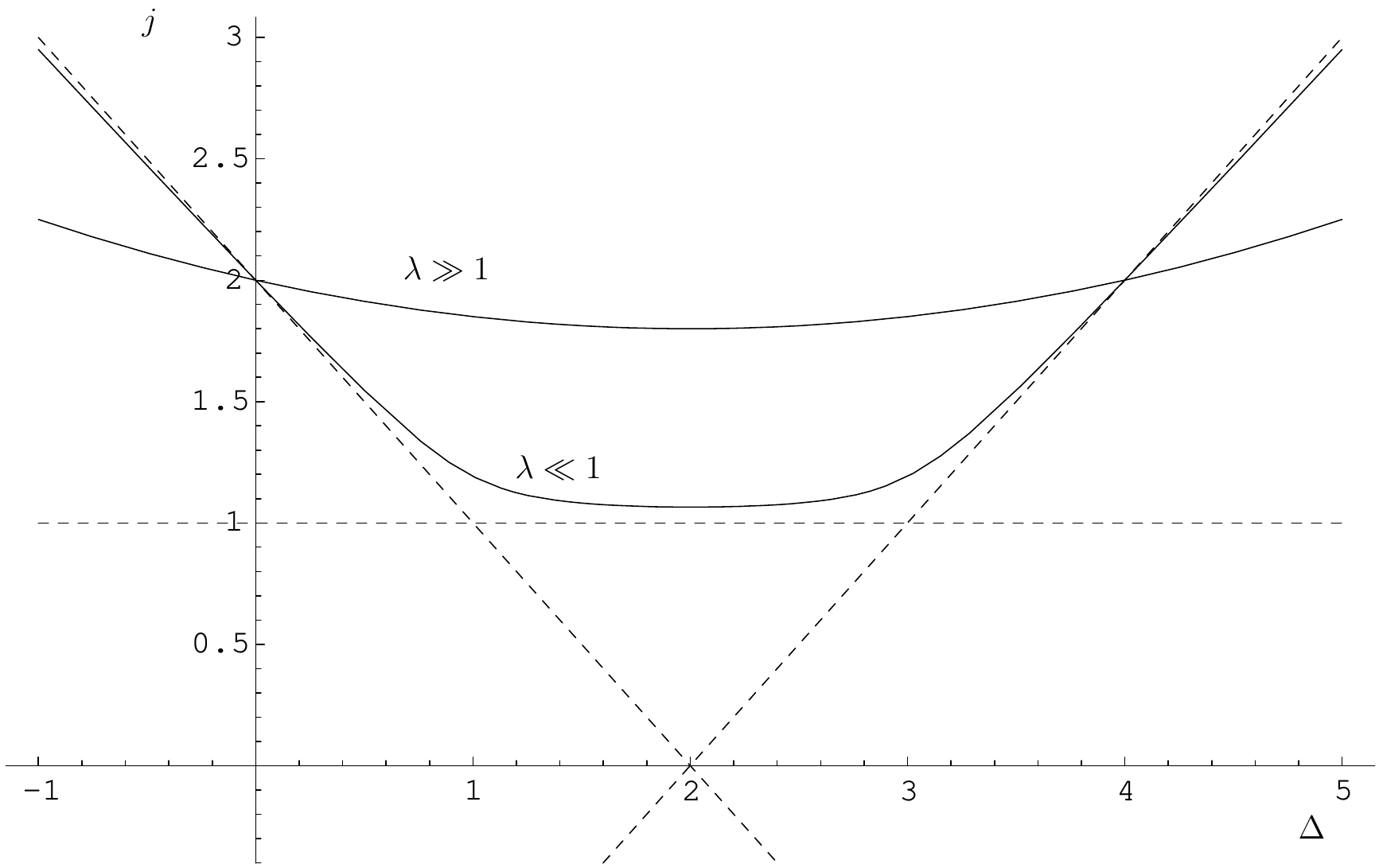}
\caption{Schematic form of the $\Delta-j$ relation for twist-2 spectral curve at  weak ($\lambda\ll 1$) and strong coupling
($\lambda\gg 1$),  reproduced from Ref.~\cite{Brower:2006ea}.  Symmetry about $\Delta=2$ follows from conformal invariance.
}
\label{fig:Delta-J}
\end{center}
\end{figure}

A standard analysis  leads to an expansion for $\widetilde F_\alpha$ in $x^{-1}$, valid in the limit of large $Q^2$.   Through a dispersion relation, the coefficients $M^{(\alpha)}_n(Q^2)$ of this expansion  can be expressed as ``moments" over its discontinuity across $0<|x|<1$, i.e., $
M^{(\alpha)}_n(Q^2) =\int dx x^{n-\alpha} F_\alpha(x,Q^2)$.   In the large $Q^2$ limit, these coefficients are given approximately by
\be
M^{(\alpha)}_n(Q^2) \sim ({Q^2})^{-\gamma_n}.\label{eq:moments}
\ee
Here, the $\gamma_n$ are the anomalous dimensions of twist-2 operators with even integer spin $j=n$, defined by 
\be
\gamma (j)= \Delta(j)- j-2.
\ee
For $j=2$, we have $\gamma_2=0$ due to energy momentum conservation~\footnote{We are restricting ourselves here to the so-called single-trace conformal primaries. For other related discussion, see \cite{Komargodski:2016gci} and references therein. See Sec. \ref{sec:deviations}  and also Appendix \ref{sec:BPST}
 for  going beyond this restriction. }.  For $j\neq 2$, anomalous dimensions do not vanish, which  leads us directly to  CFT dynamics. We will frequently treat $\Delta= j+2 +\gamma (j)$ as a function of $j$, or, equivalently, its inverse, $j(\Delta)$, as a continuous function of $\Delta$, as shown in Fig. \ref{fig:Delta-J}. That is, by treating the structure functions as  discontinuities, one can explore anomalous dimensions  through a $t$-channel OPE, which can serve as a spring-board for introducing stringy effects via AdS/CFT~\footnote{Conformal invariance forces $j(\Delta)$ to be symmetric about $\Delta=2$, with $j(\Delta)$ having  a minimum at $\Delta=2$.   The intercept $j_0$ of the Pomeron~\cite{Brower:2006ea}, which obeys $2>j_0=1+\epsilon>1$, can be  found by  demanding $\Delta(j_0)=2$.   See also Appendices \ref{sec:BPST} and \ref{sec:ConformalRegge}.}. In particular, at  large 't Hooft coupling $\lambda$,  by exploring Regge behavior, one has $F_s\sim x^{-(2-s) -j_0}$ for $2>j_0 >1$, where $\Delta(j_0)=2$ and  is identified with the Pomeron intercept. In the strong coupling limit, $j_0\simeq 2 -2/\sqrt \lambda$.  Clearly, exploring this holographically requires going beyond the SUGRA limit of $\lambda\rightarrow \infty$, which we turn to next.

\section{AdS/CFT Scattering and the BPST Program}\label{sec:BPST} 

We provide here further details of scattering in the AdS/CFT and give a brief summary of the BPST program~\cite{Brower:2006ea,Brower:2007qh,Brower:2007xg}, which constitutes the steps leading to Eqs. (\ref{eq:adsPomeronScheme}) and (\ref{eq:adsDoublePomeronScheme}). In \cite{Brower:2006ea,Brower:2007xg}, AdS/CFT is implemented by starting  first with flat-space string theory.  Alternatively, the construction of the BPST Pomeron can be initiated with a CFT OPE, and the corresponding Witten diagram expansion in the supergravity theory, and from there incorporate stringy effects~\cite{Brower:2014wha,Cornalba:2007fs,Cornalba:2008qf,Cornalba:2009ax,Costa:2012cb}. The two approaches are equivalent, and provide separate intuitive frameworks. Here and in Appendix \ref{sec:ConformalRegge} we will integrate both approaches.

\paragraph{$t$-Channel OPE and Witten Diagrams:}
For 2-to-2 scattering at high energy, with $s=(p_1+p_2)^2\rightarrow \infty$ and $t=(p_3-p_1)^2<0$,  the simplest Witten diagram that appears in the $t$-channel OPE is that from a single scalar  exchange. In a momentum-space representation, up to a constant it is given by 
\be
T_{dilaton}(s,t,p_i^2) = \int d\mu(z) d\mu(z') \Phi_1(z,p_1^2)\Phi_3(z,p_3^2) G_F(z,z',t) \Phi_2(z',p_2^2)\Phi_4(z',p_4^2)\, , \label{eq:adsDilaton}
\ee
where $d\mu(z)=dz \sqrt {-g}$ is the $AdS_5$ measure and $G_F(z,z',t) $ is the scalar bulk-to-bilk propagator given in Eq. (\ref{eq:AdSpropagator}).   In anticipation of the confinement deformation we will later introduce, we will replace bulk-to-boundary propagators $\Phi_i(z,p^2)$ with normalizable physical wave functions $\phi_i(z)$  in what follows.

In a Minkowski setting, the exchange of a spin $J$ excitation leads to a contribution whose growth is bounded from above by $s^J$.  Therefore, the $t$-channel scalar exchange in Eq. (\ref{eq:adsDilaton}) is independent of $s$.   
For ${\cal N} = 4$ SYM, in the {extreme limit} $\lambda = g^2N_c \rightarrow \infty$ of large  't Hooft coupling, the dominant Witten diagram comes from the exchange of one $J=2$ graviton.  This diagram has a form similar to that in Eq. (\ref{eq:adsDilaton}) but with the scalar bulk-to-bulk propagator replaced with a tensor propagator $G_{MNM'N'}(z,z',t)$, and the factors of the coupling at the vertices replaced by the conserved energy-momentum tensors $T^{MN}$ and $T^{M'N'}$. At large $s$ and fixed $t$, the dominant contribution comes from the $(++--)$ helicity component, which couples to the large light cone momenta $(zz')^4(p_1^+)^2(p_2^-)^2$. Thus, in this limit we have
\be
T_{graviton}(s,t,p_i^2) \simeq  \int d\mu(z) d\mu(z') \Phi_{13}(z) \,\widetilde {\cal K}_G (s, t, z,z')  \Phi_{24}(z')\, , \label{eq:adsGraviton}
\ee
where the graviton kernel can be expressed in terms of scalar propagator $G_F(z,z',t)$ and the red-shifted energy invariant $\widetilde s$ as $
\widetilde {\cal K}_G =  G_{++--} \, \widetilde s^2= (zz')^{-2} G_F(z,z',t)\, \widetilde s^2$. 
 We have also defined vertex factors 
 $\Phi_{13}(z)=z^2\phi_1(z,p_1^2)\phi_3(z,p_3^2)$ and $\Phi_{24}(z') ={z'}^2 \phi_2(z',p_2^2)\phi_4(z',p_4^2)$.   We therefore see that the amplitude scales as $s^2$, as expected. Schematically, we write this as 
\be
T_{graviton}(s,t)  = \Phi_{13}*\widetilde {\cal K}_G * \Phi_{24}  \, . \label{eq:bulk-integral}
\ee 
where $*$ corresponds to integration over the AdS bulk. 

\paragraph{Ultralocal Scattering and the Polchinski-Strassler Mechanism}\label{sec:PS}

It has been stressed in \cite{Polchinski:2001tt} that scattering amplitudes  in gauge theories with a good string dual description can often be simplified, since the dual ten-dimensional string scattering on $AdS_5\times S^5$  is effectively local. This simplification is particularly applicable in the limit of fixed angle-scattering when all four-dimensional Mandelstam invariants are large and {\it of the same order}.  In this limit, gauge theory amplitudes can be expressed as a coherent sum of local scattering in the AdS bulk, where again we ignore fluctuations in $S^5$ throughout ~\cite{Polchinski:2001tt}.   As an effective  five-dimensional scattering process, the momenta $p^\mu$ for external states are seen by local observers in the AdS bulk to be red-shifted, with large components $\widetilde p^\mu$ along $p_i^\mu$ where
\be
\widetilde p^\mu_i\simeq (z /R) p^\mu_i\, .\label{eq:redshift}
\ee
We are interested in a strongly coupled boundary theory, so as above we take the AdS radius $R$ large compared to the string scale. In what follows, we shall set $R=1$.

In this limit, a 4-D scattering amplitude reduces to a coherent sum over local scattering in the AdS bulk, so that
\be
T_n(p_1,p_2, \cdots) =\int\, \frac{dz}{z} \,   {\cal T}_n (\tilde p_1, \tilde p_2,  \cdots)  \Pi_i\{\phi_i (z)\}\, , \label{eq:ultralocal}
\ee
where $ {\cal T}_n$ corresponds to the amputated bosonic string Green's function in flat space.   In terms of invariants, the arguments for  $ {\cal T}_n$ are red-shifted, 
$s_{ij}\rightarrow z^2 s_{ij}$.  A flat-space bosonic 4-point amplitude   can be expressed  in  a Koba-Nelson representation involving an integral over a  single modulus. In the  limit of $-t \simeq (1-\cos \theta_{cm}) s\rightarrow \infty$,  the integral is dominated by a saddle point.  This leads to 
an exponential cutoff,  
\be
{\cal A}_4(s,t)\sim e^{-f(\cos \theta_{cm}) \, \alpha' s} .  \label{eq:flatfixangle}
\ee
More details are provided in Appendix \ref{sec:FlatString}.   This exponential suppression  is a generic feature of flat-space string scattering, and also holds for multi-particle scattering in  similar generalized fixed-angle limits~\footnote{Indeed, as mentioned in Sec. \ref{sec:Intro}, this feature represents a serious failure for earlier attempt in formulating gauge theories as strings.}. As stressed in \cite{Polchinski:2001tt}, the exponential suppression in Eq. (\ref{eq:flatfixangle}) allows us to restrict the domain of integration in Eq. (\ref{eq:ultralocal}) to an effective scattering region $z\in[0,z_s(s)]$, where $z_s(s)= O(1/\sqrt{s})$. We will refer to this simplification as the Polchinski-Strassler mechanism. In the scattering region, ${\cal T}_4(\widetilde s, \widetilde t)=O(1)$. This, combined with the wavefunctions in Eq. (\ref{eq:conformalWF}), leads to a power-law falloff for the cross section of the form $
\frac{d\sigma}{dt} \sim s^{-\tau_{total}}$, where $\tau_{total}$ is the sum of the twists $\tau_i = \Delta_i-J_i$ of the external particles. This is consistent with the dimensional counting rule of ~\cite{Brodsky:1973kr,Brodsky:1974vy,Matveev:1973ra}.

\paragraph{Beyond the SUGRA Limit:} 
The standard Witten expansion   involves  only propagators and vertices of  super-gravity  fields  in $AdS_5$, such as the dilaton $\phi$, metric fluctuations $h_{\mu\nu}$, and the anti-symmetric tensor $B_{\mu\nu}$.  This dramatic reduction in the number of degrees of freedom can be understood in terms of the boundary theory by the rapid increase of anomalous dimensions for all unprotected gauge-invariant local operators  in the large 't Hooft coupling limit.  Generically, their conformal dimensions grow as
	\be
	\Delta(j)=j+ 2+\gamma_j = O(\sqrt \lambda)\, ,
	\ee
so that in the $\lambda\to\infty$ limit their string duals become heavy and decouple.    In this limit of the sum can often be truncated so that it is given approximately by sums of perturbative  $t$-, $s$- and $u$-channel exchange diagrams. Perturbatively each of these diagrams will  contribute only to discontinuities in their respective channel. 

\paragraph{From the Graviton to the BPST Pomeron:}
 In a $t$-channel OPE, the contribution from a conformal primary with definite spin does not lead to singularities in the cross channel invariants $s$ and $u$. Discontinuities can emerge due to re-summation of high-spin exchanges.  For finite 't Hooft coupling, incorporating the higher string modes  associated with the graviton leads to a ``reggeized AdS graviton". This in turn leads to the BPST program, where elastic amplitudes at high energy can be represented schematically in a factorizable form like that of 2-to-2 amplitude in Eq. (\ref{eq:adsPomeronScheme}).

Here the universal Pomeron kernel $\widetilde {\cal K}_P$ grows with  a characteristic power behavior at large $s>>|t|$, i.e.
$\widetilde {\cal K}_P\sim s^{j_0}$. 
The strong-coupling Pomeron intercept, at leading order in $\lambda$,  is
$
j_0 = 2 - 2 /\sqrt{\lambda} 
$,
which agrees with spin $J=2$ of the graviton in the limit of $\lambda=g^2N_c\rightarrow \infty$.  Conversely, at finite $\lambda$, the regggeized AdS graviton has its intercept lowered below $J=2$. 
More generally, this approach leads to conformal Regge theory in CFT, which we will  discuss briefly in Sec. \ref{sec:ConformalRegge}.  Holographic descriptions of scattering data agree with a Pomeron intercept near $j_0\simeq 1.3$ in a strongly coupled regime~\cite{Brower:2010wf,Brower:2012mk}.

At   finite $\lambda$,  one can incorporate  higher string modes through a Pomeron vertex operator via a world-sheet OPE. More directly, one can  adopt a $J$-plane formalism, where    the Pomeron kernel $\widetilde {\cal K}_P$  can be given by an inverse Mellin transform, as in  Eq. (\ref{eq:Pomeron-Mellin}), 
\bea
\widetilde{\cal K}_P(s,t,z,z') &=&
-  \int_{L-i\infty}^{L+i\infty} \frac{dj}{2\pi i} (
\alpha'  \widetilde s)^{j} \frac{1 + e^{-i \pi j}}{\sin\pi j}   \widetilde G_j(t,z,z') \; ,\nonumber
\eea
with ${\rm Re}\, ( j-j_0)=L>0$.  Due to curvature of AdS, the effective spin of a graviton exchange is lowered from 2 to $j_0<2$.
The propagator
 $\widetilde G_j(z,z';t)$  can be found via a spectral analysis in either $t$ or $j$.  Let us focus  on the conformal limit. Holding $j>j_0$ real and working at leading order in $\lambda$,  the spectrum in $t$ can be seen to be continuous along its positive real axis leading to 
\bea
\widetilde G_j(z,z';t)
&= & \int_0^\infty kdk \frac{ J_{(\Delta(j)-2)}(kz)J_{(\Delta(j)-2)}(kz')}{k^2-t}\;,\label{eq:spectrumq3}
\eea
where $\Delta(j)=2+ \sqrt {2\sqrt{\lambda} (j-j_0)}$.   At $J=2$, this reduces to the graviton kernel.

An alternative spectral representation in $j$ has  also been provided in \cite{Brower:2006ea}. The leading contribution  to Eq. (\ref{eq:Pomeron-Mellin}) comes from a branch-cut at $j_0$ which corresponds to a coherent sum of contributions from string modes associated with the graviton.  
At $s$ large and $t=0$ the kernel becomes 
\bea\label{eq:forwardPom}
\widetilde{\cal K}_P(s,0,z,z') &=&
-( \frac{1 + e^{-i \pi j_0}}{\sin\pi j_0} ) ( \alpha'  \widetilde s)^{j_0} ,
\eea
  up to log corrections. This is the form we adopt for  inclusive central production.

\paragraph{Tensor Glueballs and Confinement Deformation:}
Consider next the addition of a confinement deformation, leading to a theory with a discrete hadron spectrum, e.g. tensor glueballs lying on the Pomeron trajectory.  To  gain a qualitative understanding, it is instructive to rely on the ``hard-wall" model,  where the AdS coordinate $z$ is restricted to lie in the range $[0,z_{max}]$; we take $z_{max}\sim 1/\Lambda_{QCD}$.  This model captures key features of confining theories with string theoretic dual descriptions.  
The propagator is now given by a discrete sum over allowed states as

\bea
\widetilde G_j(z,z';t)
&=&  (zz')^{-2} \sum_n \frac{ \widetilde \phi_n(z,j)\widetilde \phi_n(z',j) }{m^2_n(j) -t} \;,\label{eq:ReggeProp}
\eea

where  $\widetilde \phi_n(z,j)$ can  be expressed in terms of Bessel functions. Eq. (\ref{eq:ReggeProp}) extends Eq. (\ref{eq:scalarB2B}) to a sum over Regge trajectories.

\paragraph{Eikonalization:} 

From a string-dual perspective, summing higher order string diagrams leads to an eikonal summation.  More generally,  eikonalization assures s-channel unitarity.  Near-forward scattering in the high energy limit is referred to in some literature simply as  the eikonal limit; this limit corresponds to $s\rightarrow \infty$ and $t$ fixed, leading to a CM frame scattering angle $\theta$ that vanishes as   $\theta \sim 1/\sqrt s$. Under plausible assumptions, it can be shown that flat-space scattering in this limit is determined by the integration of an eikonal phase, $\chi(s,\vec b)$,  over the two-dimensional space of impact parameters $\vec b$.
In this eikonal form the reduced 5-D momentum transfer squared serves as a 3-d Laplacian, 
$\widetilde t \rightarrow   \nabla^2_{AdS_\perp}; $ 
and there is a diffusion kernel in 3-dimensional  transverse space, between $(x_\perp, z)$ and $(x'_\perp,z')$. In the eikonal limit~\cite{Brower:2007qh,Brower:2007xg,Cornalba:2006xm,Cornalba:2006xk,Cornalba:2007fs,Cornalba:2008qf,Cornalba:2009ax,Costa:2012cb},  
one finds
\be
T_{1+2\rightarrow 3+4}(s,t)\simeq (-2is)\int d\vec b e^{-i\vec q_\perp\cdot \vec b} \int dz dz' \Psi_{(13)}(z) \Psi_{(24)}(z')\Big\{e^{i\chi(s,\vec b, z, z')}-1\Big\}\, , \label{eq:eikonal}
\ee
 where $\vec b= x'_\perp-x_\perp$ is the impact parameter. Expanding to first order, we thus can identify our { Pomeron kernel} with the eikonal $\chi(s,\vec b, z, z')$ as
\be
\widetilde{\cal K}_P(s,t,z,z') = 2s \int d\vec b e^{-i\vec q_\perp\cdot \vec b}\chi(s,\vec b, z, z'). \label{eq:eikonal-2}
\ee
When the eikonal becomes large, $\chi(s,\vec b, z, z')=O(1)$, multiple Pomeron exchange becomes important leading to effects like saturation.  For many purposes, for example DIS at HERA, keeping  a single Pomeron contribution is often sufficient. For p-p and p-Pb scatterings,  eikonalization  is also phenomenologically important.  This can be seen in effects like ``taming" the power increase for total cross sections with $ s^\epsilon$ to $\log^2 s$, etc.

\section{Conformal Partial-Wave and Regge Theory:}\label{sec:ConformalRegge}

The Regge limit for CFT can also be addressed more directly by analytically continuing the Euclidean OPE to Minkowski space. 
 We will now briefly discuss this approach, which will lead us to an alternate derivation of Eq. (\ref{eq:Pomeron-Mellin}).
We will focus on a four-point correlation function of primary operators ${\cal O}_i$ of dimensions $\Delta_i$.  For a $t$-channel OPE, it is customary to express the 4-point correlation function for external  scalars as 
$$
\langle 0|{\cal O}_1(x_1) {\cal O}_3(x_3){\cal O}_2(x_2){\cal O}_4(x_4)|0\rangle =
\frac{1}{(x^2_{13})^{\Delta_1}(x^2_{24})^{\Delta_2}} 
\,F(u,v)\,, 
$$
where we define $x_{ij}=x_i-x_j$  and the invariant cross ratios $
u=\frac{x_{13}^2x_{24}^2}{x_{12}^2x_{34}^2}\,$ and 
$
v=\frac{x_{14}^2x_{23}^2}{x_{12}^2x_{34}^2}.  
$
For simplicity we have assumed
$\Delta_1=\Delta_3$ and $\Delta_2=\Delta_4$.  To explore conformal invariance, one normally begins with  a  conformal partial wave expansion~\cite{Cornalba:2007fs,Cornalba:2008qf,Cornalba:2009ax,Costa:2012cb}, starting first in an Euclidean setting,  
where the connected component of the amplitude  $F(u,v)$ is given 
by a sum over conformal blocks,
\be
F(u,v)= \sum_{j}  \sum_\alpha C^{(13),(24)}_{\alpha,j}
\,  G( j,\Delta_{\alpha}(j);u,v) \, . 
\label{eq:ConformalBlock}
\ee
  For planar
${\cal N}=4$ SYM, we restrict the sum to  single-trace conformal  primary operators. 

\subsection{OPE in Minkowski Setting:}
The conformal Regge limit  corresponds to a double light-cone  limit  in a Minkowski setting. This light-cone limit for the OPE corresponds to $u \rightarrow 0$ and $v\rightarrow 1$, with   $(1 - v)/\sqrt{u}$ fixed. 
Equivalently, by introducing $u = z\bar z$ and $v = (1 - z) (1 - \bar z)$
with $z =\sigma e^{\rho}$ and $\bar z = \sigma e^{-\rho}$, the precise Regge limit
can also be specified by    
\be
\sigma \rightarrow 0, \,\,\,\, \rho \,\, \text{fixed}.
\ee
In a frame where $x_{1\perp}=x_{3\perp}$ and $x_{2\perp}=x_{4\perp}$,  this limit corresponds to approaching the respective null infinity
while keeping the relative impact parameter
$
b_\perp= x_{1\perp}-x_{2\perp}\,
\label{eq:Impact}
$ fixed.

 To make contact with Regge theory, it is useful to adopt a more 
general starting point. We introduce a basis ${\cal G} (j,\nu ; u,v)$ of functions for the  principle unitary conformal  representation of the four-dimensional conformal group $SO(5,1)$ and then expand   $F(u,v)$ in terms of this basis as
\be
F(u,v)= \sum_{j} \int^{\infty}_{-\infty} \,\frac{d \nu}{2\pi } \, a  (j, \nu) \, {\cal G} (j,\nu ; u,v)\,. 
\label{eq:groupexpansion}
\ee 
The conformal harmonics ${\cal G} (j,\nu ; u,v)$ are eigenfunctions  of the quadratic Casimir operator of $SO(5,1)$. Eq. (\ref{eq:groupexpansion})  combines a discrete sum in the spin $j$ and a  Mellin transform in  a complex $\Delta$-plane,
with $\Delta = 2 + i\nu$.  To recover the standard conformal block expansion, one can  close   the contour in the  $\nu$-plane~\cite{Mack:2009mi},  picking up 
dynamical poles in $a(j, \nu)$, at $\nu(j)=-i( \Delta(j)-2)$, 
thus arriving at Eq. 
(\ref{eq:ConformalBlock}).    These
dynamical poles correspond to the allowed
conformal primaries ${\cal O}_{\Delta(j)}$ of spin $j$ and
dimension $\Delta(j)$.

In continuing to the Minkowski limit, it is necessary to work with conformal harmonics ${\widetilde{\cal G}} (j,\nu ; u,v)$ which are eigenfunctions of $SO(4,2)$ Casimir with two continuous indices, $\nu$ and $j$. A distinguishing feature  for the Minkowski conformal harmonics is the fact that, in the Regge limit, 
$
{\widetilde{\cal G}} (j,\nu ; u,v) \sim \sigma^{1-j}\Omega_{i\nu}(\rho)=\sigma^{1-j} \frac{1}{4\pi^2}\, \frac{\nu \sin (\nu \rho)
}{\sinh \rho}\,$,  so that the ${\widetilde{\cal G}} (j,\nu ; u,v)$ 
are more and more divergent  for  increasing  $j>1$ as  $\sigma\rightarrow 0$. It follows that the conventional discrete sum over spin would no longer converge.
As explained in \cite{Brower:2014wha},  a Sommerfeld-Watson resummation leads to a double-Mellin representation 
\be
F(u,v)= - \int_{L-i\infty}^{L+i\infty}  \frac{dj}{2\pi i}   \;    \sum_{\tau=\pm}  \frac
{1 +\tau  e^{-i \pi j } }{\sin\pi j }   \int^{\infty}_{-\infty}
\frac{d\nu}{2\pi }  \,\, a_\tau(j, \nu)\,\,  \widetilde{\cal G} (j,\nu ; u,v) \, ,
\label{eq:newgroupexpansion}
\ee
where the contour in $j$ is  to stay to the right of singularities of $a_\tau(j, \nu)$.  The factor, $\frac
{1 +\tau  e^{-i \pi j } }{\sin\pi j } $, is referred to as the ``signature factor". We will in what follows consider even signature case, $\tau=+$. For more discussions, see \cite{Brower:2014wha}.

\subsection{Conformal Regge Theory and Eikonal:}\label{sec:eikonal}
Conformal Regge theory assumes that $a(j,\nu)$ is meromorphic   in the $\nu^2-j$ plane, with
poles specified by the collection of allowed spectral curves
$\Delta_\alpha(j)$.  An example of such an $a$ is
$
a(j,\nu) =\sum_\alpha \frac{r_\alpha(j)}{\nu^2 + (\Delta_\alpha(j) -2)^2 }
$.
In the Regge limit, for even signature, $\tau=+$, the  spectral curve associated with the energy-momentum tensor plays the dominant role.  Here $\Delta_P(2)=4$ and this spectral curve is where the Pomeron singularity lies, as in Fig. \ref{fig:Delta-J}.  Keeping this contribution only leads directly to the Pomeron kernel in Eq. (\ref{eq:Pomeron-Mellin}).
For more discussions, see~\cite{Brower:2010wf,Costa:2013uia} and \cite{Brower:2006ea,Brower:2014wha}.

In flat-space, by expanding Eq. (\ref{eq:eikonal}) to first order in $\chi$ and applying Eq. (\ref{eq:Optical4}), one can see that exchanging the eikonal once contributes to the total cross section, so that $
\sigma_{total}(s)  \simeq 2 \int d\vec b \, \chi_I(s,\vec b) + O(\chi^2)$, where $\chi_I>0$ is the imaginary part  of the eikonal.
With AdS/CFT, it is possible to associate  the eikonal with the leading $t$-channel exchange, as is done in \cite{Brower:2006ea}. The result is the leading (Pomeron)  kernel, given by Eq. (\ref{eq:eikonal-2}), and repeated here, 
\be
\widetilde {\cal K}_P(\widetilde s,\widetilde t, z, z')= \int d\vec b e^{i\vec q_\perp\cdot \vec b} {\cal K}_P(s,\vec b, z, z')\simeq  (2 s)  \int d\vec b e^{i\vec q_\perp\cdot \vec b} \chi(s,\vec b, z, z')\, , \label{eq:PomeronKernel}
\ee
where $ \widetilde s$ and $\widetilde t$ are redshifted holographic invariants, as in Eq. (\ref{eq:redshift}). In the conformal limit, Eq. (\ref{eq:PomeronKernel}) provides a representation for a general scattering kernel. 
The eikonal ${\chi}$ encodes all dynamical information and, due to conformal symmetry, depends only on 
$
\widetilde s=zz' s $ and $
\cosh \xi=\frac{z^2+z'^2+b_\perp^2}{2zz'}$,
where $\cosh \xi$ corresponds to a transverse chordal distance. The Regge limit is now $\widetilde s\to \infty$ with fixed $\xi$.
It is important to note that the conformal  representation (\ref{eq:PomeronKernel})  is valid for any value of the coupling constant, since it relies
only on conformal invariance.    We end by providing a {Regge Dictionary for CFT:} 
\be
F(u,v) \leftrightarrow{\chi}(\widetilde s, \xi)\,; \quad 
\sigma =\sqrt u \,\, \leftrightarrow \,\, {\widetilde s}^{-1}\,; \quad 
\cosh \rho \approx \frac{1-v}{2\sqrt u} \,\, \leftrightarrow \,\, \cosh \xi = \frac{b_\perp^2 + z^2 + {z'}^2}{2 z z'}\,.  \label{eq:CFT-Regge-dictionary}
\ee 
For more details, see \cite{Brower:2014wha}.

\section{Flat-Space String Amplitudes}\label{sec:FlatString}

Here we describe and evaluate some flat-space string amplitudes. As an illustration, we will begin with tree-level amplitudes for tachyons in bosonic string theory. The four-point open-string tachyon amplitude, known as the Veneziano amplitude, can be expressed in a Koba-Nielson form as

\be
{\cal A}_0(s,t) =\int^1_0 dw\; (1-w)^{ - 2 - \alpha's}
\;w^{ -2 -  \alpha't }. \label{eq:veneziano}
\ee

This is a planar-ordered amplitude, with singularities in $s$ and $t$ only. The full amplitude is given as a sum of three planar amplitudes, with singularities in $(t,u)$ and $(u,s)$ respectively, 
$
{\cal A}_{open}(s,t)= {\cal A}_0(s,t)+{\cal A}_0(t,u)+{\cal A}_0(u,s).
$
Since the external particles are tachyons, we have $\alpha'(s+t+u) = -4\,$. The corresponding 4-point closed-string tachyon amplitude is the Virasoro amplitude, given by
\be
{\cal A}_{closed}(s,t) =  \int d^2w\, |w|^{-4 - \alpha' t/2} |1 - w|^{-4-\alpha' s/2} \ ,
\label{eq:virasoro}
\ee
where for closed strings $\alpha'(s+t+u) = -16$. Unlike the Veneziano amplitude, the Virasoro amplitude contains singularities in all three channels. There exists  closed-form expressions for these integrals in terms of $\Gamma$-functions.

\paragraph{Fixed-Angle Limit for 4-Point Amplitudes}
For four-point scattering, the limit of fixed angle scattering is given by large $s$ and $t$, with $s/t$ held fixed. In the CM frame, we then have
\be
t\simeq - s (1-\cos\theta)/2.
\ee
It is possible to read off the behavior for the Veneziano formula directly, but it is more instructive to work with the Koba-Nielson representation.  Consider the open-string amplitude. When $s$ and $t$ are both large, with $t/s$ fixed, the integrand has a saddle point at $w^* = t/(s+t)$.  When the integral is appropriately defined by analytic continuation, this saddle-point indeed dominates~\cite{Gross:1987kza}, and we then have
\be
{\cal A}_0(s,t) \sim e^{-f(\theta) (\alpha' s)}  \ .\label{eq:fixedangleflat}
\ee
A similar analysis applies to all three terms for ${\cal A}_{open}(s,t)$. This property is clearly also shared for closed string amplitudes. It can be shown that the integral for ${\cal A}_{closed}(s,t) $  is again dominated by the saddle-point  at $w^* = t/(s+t)$, thus leading to an expression like that in Eq. (\ref{eq:fixedangleflat}) but with $\alpha'$ replaced by $\alpha'/2$.   This represents a generic property, which also applies to multiparticle amplitudes: in the fixed-angle limit where all invariants are large with relative ratios fixed, all flat-space string amplitudes are exponentially suppressed. 
 
\paragraph{Regge Limit for 4-Point Amplitudes:}

In the Regge limit of $s\rightarrow \infty$ with $t$ fixed, the saddle-point $w^*$ moves to one of the end-points of the domain of integration, $w=0$, and the amplitude can no longer be evaluated at $w^*$. Instead, we must sum the contributions from $w=O(1/s)$. In \cite{Brower:2010wf}, it was shown that this summation corresponds to a world-sheet OPE, and can be represented by a Reggeon vertex operator. More directly, one finds,  

\begin{equation}
{\cal A}_0(s,t)  \simeq (-\alpha' s )^{1+\alpha' t} \int_0^\infty d z z^{ -2 -  \alpha't } e^{-z} 
=  \Gamma(-1 - \alpha't  ) ( e^{- i\pi}
\alpha' s)^{1 + \alpha't } \; .  
\end{equation}

Consider next ${\cal A}_0(t,u)$ and ${\cal A}_0(u,s)$.  For ${\cal A}_0(u,s)$, this corresponds to a fixed-angle limit  and its contribution  is exponentially suppressed.  For ${\cal A}_0(t,u)$, it leads to $\Gamma(-1 - \alpha't  ) ( e^{- i\pi}\alpha' u)^{1 + \alpha't } \simeq \Gamma(-1 - \alpha't  ) ( \alpha' s)^{1 + \alpha't } $, leading to a total contribution that can be expressed as
\bea
{\cal A}_{open}(s,t)&\simeq &\pi \Gamma( \alpha't  ) \frac{( 1+  e^{- i\pi(1 + \alpha't )})}{\sin \pi (1 + \alpha't )} (\alpha' s)^{1 + \alpha't } .
\eea
In the physical region where $s>0$ and $t<0$, the discontinuity formula corresponds to  
\be
{\rm Im}\, {\cal A}_{open}(s,t)\simeq \pi \Gamma( \alpha't  )  (\alpha' s)^{1 + \alpha't } \, .
\ee

The same analysis can also be carried out for the closed-string amplitude.  For large $s$ at fixed $t$, the region $w=O(s^{-1})$ dominates, leading to Regge behavior 
\be
{\cal A}_{closed} \sim 
2 \pi \frac{\Gamma(-1-\alpha't/4)}{\Gamma(2+\alpha't/4)} (e^{-i\pi/2} \apm s/4)^{2 + \apm t/2}.
\ee

\paragraph{Double-Regge Limit for 5-Point Amplitude:} 

We  will be interested five-point string scattering, shown in Fig.  \ref{fig:2to3}, in the double-Regge limit, where we take  $s=(p_1+p_2)^2$, $s_1=(p_3+p_c)^2$, and $s_2=(p_5+p_c)^2$, large, with $t_1=(p_3-p_1)^2$, $t_2=(p_2-p_5)^2$ and $\kappa \equiv \frac{s_1\, s_2}{s}$
fixed.  Consider a planar order amplitude $V_5$ with   planar ordering $(13452)$. For exploring the double-Regge limit, it is best to use the Koba-Nielson representation 
\be
V_5= \int_0^1 \frac{d u}{u^{1 + \alpha(t_1)} (1-u)^{ 1 + \alpha(s_1)}}  \int_0^1 \frac{d v }{v^{1 + \alpha(t_2)} (1-v)^{ 1 + \alpha(s_2)} } \, \,      (1-uv)^{\alpha(s)-\alpha(s_1)-\alpha(s_2)}\, .
\ee
Now we take the limit $s_1\rightarrow -\infty$, $s_2\rightarrow -\infty$ and $s\rightarrow -\infty$, with $\kappa=s_1s_2/s$ fixed, to find that 
$
V_5\simeq  (-\alpha' s_1)^{\alpha( t_1) } V_c(\alpha'\kappa,t_1,t_2) (-\alpha' s_2)^{\alpha(t_2)}
$,  where we have defined $\alpha(t)=1+\alpha' t$ as well as a central vertex coupling
\be
V_c (t_1,t_2,x) = \int_0^\infty dy_1\int_0^\infty dy_2 \,  y_1^{-\alpha't_1-2}y_2^{-\alpha't_2-2} e^{-y_1-y_2+\frac{ y_1y_2}{x}}, \label{eq:Vc}
\ee  
with $x = \frac {s_1s_2}{\alpha' s}$.  
This representation is valid for $\kappa<0$, and the physical region $\kappa>0$ is to be reached via analytic continuation. From Eq. (\ref{eq:Vc}), one observes that $V_c(x,t_1,t_2)$ is real-analytic, with a branch-cut over $0<x<\infty$.  For $x>0$, one  finds that
\bea\label{eq:5pt-kappa}
{\rm Im}\, V_c(x,t_1,t_2)  &=& \frac{\pi x^{-(\alpha_1+\alpha_2+1)} e^{-x} }{\Gamma(\alpha_1+1)\Gamma(\alpha_2+1)} \int_0^\infty du  \int_0^\infty dv \, u^{\alpha_1}v^{\alpha_2} \, e^{-(u+v +  \frac{u v}{x} )}
\nn
&=&\pi e^{-x} \Psi (\alpha_2+1,-\alpha_1+\alpha_2+1;x)\, ,
\eea
where $\Psi$ is the confluent hypergeometric function and we have also abbreviated $\alpha(t_i)$ by $\alpha_i$, $i=1,2$. Most importantly, for $x>0$ and $x\rightarrow \infty$,  ${\rm Im} V_c$ vanishes exponentially,
\be
{\rm Im}\, \,V_c(x,t_1,t_2)\simeq\pi x^{-\alpha_1-\alpha_2-1} e^{-x}, \label{eq:5pt-kappa-disc}
\ee
so that in this limit $V_c$ becomes real and factorizable,
$
V_c(x,t_1,t_2) \rightarrow \Gamma (\alpha_1) \Gamma(\alpha_2)\ .
$

We have considered so far only a particular planar ordering for the amplitude; to obtain the full amplitude, we need to sum over all other orderings, each of which we expect to have a similar double-Regge limit. A similar expression holds for  closed strings~\cite{Clavelli:1978tt,Ader:1977qy}.  In AdS/CFT,  the central vertex takes on the form $V_c (\widetilde t_1,\widetilde t_2,\widetilde x)$ with all invariants redshifted \cite{Herzog:2008mu} as appears in Eq. (\ref{eq:adsDoublePomeronScheme}).

\paragraph{The Six-Point String Amplitude}
\label{sec:t6}
From \cite{Detar:1971dj}, the six-point amplitude  depicted in Fig. \ref{fig:feyn6} is given by
\bea
V_6
&=& \int_0^1 \frac{d u }{u^{\alpha_1+1} (1-u)^{\alpha_{a\bar c}+1}} \int_0^1 \frac{d v }{v^{\alpha_2+1} (1-v)^{\alpha_{\bar b c}+1}} \int_0^1 \frac{d w }{w^{\alpha_{\bar a a \bar c}+1} (1-w)^{\alpha_{b\bar b c}+1}  }\nn
&& \times\,  (1-(u+v)w + uv w)^{\alpha_{a\bar a c}}\Big [\frac{1-(u+v)w + uv w} {(1-uw)(1-vw)}    \Big]^{\alpha(M^2)-\alpha_{a\bar c} - \alpha_{b\bar c}}.
\eea
In the double Regge limit, we have
$s_{a\bar c}\simeq s_{\bar a c}\rightarrow -\infty$, $s_{b\bar c}\simeq s_{\bar b c}\rightarrow -\infty$ with
$
M^2=(p_a+p_b-p_c)^2\simeq s=(p_a+p_b)^2\ ,$
and
$
\kappa \equiv \frac {s_{a\bar c}s_{b\bar c}}{M^2}
$
fixed.  In this limit, the dominant contribution comes from $u=O(1/s_{a\bar c})$ and $v=O(1/s_{b\bar c})$, and one finds
$
V_6 \rightarrow  (-\alpha_{a\bar c})^{\alpha_1}\, {\cal V}_{c\bar c}\, (-\alpha_{b\bar c})^{\alpha_2}
$,
where 
\be
{\cal V}_{c\bar c}= \int_0^1 \frac{d y }{y^{\alpha_{\bar a a \bar c}+1} (1-y)^{\alpha_{b\bar b c}+1}  } 
 V_c(t_1,t_2, \frac{\alpha' \kappa}{y(1-y)}) \, ,   
\ee
with the usual identification of invariants with it linear trajectory function, e.g., $\alpha(t) = \alpha' t +1$. The discontinuity in $M^2$, which now enters through $\kappa$, is given by
\be
{\rm Im}\,\, {\cal V}_{c\bar c}(\kappa, t_1,t_2) =\int_0^1 \frac{d y }{y^{\alpha_{\bar a a \bar c}+1} (1-y)^{\alpha_{b\bar b c}+1}  }  {\rm Im} \,V_c(t_1,t_2, \frac{\alpha' \kappa}{y(1-y)})\, . 
\ee
 For external tachyons, at  $t_1=t_2=0$ we have $\alpha_{\bar a a \bar c}(0)=\alpha_{b\bar b c}(0)=0$,  with ${\rm Im}\, {\cal V}_{c\bar c}(0,0,\kappa)$ finite.

\section{Fit Validation and Parameter Stability}
\label{sec:Validation}

Here we provide more details on the fits to data presented in Sec. \ref{sec:Test}, specifically with respect to implementing a cut-off and the stability of parameters; the discussion here will be focused on technical details, and physical interpretation will be left in Sec. \ref{sec:Test} and \ref{sec:discussion}. 
\subsection{Power-Law Behavior}
As discussed above, the arguments of Sec. \ref{sec:CentralProduction} suggest that the cross section should behave as \be\frac{d^3\sigma}{d^2p_\perp d\eta} \sim p_{\perp}^{-2\Delta},\label{eq:dsig}\ee where $\Delta$ is the conformal weight of the particle mediating production in the bulk. This naively indicates that we should fit to data a power-law curve of the form \be\frac{d^3\sigma}{d^2p_\perp d\eta}  = \frac{A}{p_\perp^B},\label{eq:2param}\ee where the overall normalization $A$ and exponent $B$ are floated.  

However, this formula is only expected to be true asymptotically as $p_\perp\to\infty$. In general, there are expected to be small-$p_\perp$ effects that are not visible to our analysis. This can be easily seen by noticing that the cross section diverges as $p_\perp\to0$. There are several ways one could imagine modifying Eq. (\ref{eq:dsig}) to include these effects. One particularly obvious way to avoid these effects is to fit a sum of the power-law curve and \emph{some other} curve to data; in this approach, the second curve is intended to model directly the low-$p_T$ physics. Such an approach was recently advocated in \cite{Bylinkin:2014qea,lhcfit}. Although we are not interested here in this region we comment on it below in Section ~\ref{sec:smallpt}. Given that we are not interested in these non-universal effects at small momenta, we have no principled reason to prefer any one form of this low-$p_\perp$ curve over any other. Especially given that introducing such an extra curve would drastically increase the number of floated parameters, and hence potentially lead to overfitting, it is best to be more agnostic as to the form of the low-$p_\perp$ effects. 

We will therefore consider simpler ways to remove low-$p_\perp$ effects. Perhaps the most obvious solution would be to simply introduce a lower cuttoff $p_{\text{min}}$ on the allowed $p_\perp$, and therefore only fit to a subset of each data sample. Another approach is to allow a small offset in the momentum that appears in the power law curve, i.e. to fit a three-parameter curve of the form \be \frac{d^3\sigma}{d^2p_\perp d\eta}  = \frac{A}{\left(p_\perp+C\right)^B}\, , \label{eq:3param} \ee instead of the two-parameter form presented in Eq. (\ref{eq:2param}). This form has two advantages. First, for $C>0$, the numerical singularity at $p_\perp=0$ is directly removed; additionally, as $p_\perp\to\infty$, it is readily seen to agree with Eq. (\ref{eq:dsig}). One could imagine adding in a lower cutoff to this form of the curve as well. 

Without a handle on the small-$p_\perp$ physics, we have no theoretical reason to prefer one of these approaches over the other. We will therefore fit both forms to data, both with and without a cutoff, and choose the approach that gives the quantitatively best overall results, as quantified by $\chi^2/$NDF. In the following pages, we will present a thorough evaluation of these two methods. For each of the five datasets discussed in the main text, we will present the results of twenty-two fits to data, corresponding to eleven different cutoffs $p_{\text{min}}$ for each of the two fit functions in Eqs. (\ref{eq:2param}) and (\ref{eq:3param}). We will also display some characteristic plots, to facilitate a visual analysis of the results.  

From the fit results in Tables \ref{tab:AppendixFirst} through \ref{tab:AppendixLast}, we can immediately exclude the proposal to fit Eq. (\ref{eq:2param}) directly to data. For all values of the cutoff tested, the $\chi^2$/NDF is unacceptable, being extremely high at small or no cutoff, and then rapidly falling to below one at large cutoff. This leads us to consider instead Eq. (\ref{eq:3param}), and leaves only the question of whether or not to institute a cutoff, and if so what value of the cutoff to use. For much the same reasons as above, we dispense with the possibility of a large cutoff. For cutoffs between 0 and 1.5 GeV, the gains in $\chi^2$/NDF are minimal for removing the low-$p_\perp$ data. Thus, to be conservative, and to minimize the overall statistical uncertainties, we will fit Eq. (\ref{eq:3param}) to data directly, without a cutoff. These are the results given in Section \ref{sec:Test}. 

\FloatBarrier

\clearpage

\begin{table}[ht]
\begin{centering}
\begin{tabular}{|c|c|c|c|}\hline
$p_{\text{min}}$/(1 GeV) & A/10 (GeV$^{-2}$) & B  & $\chi^2$/NDF \\\hline
0 & 0.0516 $\pm$ 0.00687 & 5.02 $\pm$ 0.164 & 51.2 \\\hline

0.5 & 0.0575 $\pm$ 0.00718 & 5.15 $\pm$ 0.148 & 29.8 \\\hline

1.0 & 0.0943 $\pm$ 0.0140 & 5.60 $\pm$ 0.139 & 3.21 \\\hline

1.5 & 0.153 $\pm$ 0.0585 & 5.88 $\pm$ 0.231 & 0.135 \\\hline

2.0 & 0.183 $\pm$ 0.131 & 5.97 $\pm$ 0.368 & 0.0412 \\\hline

2.5 & 0.199 $\pm$ 0.247 & 6.01 $\pm$ 0.578 & 0.0337 \\\hline

3.0 & 0.205 $\pm$ 0.291 & 6.027 $\pm$ 0.646 & 0.0316 \\\hline

3.5 & 0.218 $\pm$ 0.348 & 6.05 $\pm$ 0.712 & 0.0258 \\\hline

4.0 & 0.233 $\pm$ 0.416 & 6.07 $\pm$ 0.770 & 0.0189 \\\hline

4.5 & 0.253 $\pm$ 0.518 & 6.10 $\pm$ 0.846 & 0.0127 \\\hline

5.0 & 0.150 $\pm$ 0.736 & 5.93 $\pm$ 1.70 & 0.000621 \\\hline

\end{tabular}
\caption{Fitted values of parameters in Eq. (\ref{eq:2param}) for the ATLAS dataset at $\sqrt{s}=8$ TeV \cite{Aad:2016xww}.}
\label{tab:AppendixFirst}
\end{centering}
\end{table}

\begin{table}[ht]
\begin{centering}
\begin{tabular}{|c|c|c|c|c|}\hline
$p_{\text{min}}$/(1 GeV) & A/10 (GeV$^{-2}$) & B  & C/(1 GeV) & $\chi^2$/NDF \\\hline

0 & 4.46 $\pm$ 2.60 & 7.04 $\pm$ 0.264 & 1.07 $\pm$ 0.123 & 1.19 \\\hline

0.5 & 3.81 $\pm$ 2.98 & 6.98 $\pm$ 0.326 & 1.03 $\pm$ 0.182 & 1.08 \\\hline

1.0 & 1.16 $\pm$ 1.72 & 6.57 $\pm$ 0.554 & 0.689 $\pm$ 0.405 & 0.279 \\\hline

1.5 & 0.387 $\pm$ 1.06 & 6.21 $\pm$ 0.959 & 0.311 $\pm$ 0.917 & 0.0314 \\\hline

2.0 & 0.309 $\pm$ 1.22 & 6.14 $\pm$ 1.29 & 0.212 $\pm$ 1.573 & 0.0255 \\\hline

2.5 & 0.369 $\pm$ 2.15 & 6.19 $\pm$ 1.74 & 0.312 $\pm$ 2.89 & 0.0241 \\\hline

3.0 & 0.448 $\pm$ 3.10 & 6.24 $\pm$ 1.97 & 0.429 $\pm$ 3.74 & 0.0228 \\\hline

3.5 & 0.417 $\pm$ 3.76 & 6.22 $\pm$ 2.47 & 0.383 $\pm$ 5.29 & 0.0228 \\\hline

4.0 & 0.191 $\pm$ 2.03 & 6.02 $\pm$ 2.87 & -0.126 $\pm$ 6.64 & 0.0187 \\\hline

4.5 & 0.0465 $\pm$ 0.483 & 5.66 $\pm$ 2.81 & -1.09 $\pm$ 6.72 & 0.00502 \\\hline

5.0 & 0.325 $\pm$ 5.36 & 6.13 $\pm$ 4.13 & 0.629 $\pm$ 13.7 & 0.000189 \\\hline

\end{tabular}
\caption{Fitted values of parameters in Eq. (\ref{eq:3param}) for the ATLAS dataset at $\sqrt{s}=8$ TeV \cite{Aad:2016xww}.}
\end{centering}
\end{table}


\begin{figure}[!tbt]
\begin{center}
\hspace{-3cm}
\begin{subfigure}[b]{.45\textwidth}
\begin{flushleft}
\includegraphics[scale=0.5]{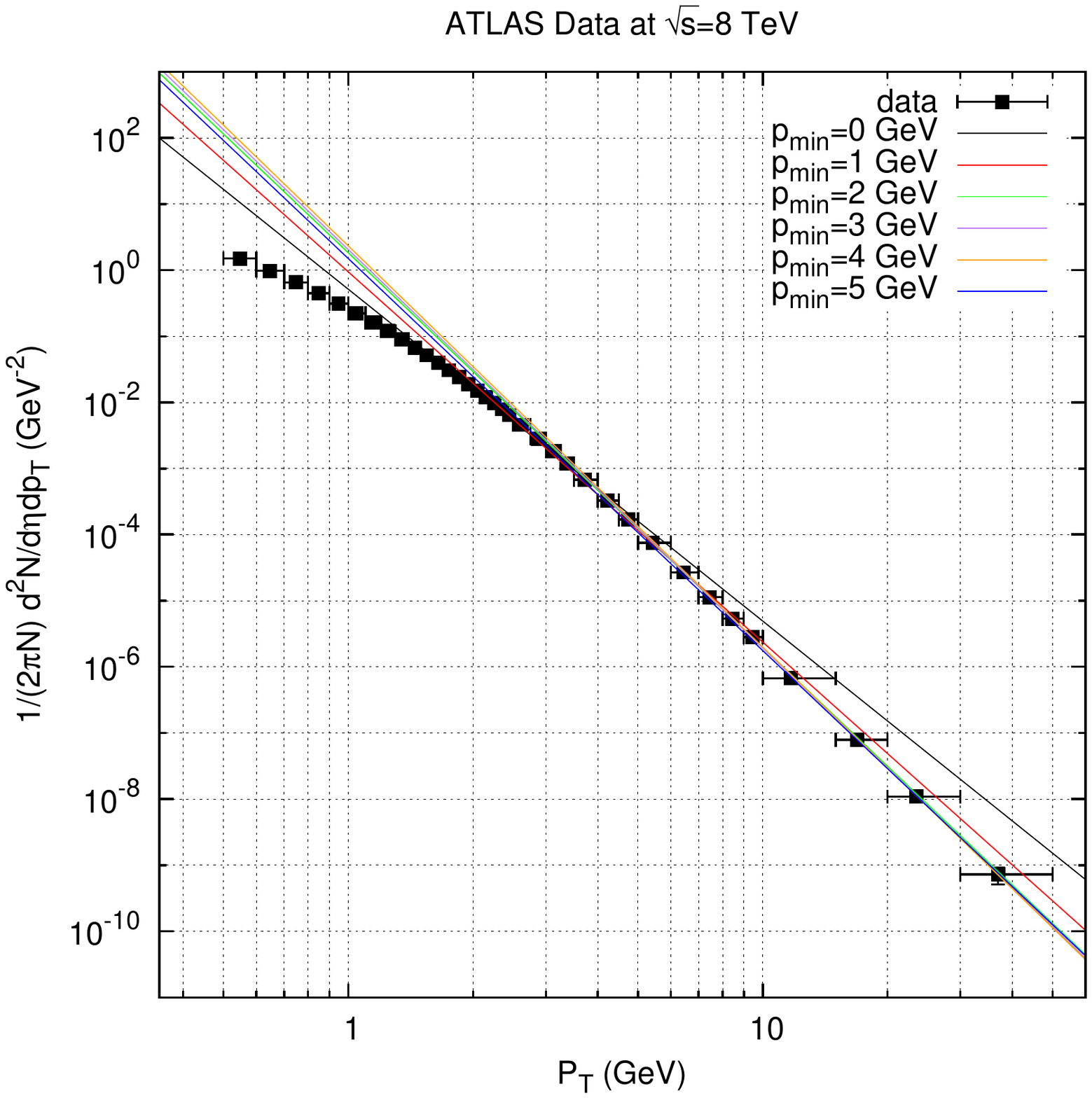}
 \vspace{-4cm}
\caption{}
\end{flushleft}
\end{subfigure}
~
\hspace{1cm}
\begin{subfigure}[b]{.45\textwidth}
\begin{flushright}
\includegraphics[scale=0.5]{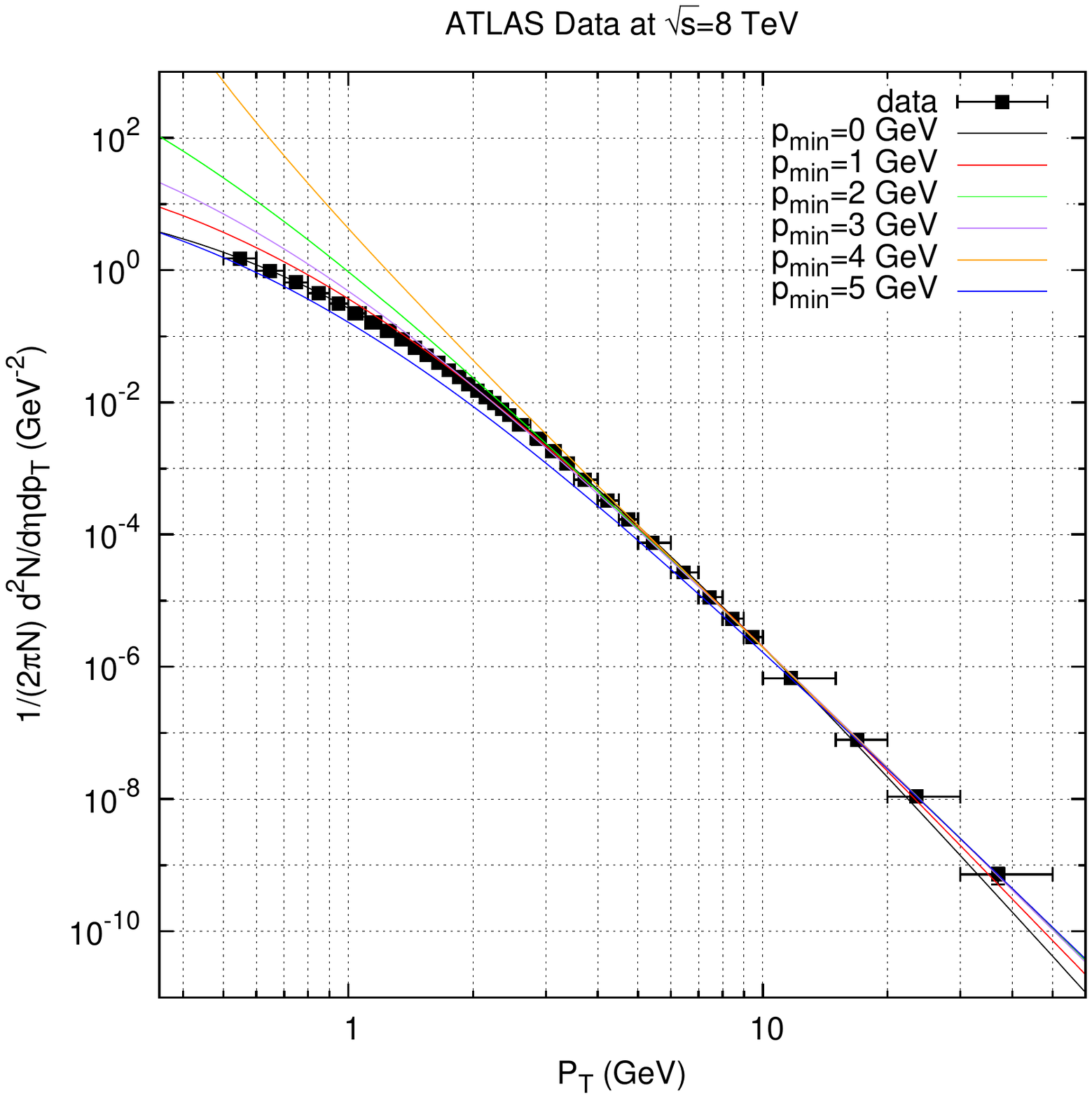}
 \vspace{-4cm}
\caption{}
\end{flushright}
\end{subfigure}
\caption{Fit of the two parameter ansatz in Eq. (\ref{eq:2param}) (left) and the three parameter ansatz in Eq. (\ref{eq:3param}) to the ATLAS dataset at $\sqrt{s}=8$ TeV \cite{Aad:2016xww} with various data cutoffs. }
\label{fig:atlasfigs2param}
\end{center}
\end{figure}

\ignore{
\begin{figure}[!tbt]
\begin{center}
\begin{subfigure}[b]{.45\textwidth}
\begin{centering}
\includegraphics[scale=.4]{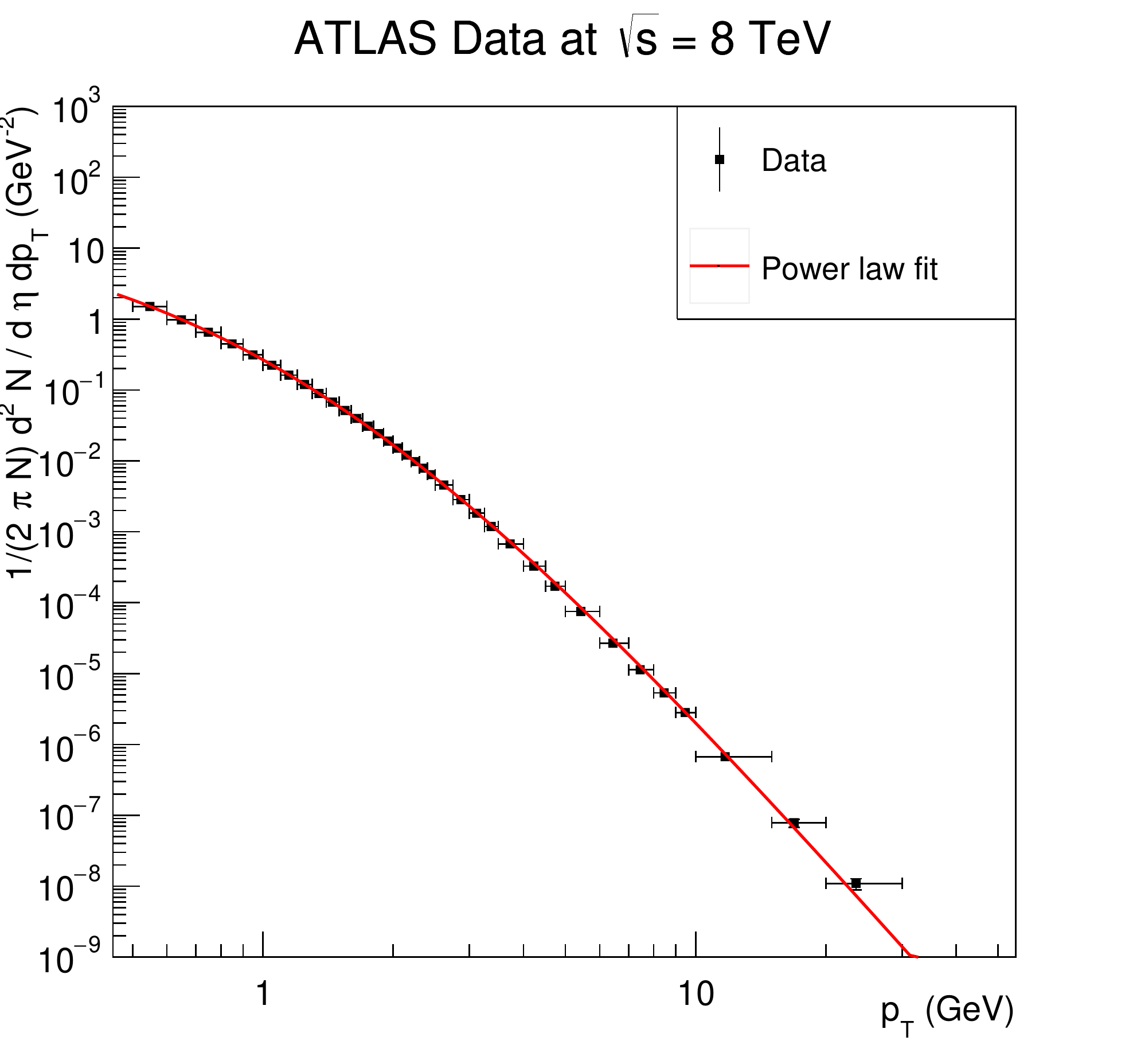}
\caption{}
\end{centering}
\end{subfigure}
~
\begin{subfigure}[b]{.45\textwidth}
\begin{centering}
\includegraphics[scale=.4]{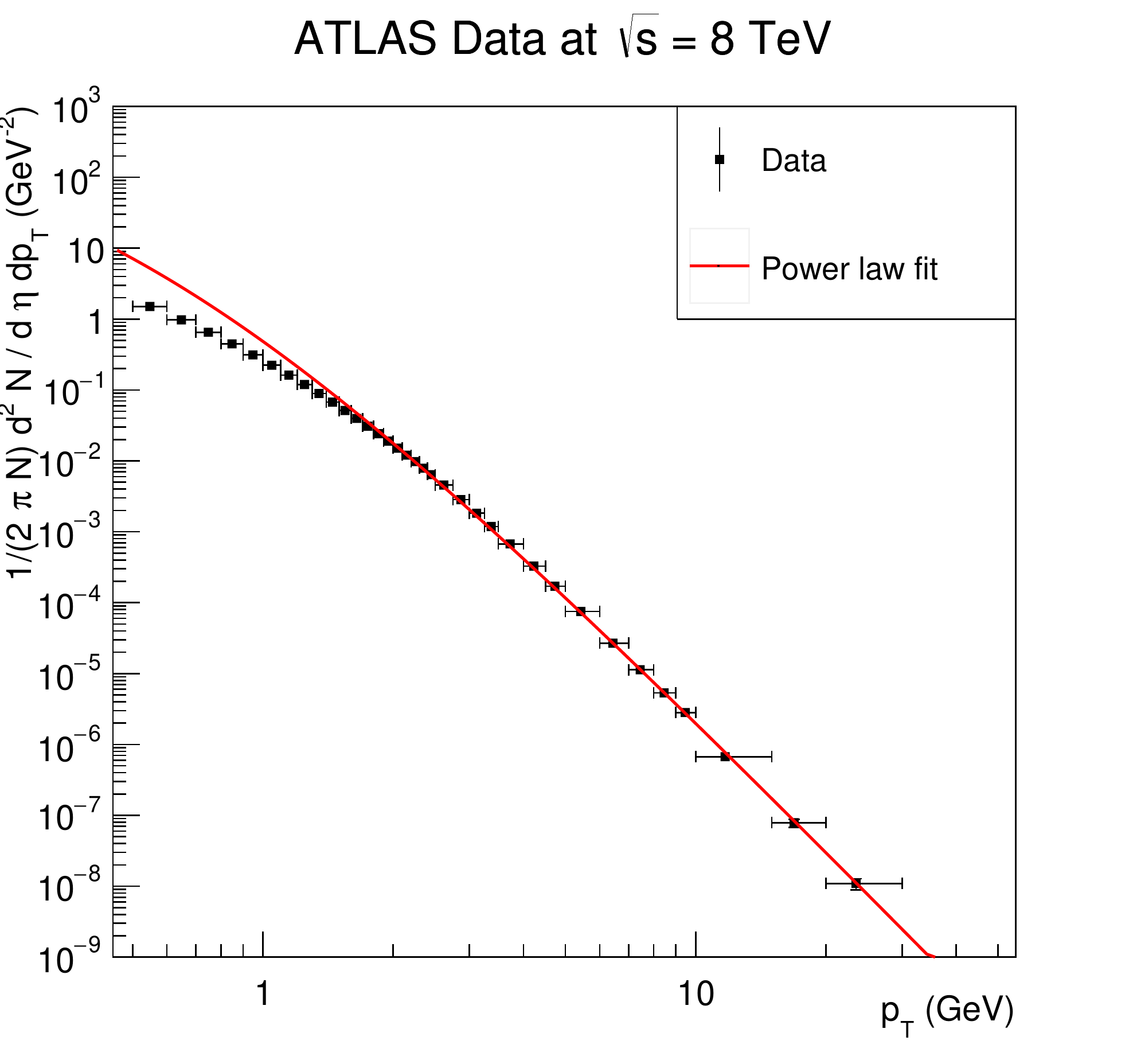}
\caption{}
\end{centering}
\end{subfigure}
\caption{Fit of the curve in Eq. (\ref{eq:3param}) to the ATLAS dataset at $\sqrt{s}=8$ TeV \cite{Aad:2016xww} with cutoffs $p_{\text{min}}$ = 0 (left) and 3 GeV (right). }
\label{fig:atlasfigs3param}
\end{center}
\end{figure}
}

\begin{table}[ht]
\begin{centering}
\begin{tabular}{|c|c|c|c|}\hline
$p_{\text{min}}$/(1 GeV) & A/10 (GeV$^{-2}$) & B  & $\chi^2$/NDF \\\hline
0 & 0.0583 $\pm$ 0.0070 & 4.85 $\pm$ 0.143 & 54.0 \\\hline

0.5 & 0.0648 $\pm$ 0.0076 & 4.97 $\pm$ 0.137 & 32.2 \\\hline

1.0 & -156. $\pm$ 0.372 & -207. $\pm$ 9.29 &  0.361 \\\hline

1.5 & 0.186 $\pm$ 0.063 & 5.79 $\pm$ 0.216 & 0.120 \\\hline

2.0 & 0.234 $\pm$ 0.170 & 5.90 $\pm$ 0.371 & 0.0489 \\\hline

2.5 & 0.270 $\pm$ 0.339 & 5.96 $\pm$ 0.583 & 0.0272 \\\hline

3.0 & 0.286 $\pm$ 0.405 & 5.99 $\pm$ 0.646 & 0.0203 \\\hline

3.5 & 0.300 $\pm$ 0.484 & 6.01 $\pm$ 0.717 & 0.0166 \\\hline

4.0 & 0.314 $\pm$ 0.574 & 6.025 $\pm$ 0.790 & 0.0141 \\\hline

4.5 & 0.305 $\pm$ 0.661 & 6.01 $\pm$ 0.897 & 0.0135 \\\hline

5.0 & 0.364 $\pm$ 2.03 & 6.07 $\pm$ 1.90 & 0.0125 \\\hline

\end{tabular}
\caption{Fitted values of parameters in Eq. (\ref{eq:2param}) for the ATLAS dataset at $\sqrt{s}=13$ TeV \cite{Aad:2016mok}.}
\end{centering}
\end{table}

\begin{table}[ht]
\begin{centering}
\begin{tabular}{|c|c|c|c|c|}\hline
$p_{\text{min}}$/(1 GeV) & A/10 (GeV$^{-2}$) & B  & C/(1 GeV) & $\chi^2$/NDF \\\hline
0 & 5.77 $\pm$ 3.39 & 6.96 $\pm$ 0.265 & 1.12 $\pm$ 0.126 & 0.852 \\\hline

0.5 & 5.22 $\pm$ 4.10 & 6.93 $\pm$ 0.328 & 1.09 $\pm$ 0.185 & 0.784 \\\hline

1.0 & 2.07 $\pm$ 2.95 & 6.61 $\pm$ 0.536 & 0.820 $\pm$ 0.396 & 0.223 \\\hline

1.5 & 0.667 $\pm$ 1.88 & 6.23 $\pm$ 0.979 & 0.431 $\pm$ 0.947 & 0.0178 \\\hline

2.0 & 0.544 $\pm$ 2.30 & 6.17 $\pm$ 1.38 & 0.341 $\pm$ 1.69 & 0.0138 \\\hline

2.5 & 0.621 $\pm$ 3.69 & 6.21 $\pm$ 1.78 & 0.418 $\pm$ 2.93 & 0.0125 \\\hline

3.0 & 0.634 $\pm$ 4.33 & 6.21 $\pm$ 1.97 & 0.429 $\pm$ 3.65 & 0.0124 \\\hline

3.5 & 0.670 $\pm$ 5.56 & 6.23 $\pm$ 2.31 & 0.463 $\pm$ 4.78 & 0.0124 \\\hline

4.0 & 0.652 $\pm$ 6.99 & 6.22 $\pm$ 2.90 & 0.445 $\pm$ 6.53 & 0.0124 \\\hline

4.5 & 2.18 $\pm$ 27.76 & 6.53 $\pm$ 3.31 & 1.26 $\pm$ 8.16 & 0.00694 \\\hline

5.0 & 45.2 $\pm$ 790. & 7.24 $\pm$ 4.24 & 3.80 $\pm$ 13.9 & 0.000184 \\\hline

\end{tabular}
\caption{Fitted values of parameters in Eq. (\ref{eq:3param}) for the ATLAS dataset at $\sqrt{s}=13$ TeV \cite{Aad:2016xww}.}
\end{centering}
\end{table}


\begin{figure}[!tbt]
\begin{center}
\hspace{-3cm}
\begin{subfigure}[b]{.45\textwidth}
\begin{flushleft}
\includegraphics[scale=0.5]{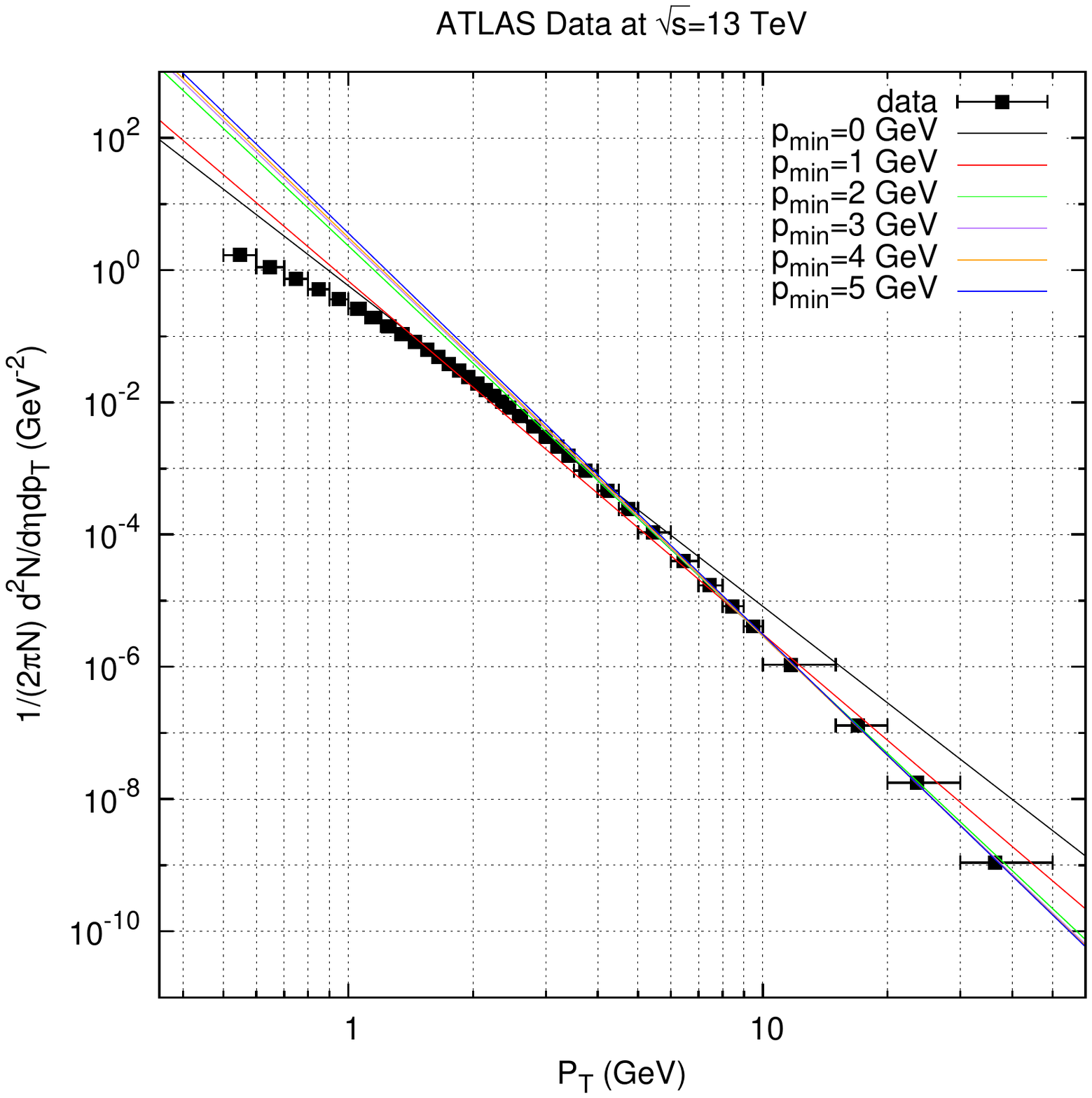}
 \vspace{-4cm}
\caption{}
\end{flushleft}
\end{subfigure}
~
\hspace{1cm}
\begin{subfigure}[b]{.45\textwidth}
\begin{flushright}
\includegraphics[scale=0.5]{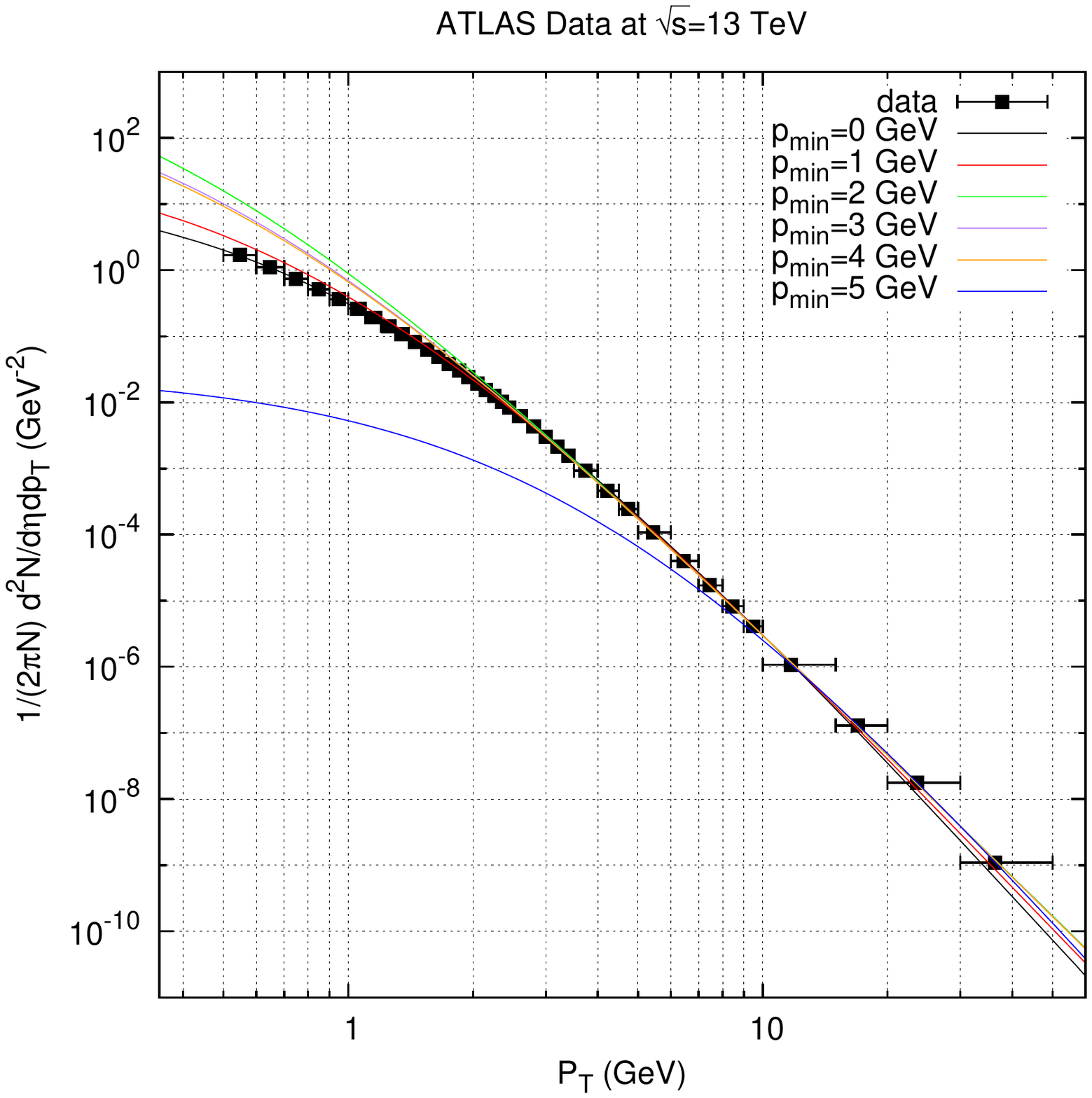}
 \vspace{-4cm}
\caption{}
\end{flushright}
\end{subfigure}
\caption{Fit of the two parameter ansatz in Eq. (\ref{eq:2param}) (left) and the three parameter ansatz in Eq. (\ref{eq:3param}) to the ATLAS dataset at $\sqrt{s}=13$ TeV \cite{Aad:2016mok} with various data cutoffs. }
\label{fig:atlasfigs2param13}
\end{center}
\end{figure}

\ignore{
\begin{figure}[!tbt]
\begin{center}
\begin{subfigure}[b]{.45\textwidth}
\begin{centering}
\includegraphics[scale=.4]{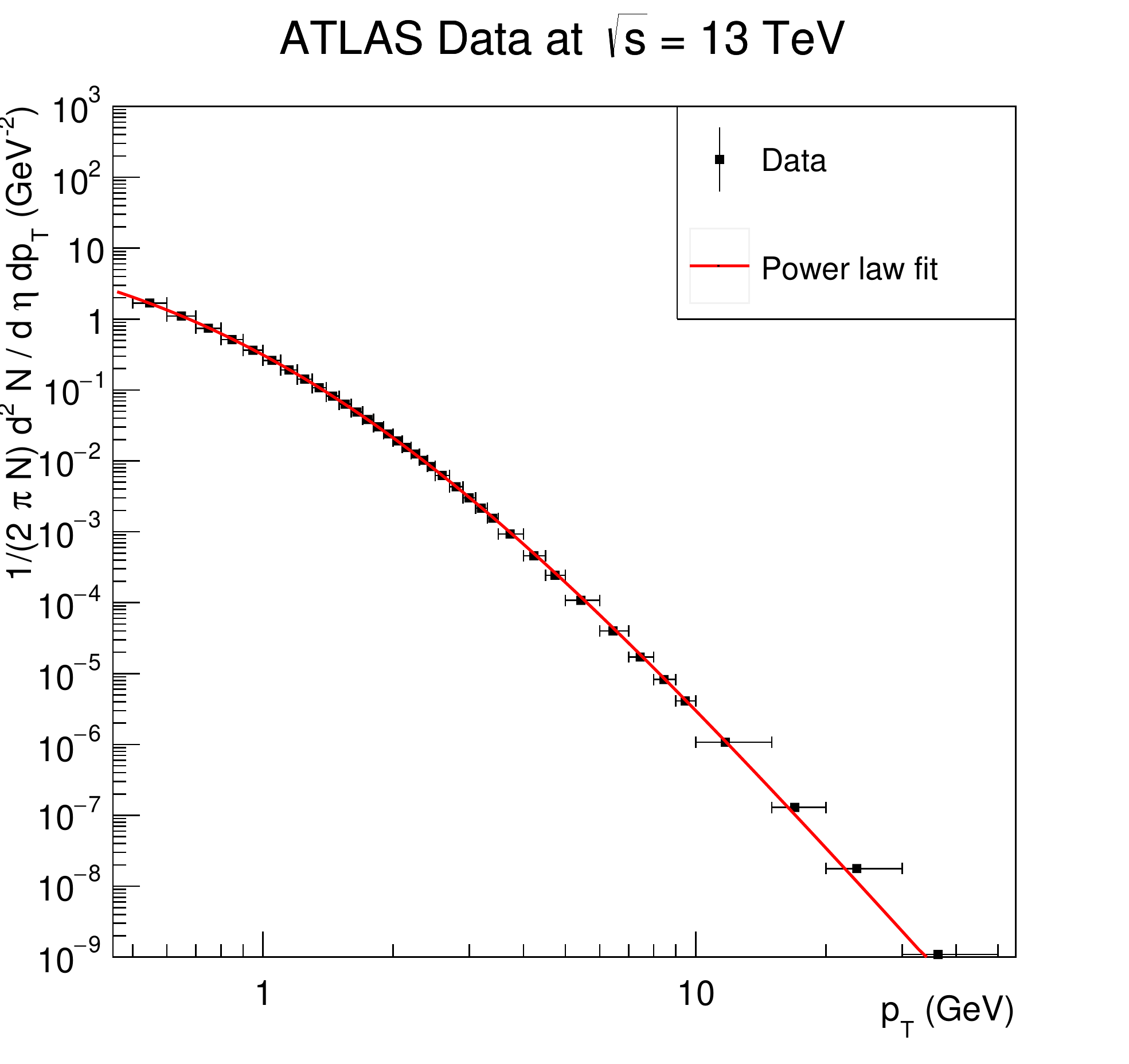}
\caption{}
\end{centering}
\end{subfigure}
~
\begin{subfigure}[b]{.45\textwidth}
\begin{centering}
\includegraphics[scale=.4]{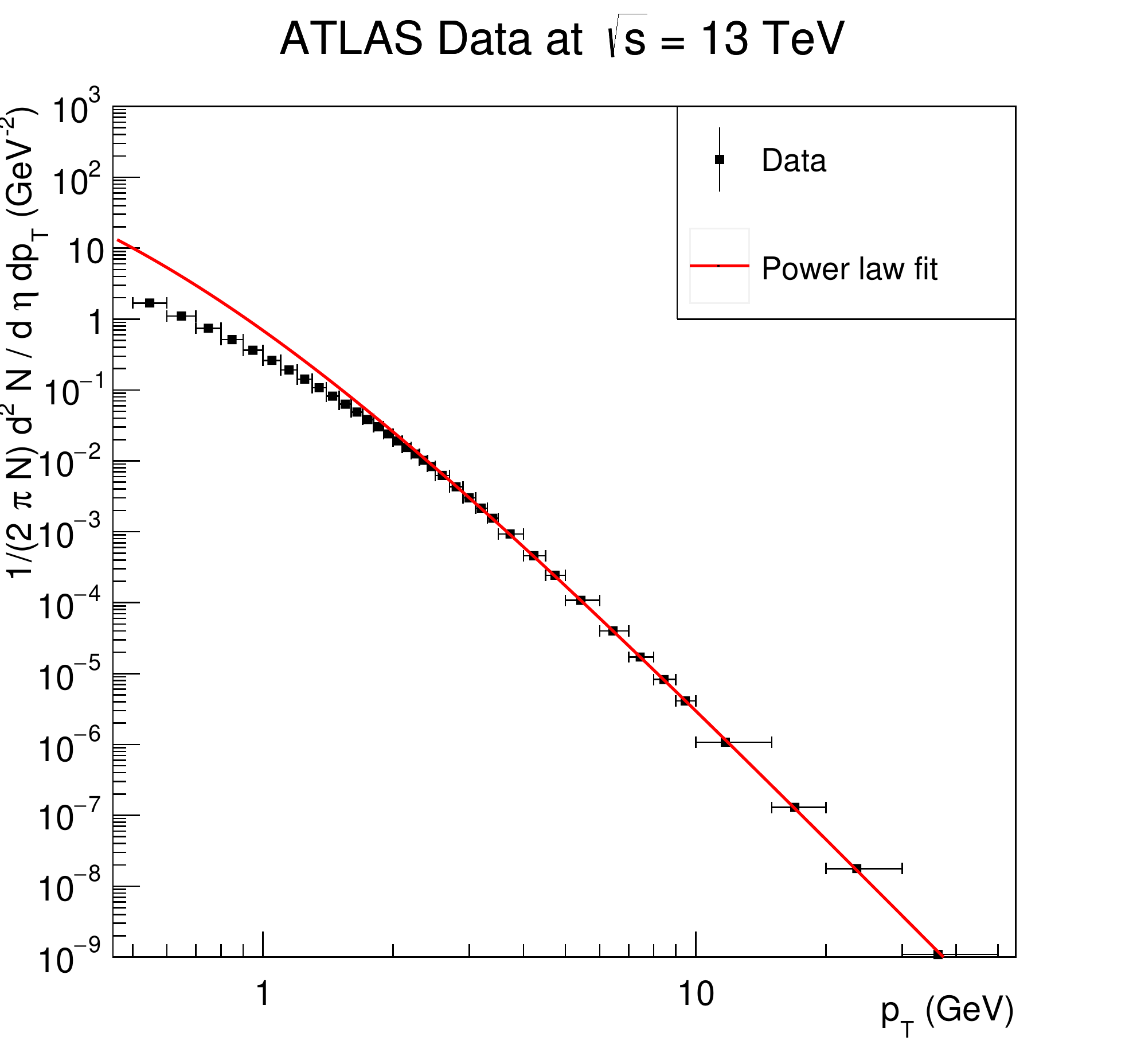}
\caption{}
\end{centering}
\end{subfigure}
\caption{Fit of the curve in Eq. (\ref{eq:3param}) to the ATLAS dataset at $\sqrt{s}=13$ TeV \cite{Aad:2016mok} with cutoffs $p_{\text{min}}$ = 0 (left) and 3 GeV (right). }
\label{fig:atlasfigs3param13}
\end{center}
\end{figure}
}

\begin{table}[ht]
\begin{centering}
\begin{tabular}{|c|c|c|c|}\hline
$p_{\text{min}}$/(1 GeV) & A/10 (GeV$^{-2}$) & B  & $\chi^2$/NDF \\\hline
0 & 0.535 $\pm$ 0.0638 & 5.87 $\pm$ 0.0710 &  352. \\\hline

0.5 & 0.528 $\pm$ 0.05674 & 5.85 $\pm$ 0.0662 &   91.1 \\\hline

1.0 & 0.868 $\pm$ 0.110 & 6.09 $\pm$ 0.0652 &  5.73 \\\hline

1.5 & 1.00 $\pm$ 0.151 & 6.15 $\pm$ 0.0720 &  0.771 \\\hline

2.0 & 1.18 $\pm$ 0.338 & 6.21 $\pm$ 0.116 &  0.100 \\\hline

2.5 & 1.19$\pm$ 0.386 & 6.21 $\pm$ 0.129 &  0.0989 \\\hline

3.0 & 1.17 $\pm$ 0.462 & 6.20 $\pm$ 0.152 & 0.0945 \\\hline

3.5 & 1.15 $\pm$ 0.647 & 6.20 $\pm$ 0.208 &  0.0890 \\\hline

4.0 & 1.13 $\pm$ 0.690 & 6.19 $\pm$ 0.222 &  0.0865 \\\hline

4.5 & 1.11 $\pm$ 0.735 & 6.19 $\pm$ 0.240 &  0.0807 \\\hline

5.0 & 1.12 $\pm$ 0.767 & 6.19 $\pm$ 0.246 &  0.0802 \\\hline

\end{tabular}
\caption{Fitted values of parameters in Eq. (\ref{eq:2param}) for the ALICE dataset at $\sqrt{s}=5.02$ TeV in the $|\eta|<0.3$ bin \cite{1405}.}
\end{centering}
\end{table}

\begin{table}[ht]
\begin{centering}
\begin{tabular}{|c|c|c|c|c|}\hline
$p_{\text{min}}$/(1 GeV) & A/10 (GeV$^{-2}$) & B  & C/(1 GeV) & $\chi^2$/NDF \\\hline

0 & 38.5 $\pm$ 8.27 & 7.23 $\pm$ 0.0853 & 1.32 $\pm$ 0.0445 & 14.7 \\\hline

0.5 & 22.0 $\pm$ 9.59 & 7.07 $\pm$ 0.146 & 1.14 $\pm$ 0.120 & 8.78 \\\hline

1.0 & 3.89 $\pm$ 2.99 & 6.55 $\pm$ 0.241 & 0.525 $\pm$ 0.261 & 1.68 \\\hline

1.5 & 1.86 $\pm$ 1.79 & 6.34 $\pm$ 0.298 & 0.232 $\pm$ 0.354 & 0.334 \\\hline

2.0 & 1.08 $\pm$ 1.63 & 6.18 $\pm$ 0.444 & -0.0403 $\pm$ 0.704 & 0.0971 \\\hline

2.5 & 0.991 $\pm$ 1.76 & 6.16 $\pm$ 0.514 & -0.0894 $\pm$ 0.886 & 0.0893 \\\hline

3.0 & 0.985 $\pm$ 2.05 & 6.16 $\pm$ 0.587 & -0.0926 $\pm$ 1.12 & 0.0882 \\\hline

3.5 & 0.946 $\pm$ 2.84 & 6.15 $\pm$ 0.808 & -0.124 $\pm$ 1.92 & 0.0853 \\\hline

4.0 & 0.979 $\pm$ 3.34 & 6.16 $\pm$ 0.904 & -0.0978 $\pm$ 2.28 & 0.0850 \\\hline

4.5 & 1.15 $\pm$ 4.51 & 6.20 $\pm$ 1.02 & 0.0252 $\pm$ 2.74 & 0.0807 \\\hline

5.0 & 1.09 $\pm$ 4.98 & 6.18 $\pm$ 1.18 & -0.0206 $\pm$ 3.33 & 0.0802 \\\hline

\end{tabular}
\caption{Fitted values of parameters in Eq. (\ref{eq:3param}) for the ALICE dataset at $\sqrt{s}=5.02$ TeV in the $|\eta|<0.3$ bin \cite{1405}.}
\end{centering}
\end{table}


\begin{figure}[!tbt]
\begin{center}
\hspace{-3cm}
\begin{subfigure}[b]{.45\textwidth}
\begin{flushleft}
\includegraphics[scale=0.5]{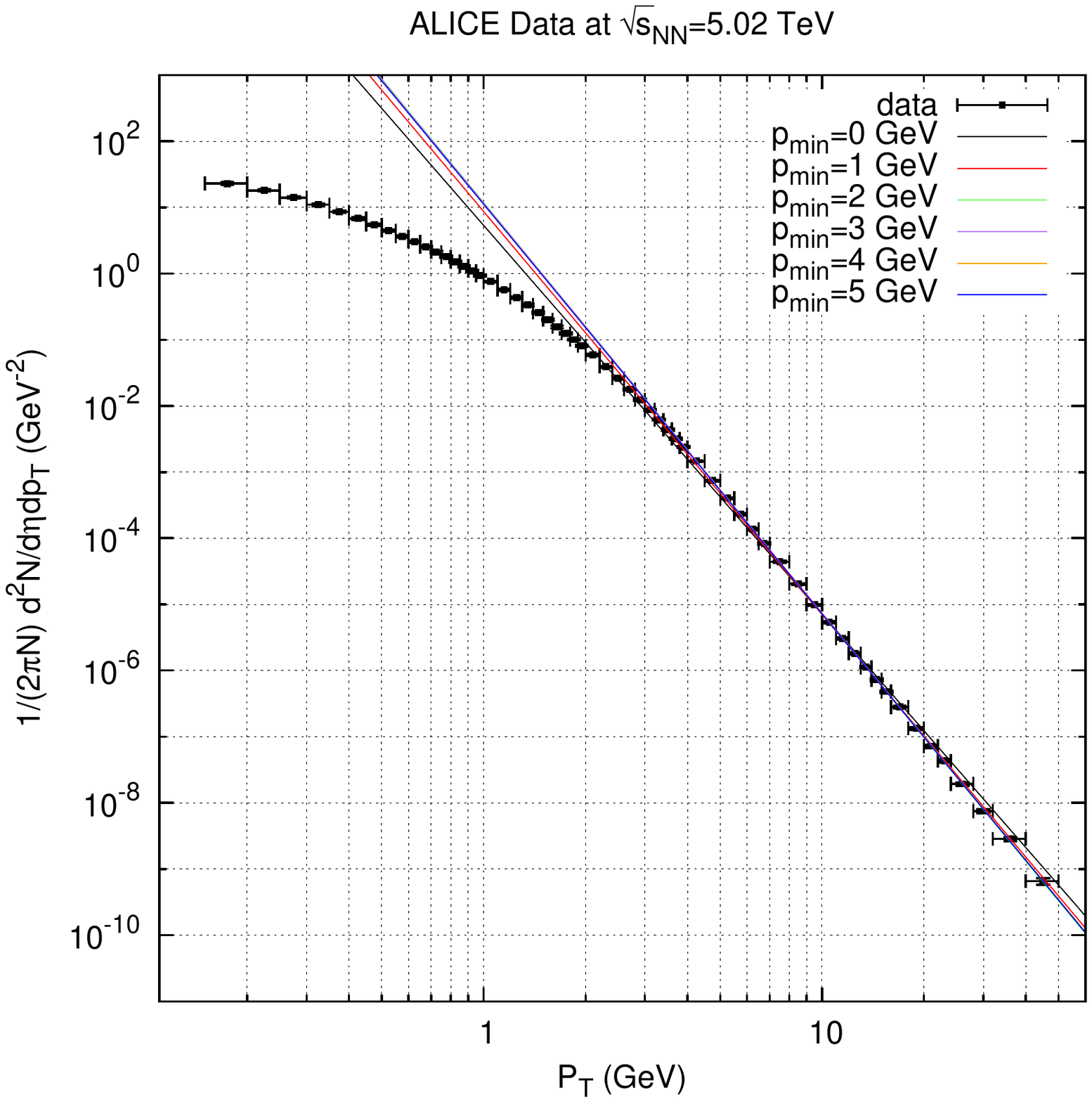}
 \vspace{-4cm}
\caption{}
\end{flushleft}
\end{subfigure}
~
\hspace{1cm}
\begin{subfigure}[b]{.45\textwidth}
\begin{flushright}
\includegraphics[scale=0.5]{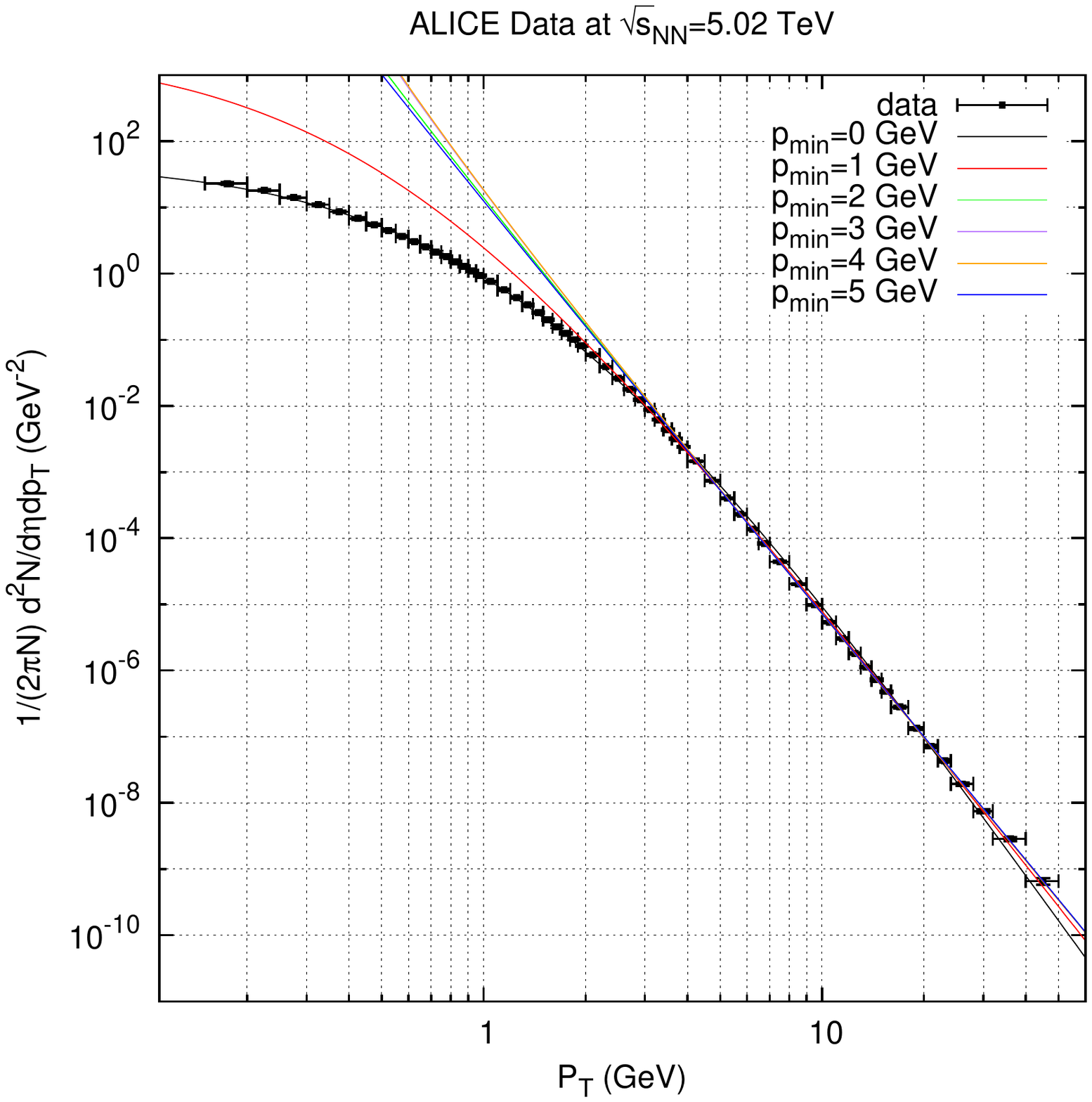}
 \vspace{-4cm}
\caption{}
\end{flushright}
\end{subfigure}
\caption{Fit of the two parameter ansatz in Eq. (\ref{eq:2param}) (left) and the three parameter ansatz in Eq. (\ref{eq:3param}) to the ALICE dataset at $\sqrt{s}=5.02$ TeV in the $|\eta|<0.3$ bin \cite{1405} with various data cutoffs. }
\label{fig:alicefigs2param}
\end{center}
\end{figure}

\ignore{
\begin{figure}[!tbt]
\begin{center}
\begin{subfigure}[b]{.45\textwidth}
\begin{centering}
\includegraphics[scale=.4]{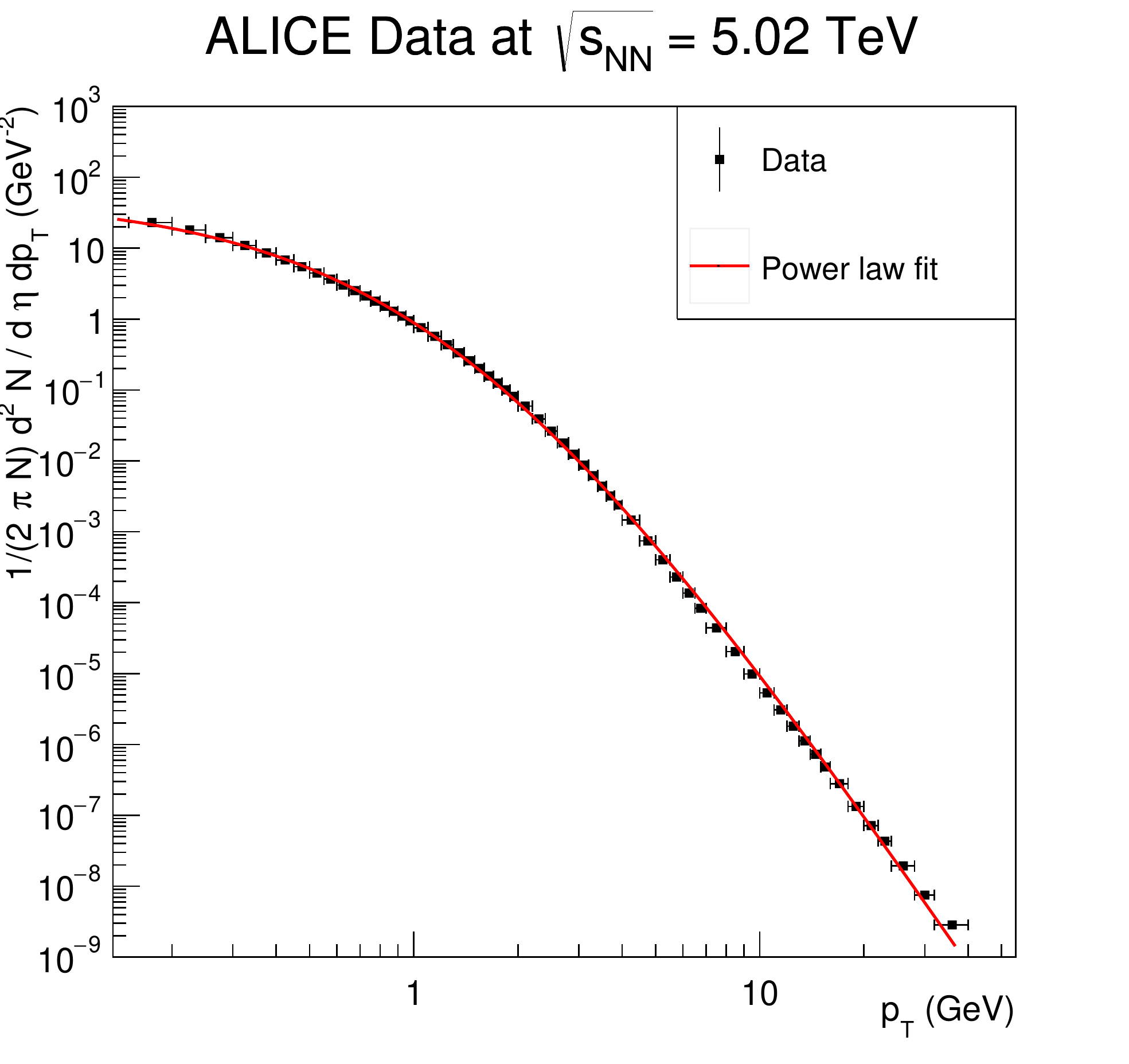}
\caption{}
\end{centering}
\end{subfigure}
~
\begin{subfigure}[b]{.45\textwidth}
\begin{centering}
\includegraphics[scale=.4]{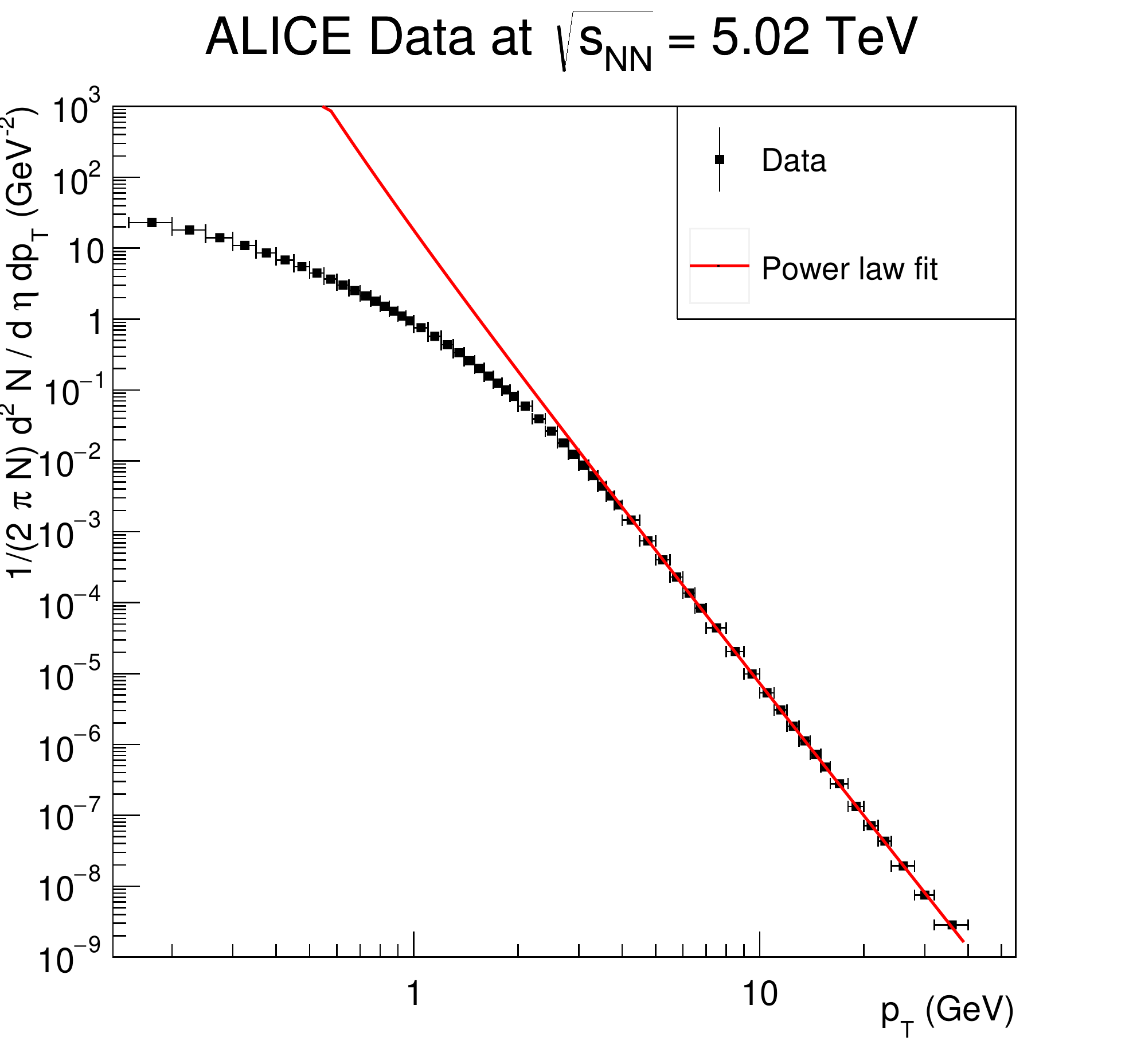}
\caption{}
\end{centering}
\end{subfigure}
\caption{Fit of the curve in Eq. (\ref{eq:3param}) to the ALICE dataset at $\sqrt{s}=5.02$ TeV in the $|\eta|<0.3$ bin \cite{1405} with cutoffs $p_{\text{min}}$ = 0 (left) and 3 GeV (right). }
\label{fig:alicefigs3param}
\end{center}
\end{figure}
}

\begin{table}[ht]
\begin{centering}
\begin{tabular}{|c|c|c|c|}\hline
$p_{\text{min}}$/(1 GeV) & A/10 (GeV$^{-2}$) & B  & $\chi^2$/NDF \\\hline

0 & 0.545 $\pm$ 0.0635 & 5.88 $\pm$ 0.0694 &  348. \\\hline

0.5 & 0.541 $\pm$ 0.0570 & 5.86 $\pm$ 0.0649 &  88.0 \\\hline

1.0 & 0.879 $\pm$ 0.110 & 6.09 $\pm$ 0.06428 & 5.12 \\\hline

1.5 & 1.00 $\pm$ 0.150 & 6.15 $\pm$ 0.0709 & 0.603 \\\hline

2.0 & 1.12 $\pm$ 0.316 & 6.19 $\pm$ 0.114 &  0.198 \\\hline

2.5 & 1.11 $\pm$ 0.356 & 6.18 $\pm$ 0.127 &  0.192 \\\hline

3.0 & 1.08 $\pm$ 0.419 & 6.17 $\pm$ 0.149 &  0.177 \\\hline

3.5 & 1.02 $\pm$ 0.567 & 6.15 $\pm$ 0.204 & 0.158 \\\hline

4.0 & 1.00 $\pm$ 0.600 & 6.15 $\pm$ 0.219 &  0.151 \\\hline

4.5 & 0.981 $\pm$ 0.638 & 6.14 $\pm$ 0.235 & 0.144 \\\hline

5.0 & 0.993 $\pm$ 0.699 & 6.15 $\pm$ 0.252 & 0.142 \\\hline

\end{tabular}
\caption{Fitted values of parameters in Eq. (\ref{eq:2param}) for the ALICE dataset at $\sqrt{s}=5.02$ TeV in the $-0.8<\eta<-0.3$ bin \cite{1405}.}
\end{centering}
\end{table}

\begin{table}[ht]
\begin{centering}
\begin{tabular}{|c|c|c|c|c|}\hline
$p_{\text{min}}$/(1 GeV) & A/10 (GeV$^{-2}$) & B  & C/(1 GeV) & $\chi^2$/NDF \\\hline

0 & 37.6 $\pm$ 7.97 & 7.22 $\pm$ 0.0841 & 1.30 $\pm$ 0.0439 & 17.6 \\\hline

0.5 & 19.6 $\pm$ 8.33 & 7.03 $\pm$ 0.142 & 1.10 $\pm$ 0.117 & 10.4 \\\hline

1.0 & 3.16 $\pm$ 2.37 & 6.49 $\pm$ 0.236 & 0.448 $\pm$ 0.256 & 2.09 \\\hline

1.5 & 1.44 $\pm$ 1.34 & 6.26 $\pm$ 0.289 & 0.135 $\pm$ 0.343 & 0.446 \\\hline

2.0 & 0.819 $\pm$ 1.19 & 6.10 $\pm$ 0.427 & -0.149 $\pm$ 0.678 & 0.152 \\\hline

2.5 & 0.754 $\pm$ 1.30 & 6.07 $\pm$ 0.497 & -0.196 $\pm$ 0.862 & 0.144 \\\hline

3.0 & 0.735 $\pm$ 1.47 & 6.07 $\pm$ 0.564 & -0.212 $\pm$ 1.09 & 0.142 \\\hline

3.5 & 0.691 $\pm$ 1.95 & 6.05 $\pm$ 0.758 & -0.256 $\pm$ 1.83 & 0.142 \\\hline

4.0 & 0.708 $\pm$ 2.30 & 6.06 $\pm$ 0.858 & -0.238 $\pm$ 2.20 & 0.142 \\\hline

4.5 & 0.766 $\pm$ 2.81 & 6.08 $\pm$ 0.956 & -0.177 $\pm$ 2.62 & 0.140 \\\hline

5.0 & 0.595 $\pm$ 2.53 & 6.01 $\pm$ 1.10 & -0.381 $\pm$ 3.15 & 0.131 \\\hline

\end{tabular}
\caption{Fitted values of parameters in Eq. (\ref{eq:3param}) for the ALICE dataset at $\sqrt{s}=5.02$ TeV in the $-0.8<\eta<-0.3$  bin \cite{1405}.}
\end{centering}
\end{table}


\begin{figure}[!tbt]
\begin{center}
\hspace{-3cm}
\begin{subfigure}[b]{.45\textwidth}
\begin{flushleft}
\includegraphics[scale=0.5]{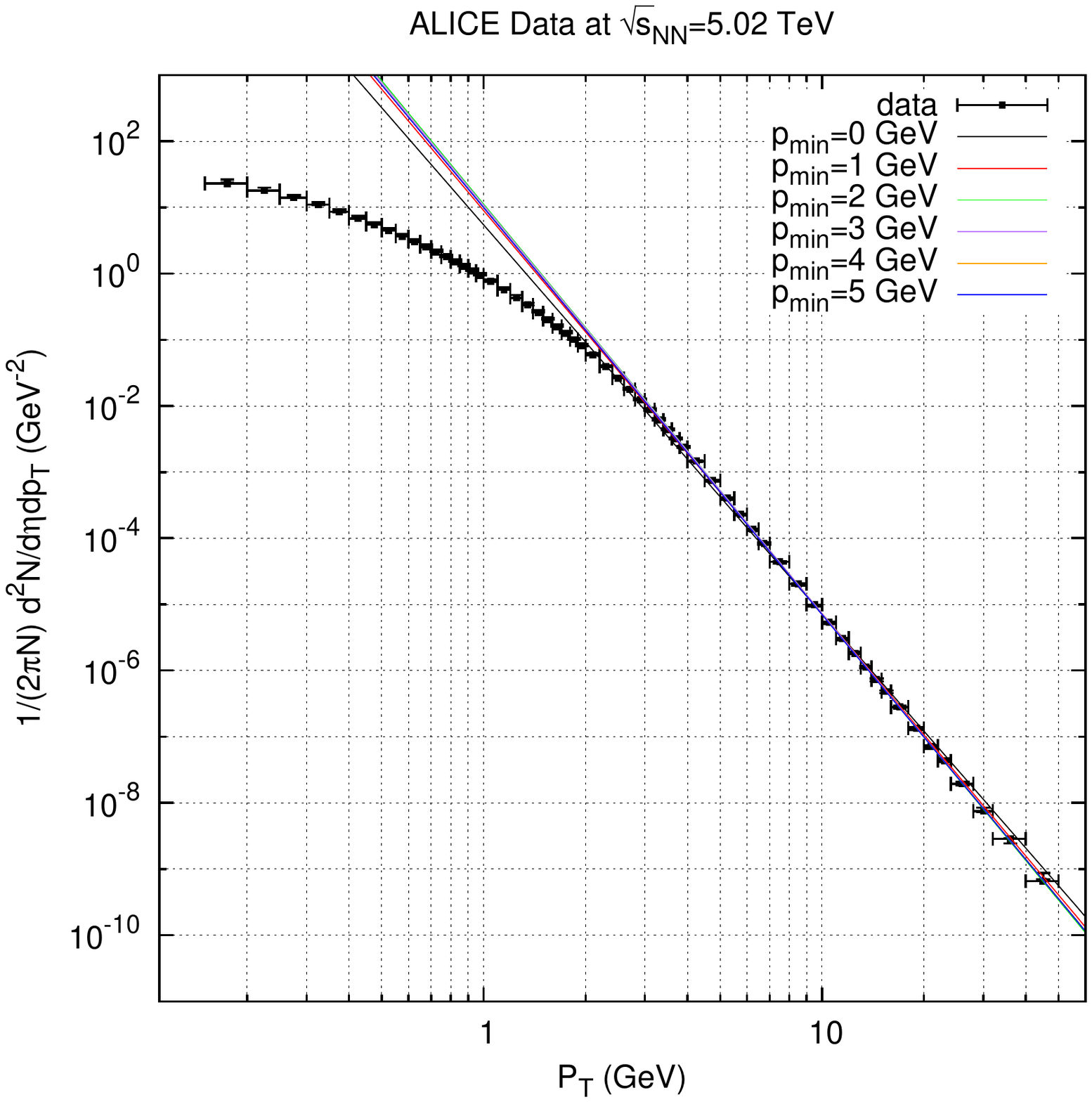}
 \vspace{-4cm}
\caption{}
\end{flushleft}
\end{subfigure}
~
\hspace{1cm}
\begin{subfigure}[b]{.45\textwidth}
\begin{flushright}
\includegraphics[scale=0.5]{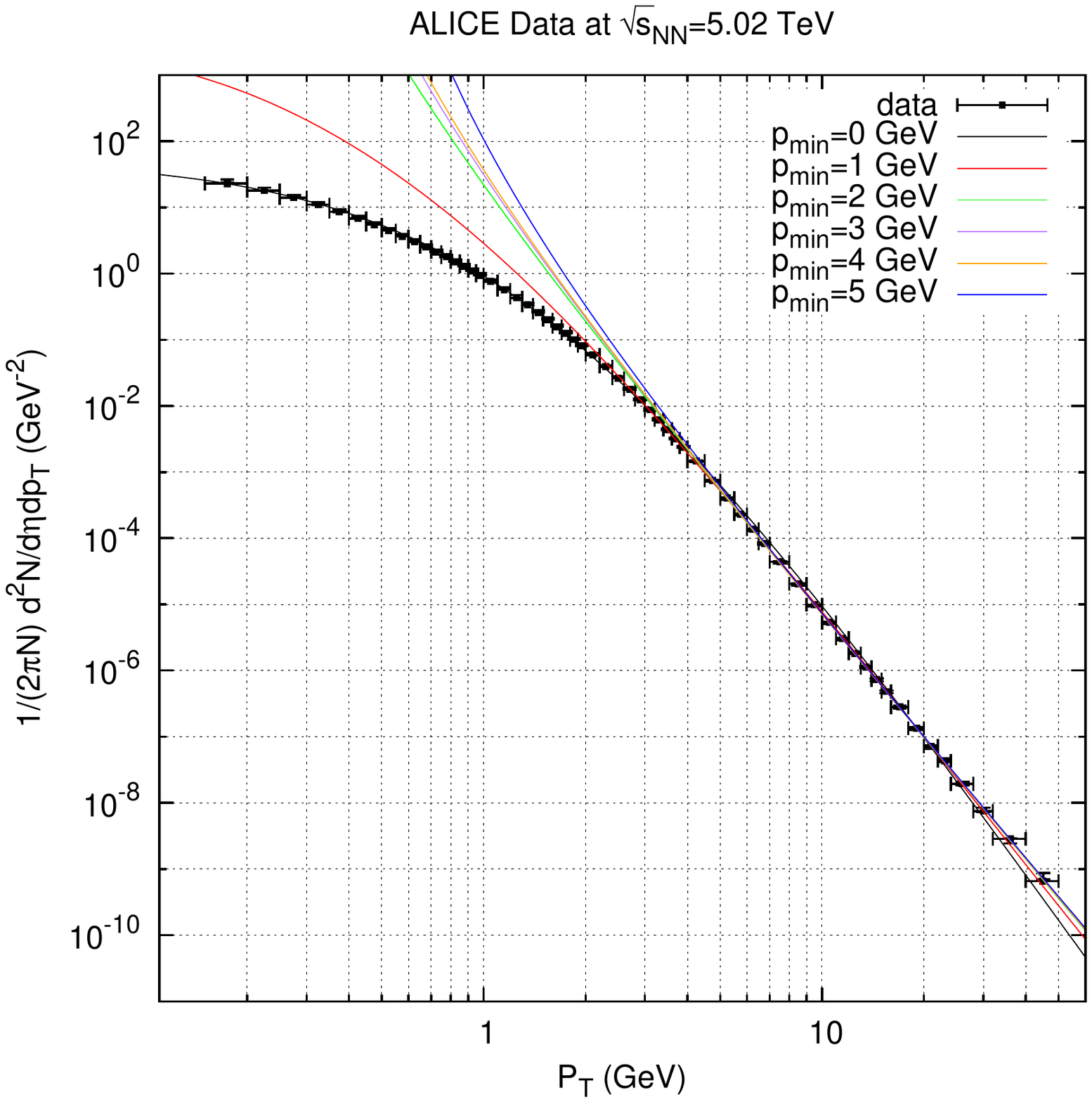}
 \vspace{-4cm}
\caption{}
\end{flushright}
\end{subfigure}
\caption{Fit of the two parameter ansatz in Eq. (\ref{eq:2param}) (left) and the three parameter ansatz in Eq. (\ref{eq:3param}) to the ALICE dataset at $\sqrt{s}=5.02$ TeV in the $-0.8<\eta<-0.3$ bin \cite{1405} with various data cutoffs. }
\label{fig:alicefigs2param2}
\end{center}
\end{figure}

\ignore{
\begin{figure}[!tbt]
\begin{center}
\begin{subfigure}[b]{.45\textwidth}
\begin{centering}
\includegraphics[scale=.4]{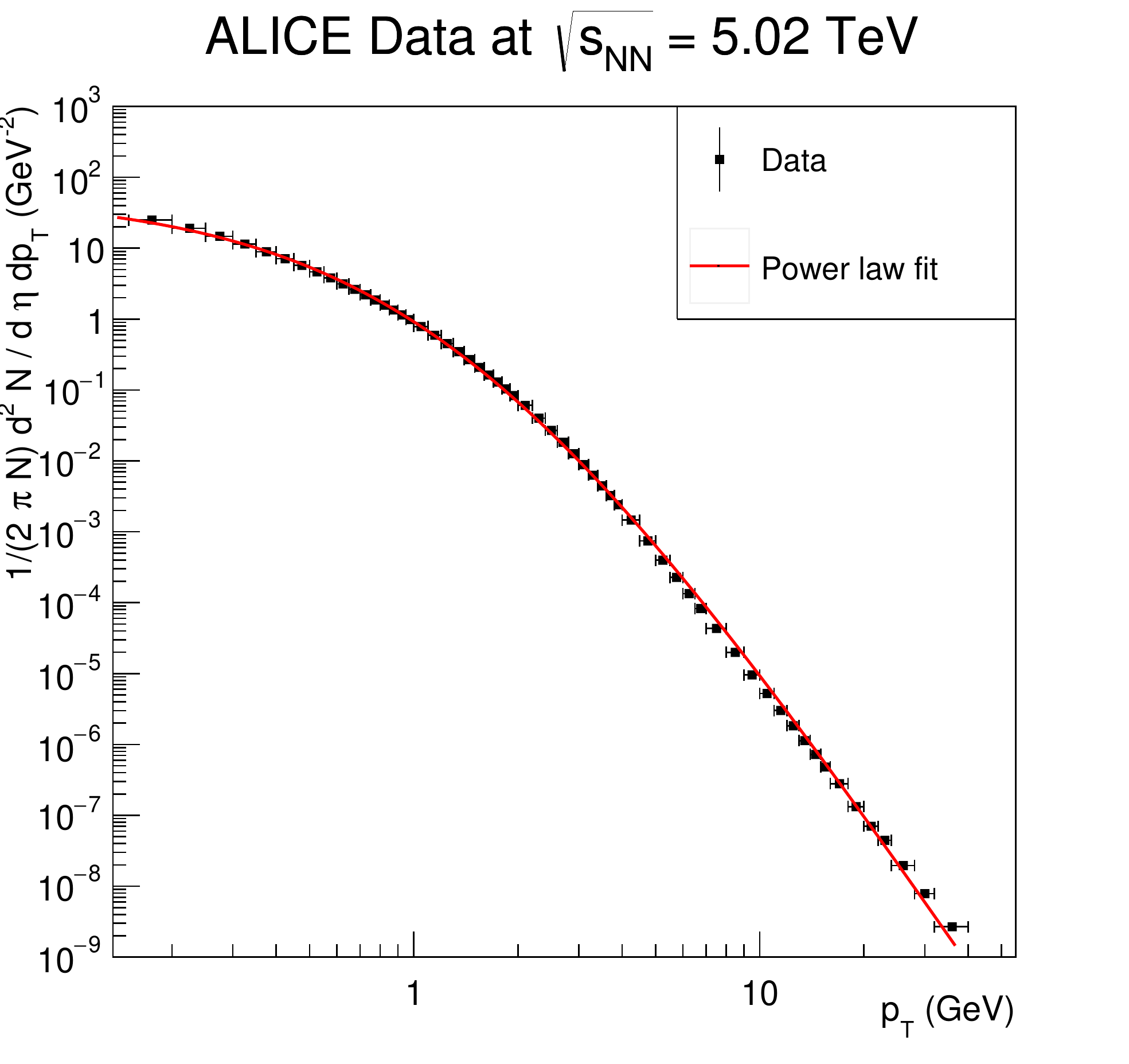}
\caption{}
\end{centering}
\end{subfigure}
~
\begin{subfigure}[b]{.45\textwidth}
\begin{centering}
\includegraphics[scale=.4]{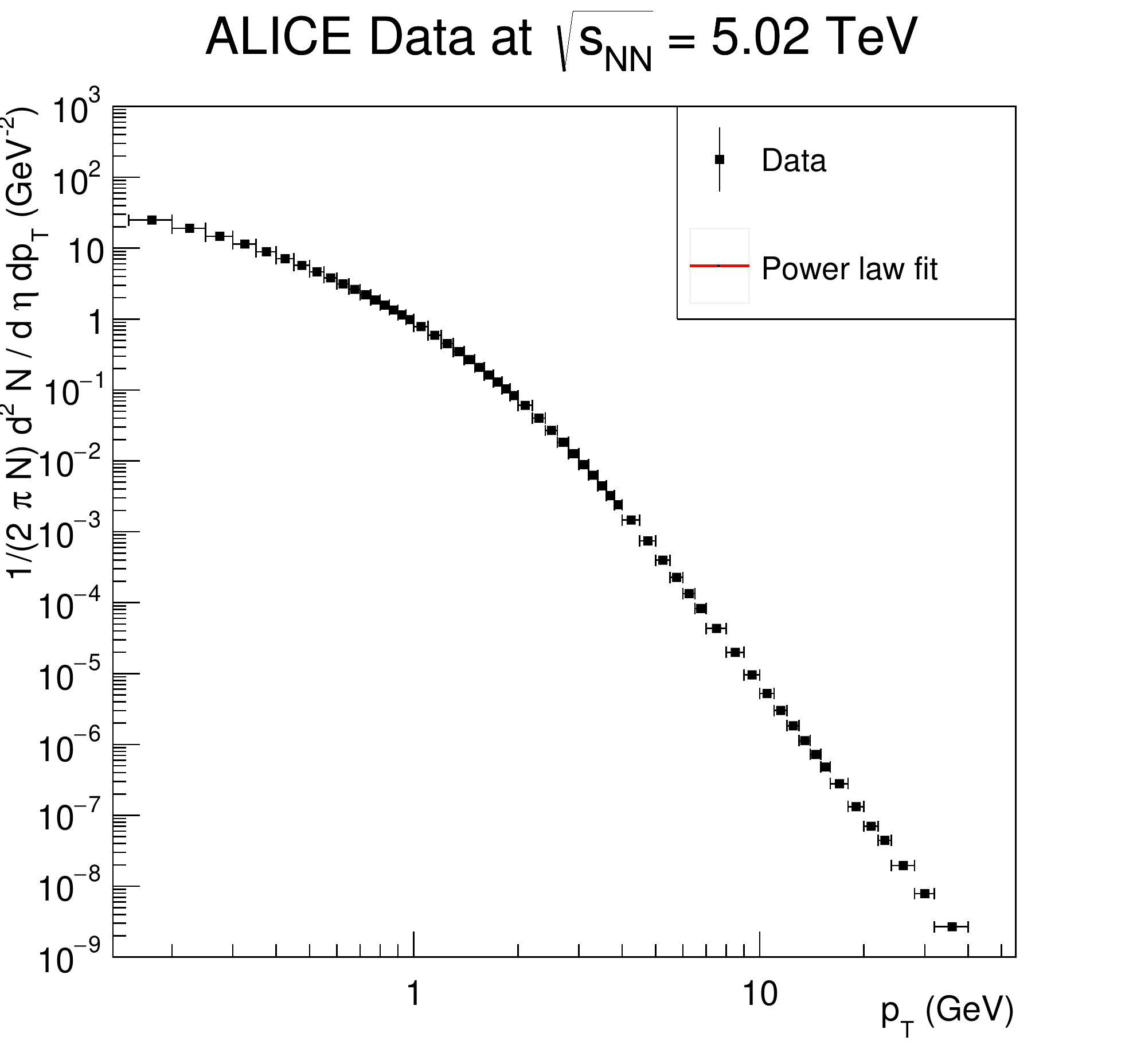}
\caption{}
\end{centering}
\end{subfigure}
\caption{Fit of the curve in Eq. (\ref{eq:3param}) to the ALICE dataset at $\sqrt{s}=5.02$ TeV in the $-0.8<\eta<-0.3$ bin \cite{1405} with cutoffs $p_{\text{min}}$ = 0 (left) and 3 GeV (right). }
\label{fig:alicefigs3param2}
\end{center}
\end{figure}
}

\begin{table}[ht]
\begin{centering}
\begin{tabular}{|c|c|c|c|}\hline
$p_{\text{min}}$/(1 GeV) & A/10 (GeV$^{-2}$) & B  & $\chi^2$/NDF \\\hline

0 & 0.592 $\pm$ 0.0708 & 5.94 $\pm$ 0.0708 &  347. \\\hline

0.5 & 0.580 $\pm$ 0.0627 & 5.91 $\pm$ 0.0665 & 89.5 \\\hline

1.0 & 0.949 $\pm$ 0.121 & 6.15 $\pm$ 0.0655 &  5.44 \\\hline

1.5 & 1.09 $\pm$ 0.166 & 6.21 $\pm$ 0.0724 &  0.745 \\\hline

2.0 & 1.25 $\pm$ 0.363 & 6.26 $\pm$ 0.117 &  0.213 \\\hline

2.5 & 1.25 $\pm$ 0.414 & 6.26 $\pm$ 0.131 & 0.213 \\\hline

3.0 & 1.241 $\pm$ 0.498 & 6.26 $\pm$ 0.154 &  0.211 \\\hline

3.5 & 1.208 $\pm$ 0.693 & 6.25 $\pm$ 0.211 & 0.207 \\\hline

4.0 & 1.19 $\pm$ 0.740 & 6.24 $\pm$ 0.226 &  0.204 \\\hline

4.5 & 1.19 $\pm$ 0.796 & 6.24 $\pm$ 0.243 &  0.204 \\\hline

5.0 & 1.16 $\pm$ 0.854 & 6.23 $\pm$ 0.263 &  0.199 \\\hline

\end{tabular}
\caption{Fitted values of parameters in Eq. (\ref{eq:2param}) for the ALICE dataset at $\sqrt{s}=5.02$ TeV in the $-1.3<\eta<-0.8$ bin \cite{1405}.}
\end{centering}
\end{table}

\begin{table}[ht]
\begin{centering}
\begin{tabular}{|c|c|c|c|c|}\hline
$p_{\text{min}}$/(1 GeV) & A/10 (GeV$^{-2}$) & B  & C/(1 GeV) & $\chi^2$/NDF \\\hline

0 & 43.0 $\pm$ 9.29 & 7.30 $\pm$ 0.0861 & 1.31 $\pm$ 0.0442 & 16.7 \\\hline

0.5 & 24.5 $\pm$ 10.7 & 7.13 $\pm$ 0.147 & 1.13 $\pm$ 0.119 & 9.29 \\\hline

1.0 & 4.01 $\pm$ 3.12 & 6.60 $\pm$ 0.244 & 0.498 $\pm$ 0.262 &  1.82 \\\hline

1.5 & 1.87 $\pm$ 1.82 & 6.37 $\pm$ 0.303 & 0.199 $\pm$ 0.355 & 0.424 \\\hline

2.0 & 1.13 $\pm$ 1.74 & 6.23 $\pm$ 0.453 & -0.0483 $\pm$ 0.709 & 0.209 \\\hline

2.5 & 1.07 $\pm$ 1.96 & 6.22 $\pm$ 0.528 & -0.0768 $\pm$ 0.902 & 0.206 \\\hline

3.0 & 1.05 $\pm$ 2.24 & 6.21 $\pm$ 0.601 & -0.0892 $\pm$ 1.14 & 0.205 \\\hline

3.5 & 1.07 $\pm$ 3.25 & 6.21 $\pm$ 0.822 & -0.0805 $\pm$ 1.92 & 0.205 \\\hline

4.0 & 1.14 $\pm$ 3.94 & 6.23 $\pm$ 0.918 & -0.0332 $\pm$ 2.28 & 0.204 \\\hline

4.5 & 1.19 $\pm$ 4.77 & 6.24 $\pm$ 1.04 & 0.00394 $\pm$ 2.74 & 0.204 \\\hline

5.0 & 1.52 $\pm$ 6.94 & 6.30 $\pm$ 1.17 & 0.195 $\pm$ 3.25 & 0.196 \\\hline

\end{tabular}
\caption{Fitted values of parameters in Eq. (\ref{eq:3param}) for the ALICE dataset at $\sqrt{s}=5.02$ TeV in the $-1.3<\eta<-0.8$ bin \cite{1405}.}
\label{tab:AppendixLast}
\end{centering}
\end{table}


\begin{figure}[!tbt]
\begin{center}
\hspace{-3cm}
\begin{subfigure}[b]{.45\textwidth}
\begin{flushleft}
\includegraphics[scale=0.5]{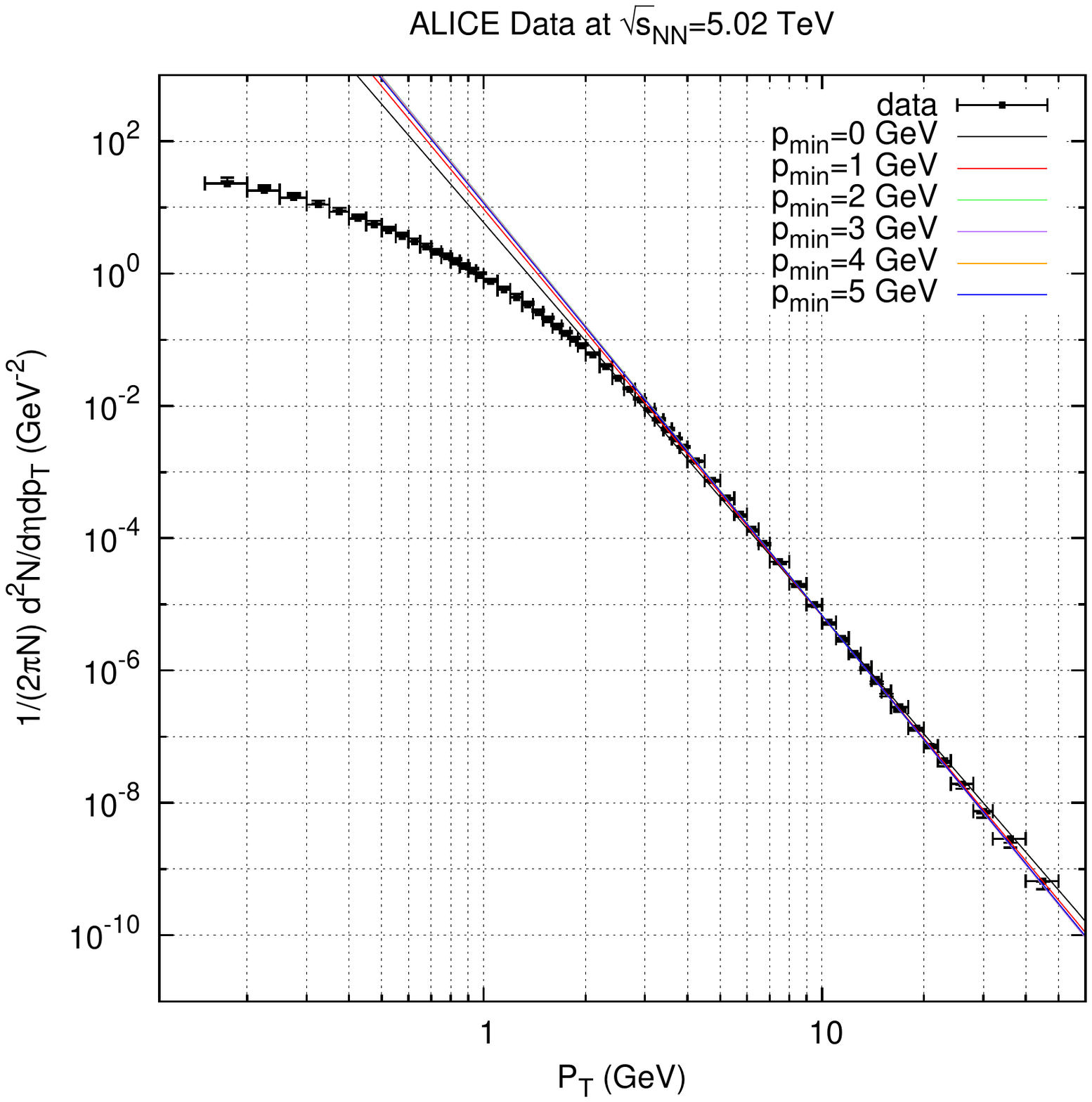}
 \vspace{-4cm}
\caption{}
\end{flushleft}
\end{subfigure}
~
\hspace{1cm}
\begin{subfigure}[b]{.45\textwidth}
\begin{flushright}
\includegraphics[scale=0.5]{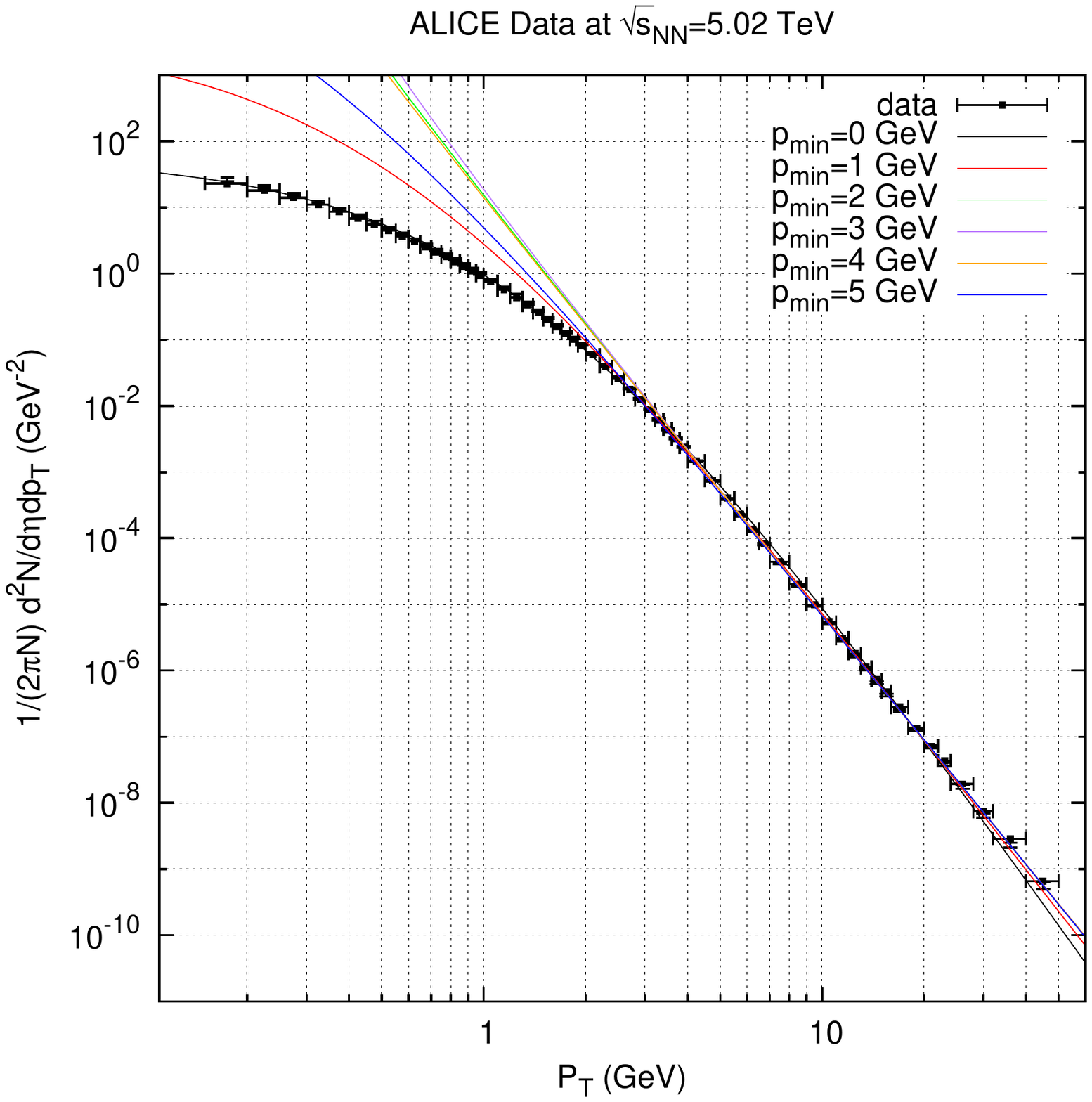}
 \vspace{-4cm}
\caption{}
\end{flushright}
\end{subfigure}
\caption{Fit of the two parameter ansatz in Eq. (\ref{eq:2param}) (left) and the three parameter ansatz in Eq. (\ref{eq:3param}) to the ALICE dataset at $\sqrt{s}=5.02$ TeV in the $-1.3<\eta<-0.8$ bin \cite{1405} with various data cutoffs. }
\label{fig:alicefigs2param3}
\end{center}
\end{figure}

\ignore{
\begin{figure}[!tbt]
\begin{center}
\begin{subfigure}[b]{.45\textwidth}
\begin{centering}
\includegraphics[scale=.4]{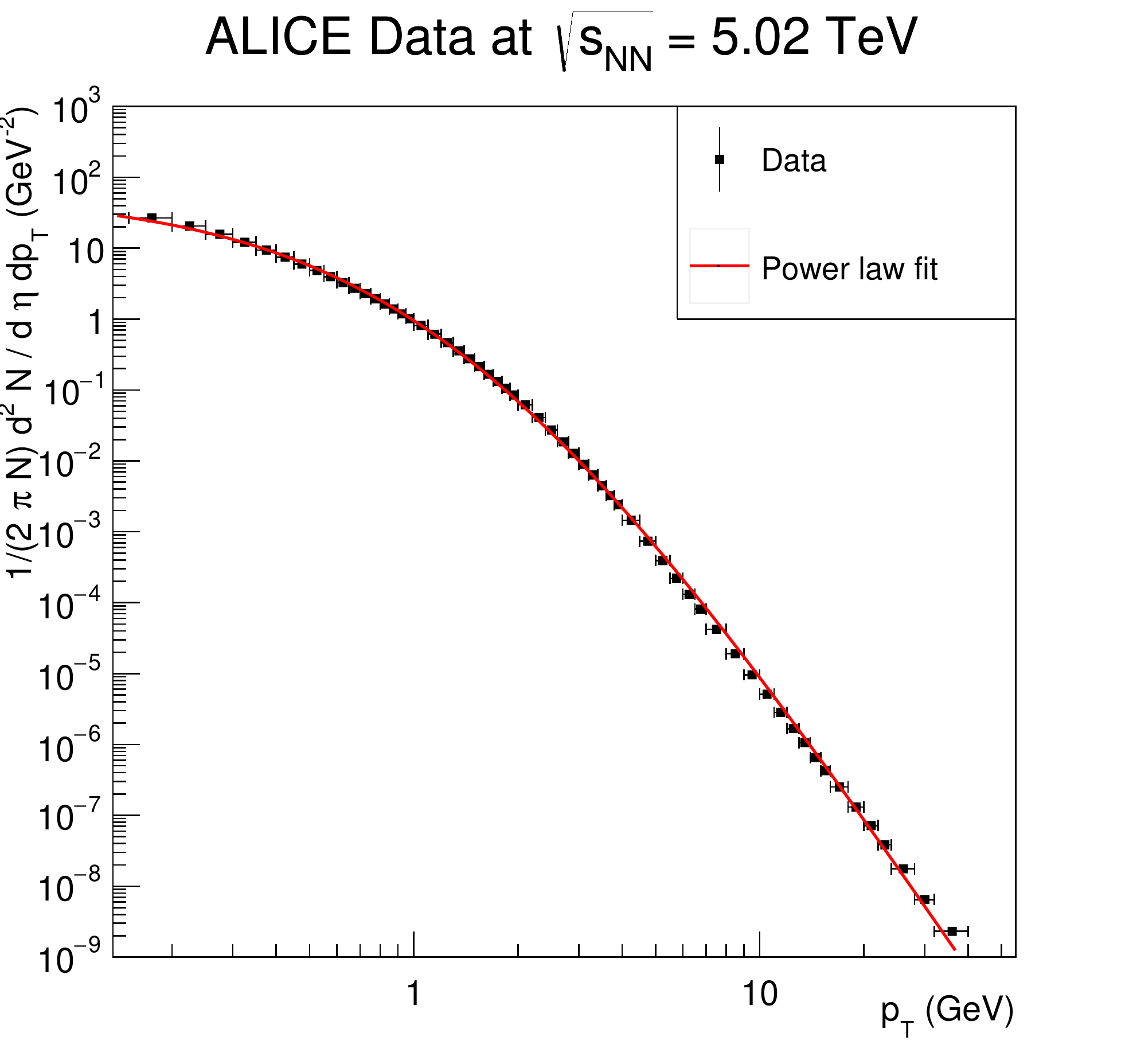}
\caption{}
\end{centering}
\end{subfigure}
~
\begin{subfigure}[b]{.45\textwidth}
\begin{centering}
\includegraphics[scale=.4]{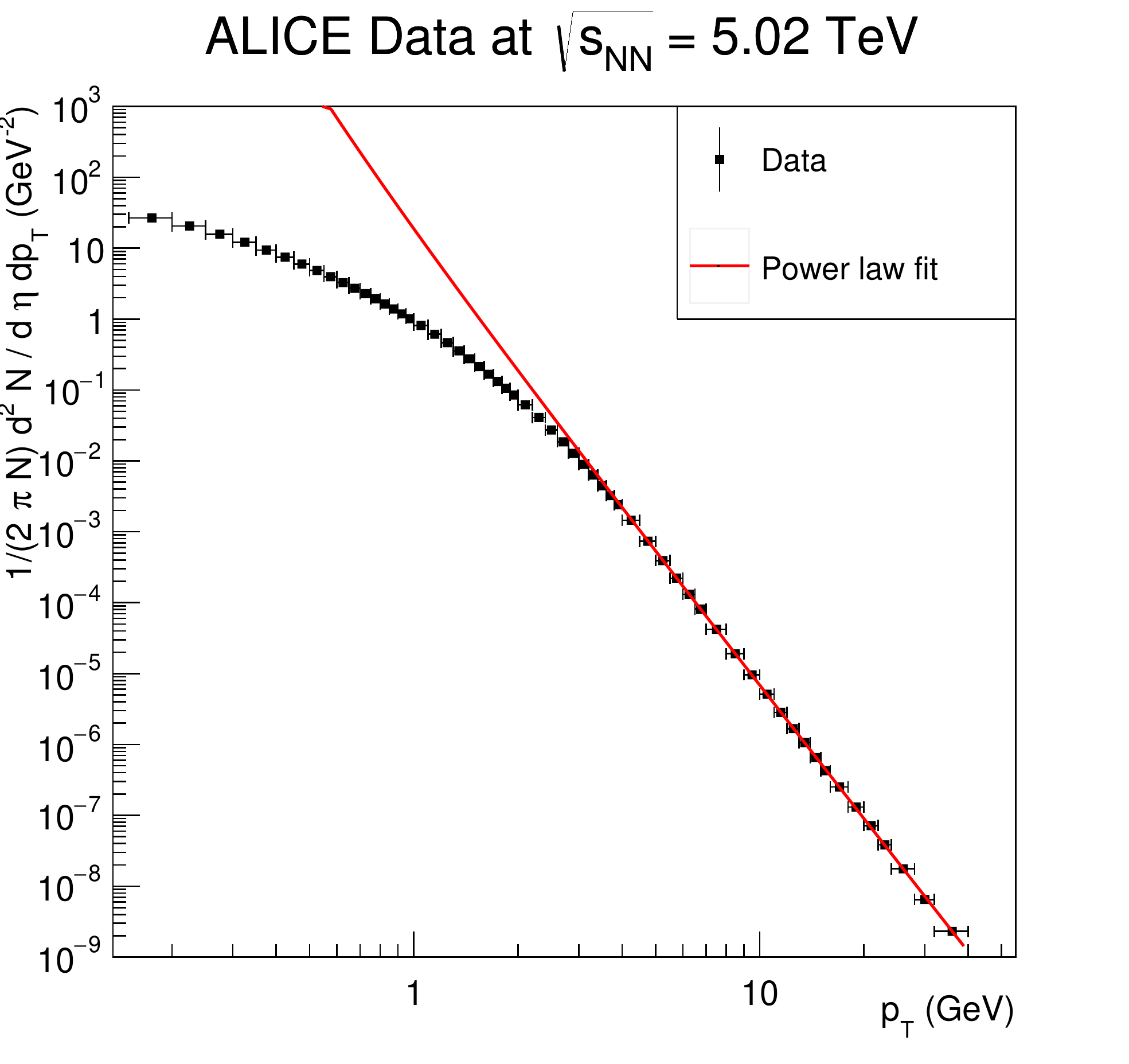}
\caption{}
\end{centering}
\end{subfigure}
\caption{Fit of the curve in Eq. (\ref{eq:3param}) to the ALICE dataset at $\sqrt{s}=5.02$ TeV in the $-1.3<\eta<-0.8$ bin \cite{1405} with cutoffs $p_{\text{min}}$ = 0 (left) and 3 GeV (right). }
\label{fig:alicefigs3param3}
\end{center}
\end{figure}
}

\FloatBarrier
\subsection{Small \texorpdfstring{$p_T$}{pT}} \label{sec:smallpt}
Here we comment on the small $p_T$ behavior.  We are focused here on $p_T<1$ GeV where confinement effects, ekonalization, and saturation effects should all play a major roll.  From the AdS/CFT perspective, many models display Regge phenomenology that are insensitive to these details at large enough energy scales ~\cite{Brower:2015hja,Brower:2014sxa}.

Although we are not focused on small $p_T$ behavior, we can however comment on two possible predictions.  The first is that of flat space string theory which predicts a fall off

\be \label{eq:flatlowpt}
d\sigma \sim A\,e^{-p^2_T/B}.
\ee

Recently it was advocated that a two parameter model applies at small $p_T$ \cite{Bylinkin:2014qea,lhcfit} of the form

\be \label{eq:by2}
d\sigma \sim A\,e^{-p_T/B}.
\ee

\begin{figure}
\begin{tabular}{ccc}
\multicolumn{3}{c}{\hspace{-10mm} \includegraphics[width=85mm]{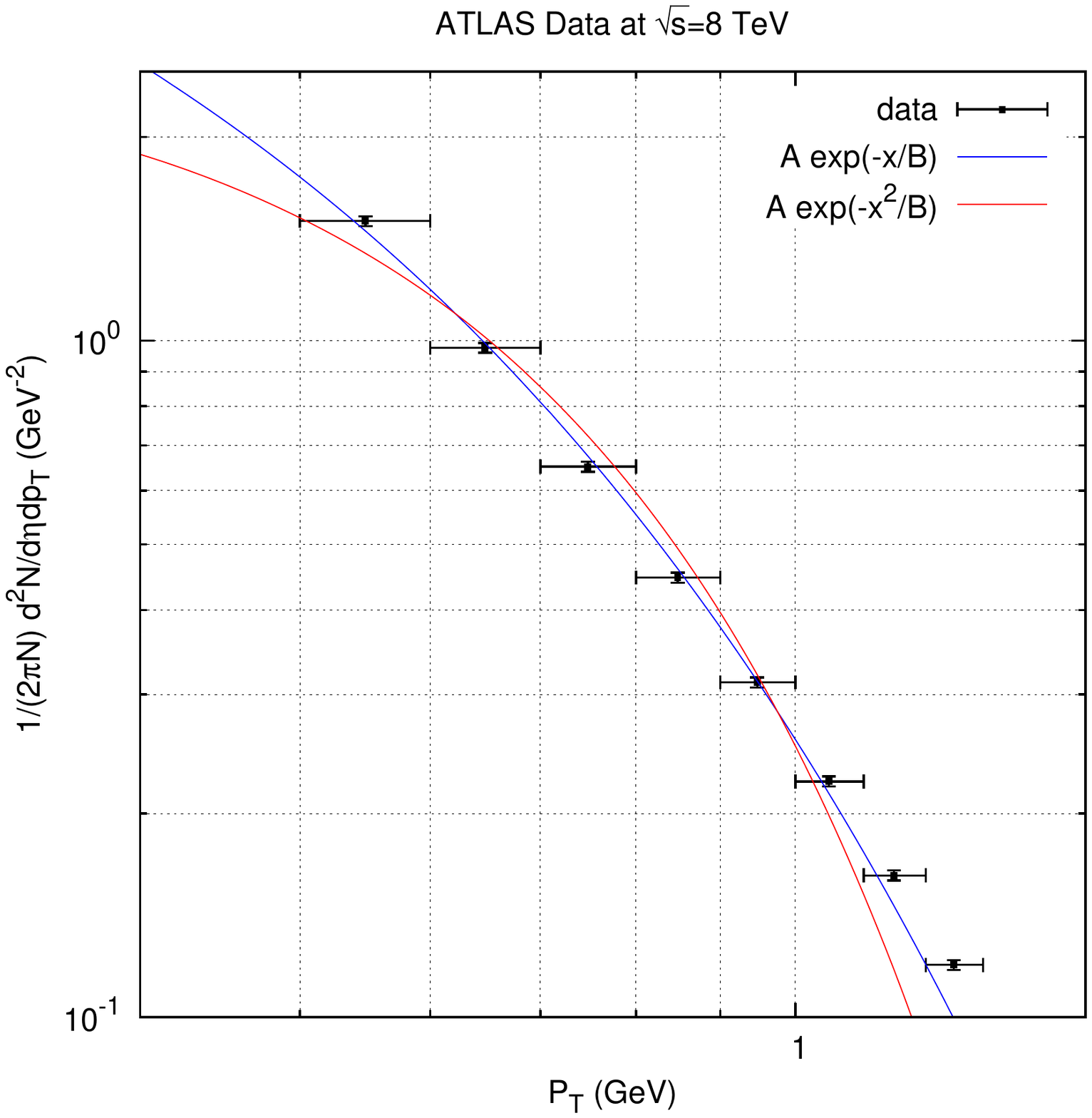} \includegraphics[width=85mm]{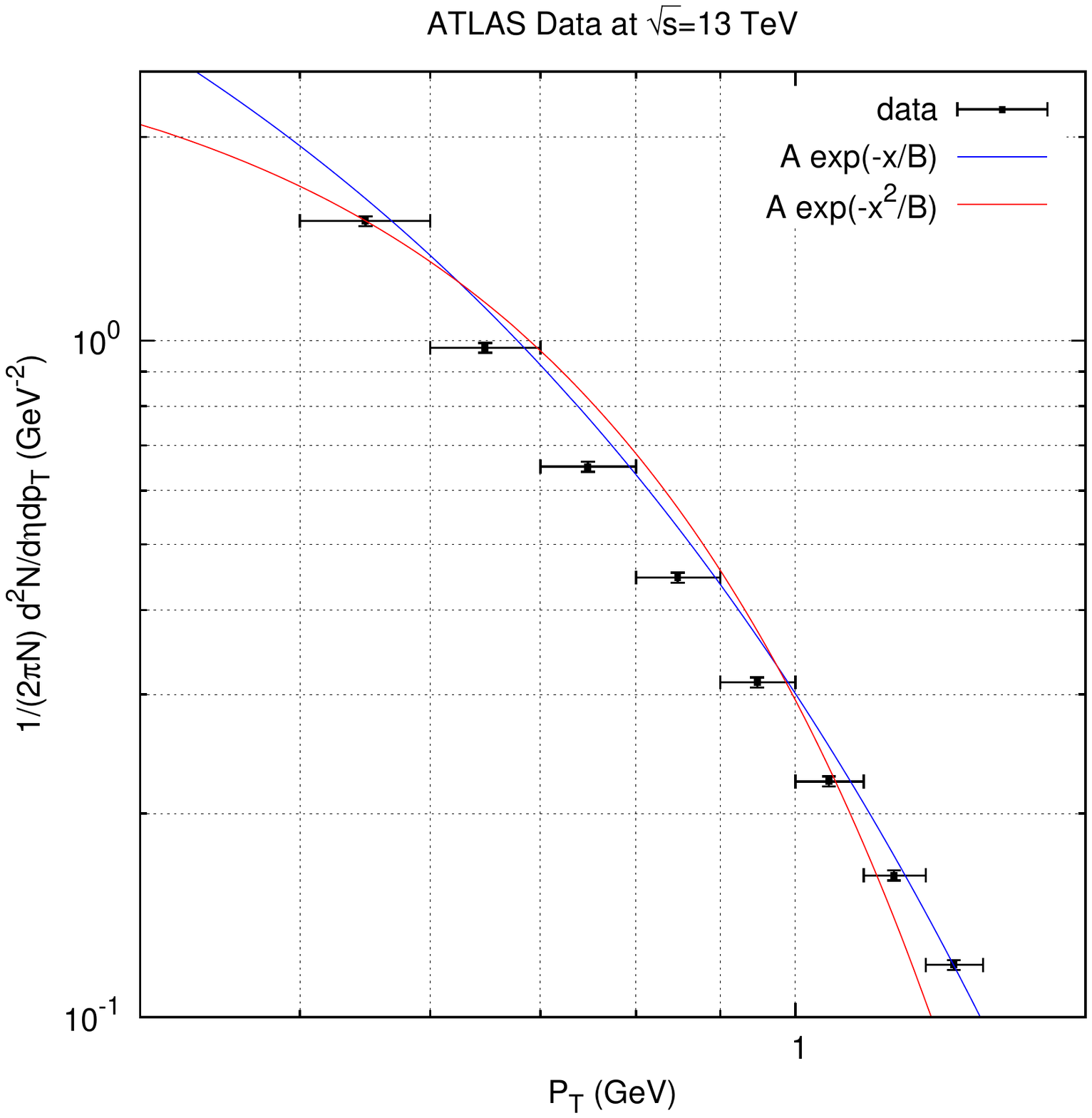}}\vspace{-20mm}\\
\multicolumn{3}{c}{\vspace{-10mm}\hspace{-15mm}(a)\hspace{6cm}  (b)}\\
 \hspace{-20mm} \includegraphics[width=65mm]{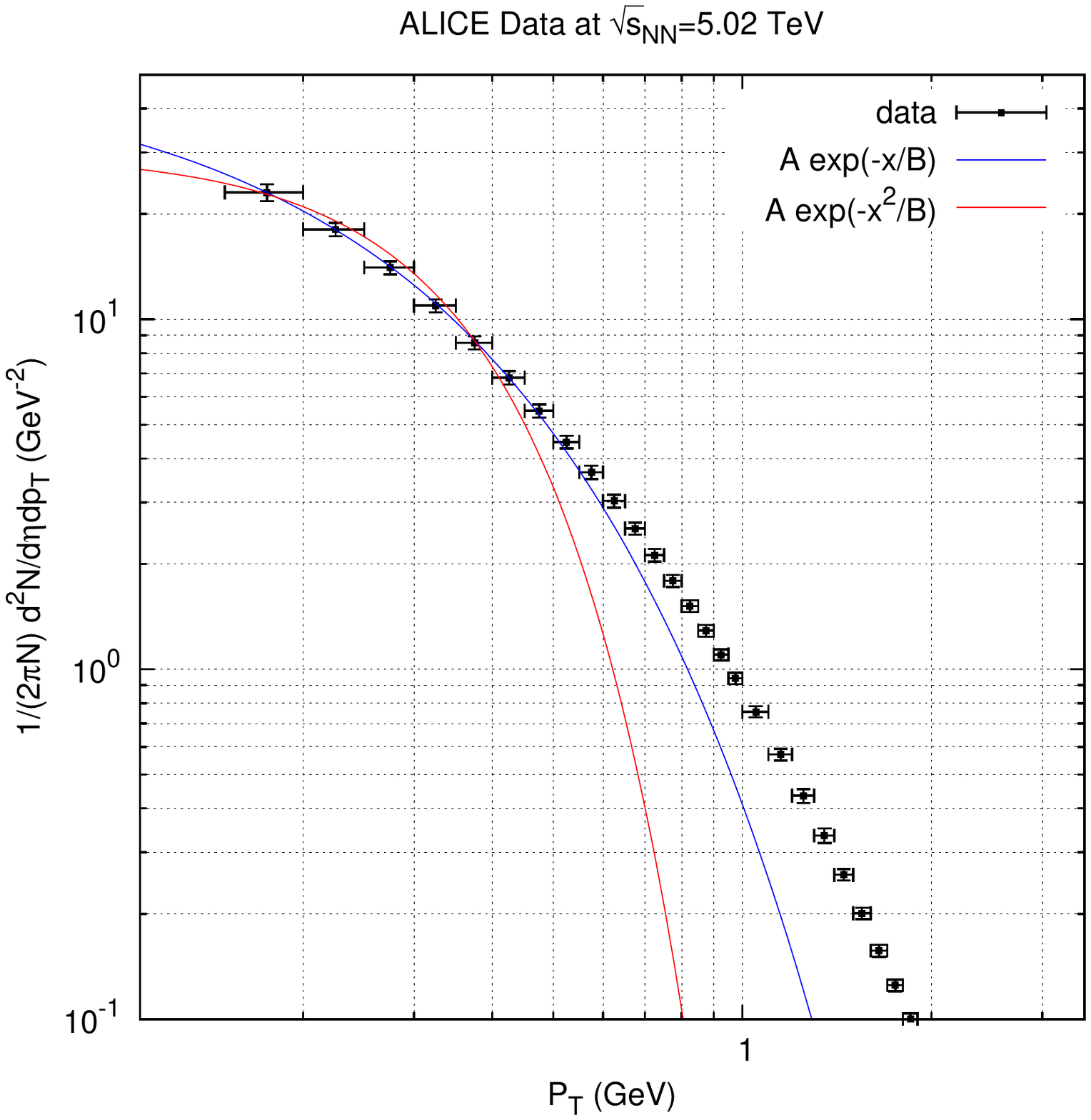} &   \hspace{-10mm}\includegraphics[width=65mm]{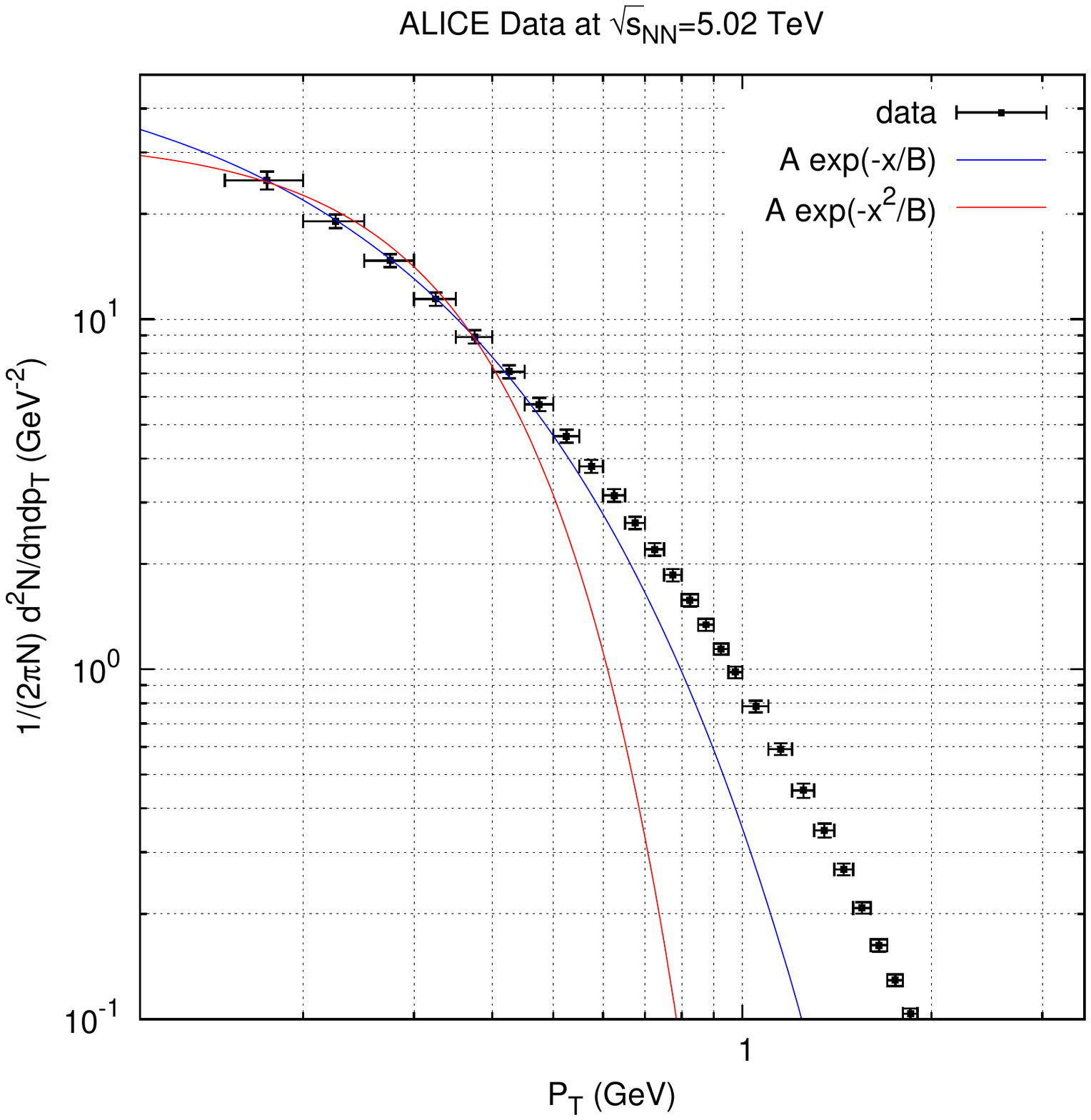} &
\hspace{-10mm} \includegraphics[width=65mm]{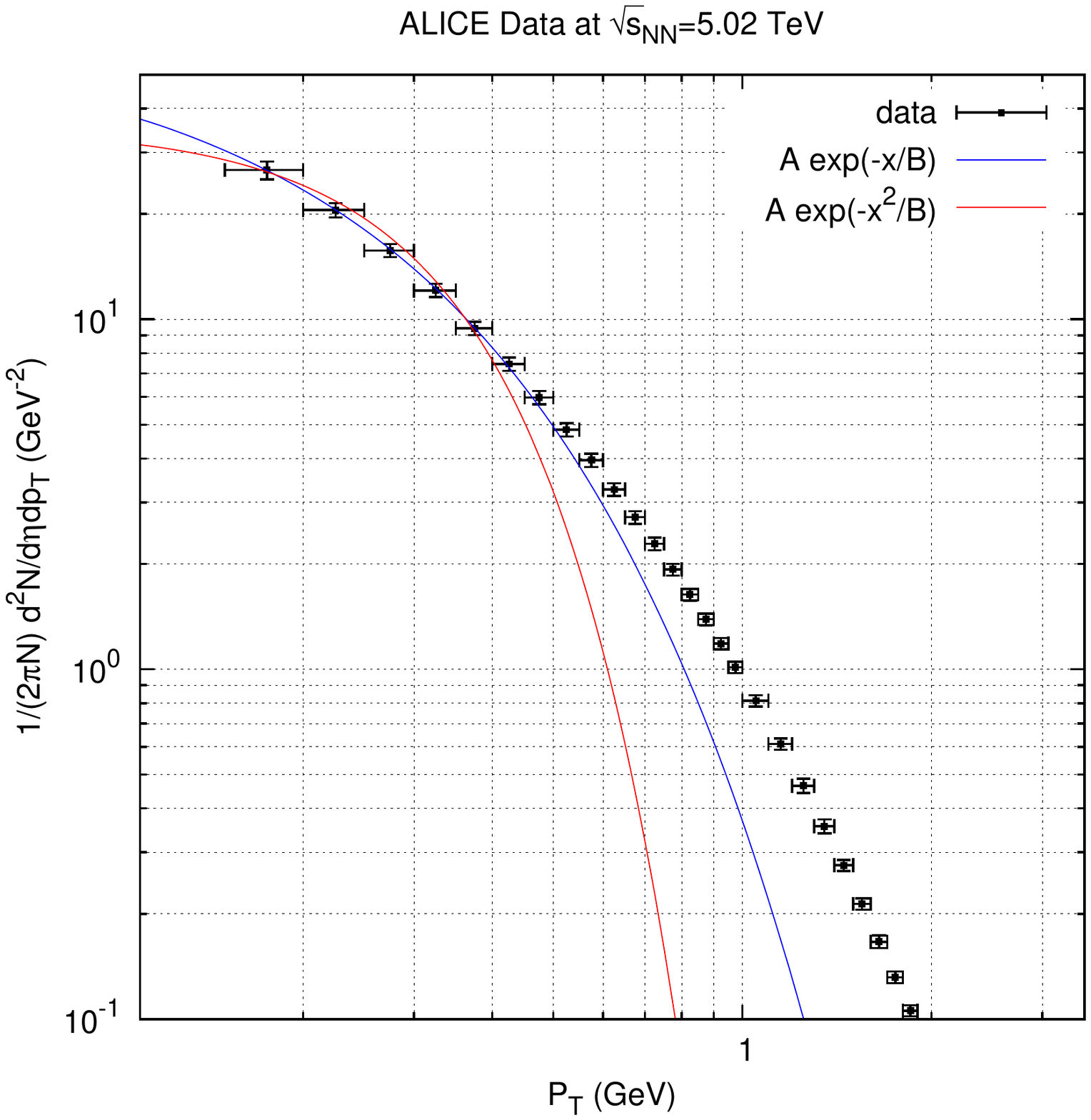} \vspace{-15mm}\\
\hspace{-15mm}(c)  & (d)  & (e)
\end{tabular}
\caption{ \label{fig:explowpt} Fits of $A\,exp(-x/B)$ and $A\,exp(-x^2/B)$ to all data below $p_T=1 GeV$ for the ATLAS $\sqrt{s}=8 TeV$ data set (a), the ATLAS $\sqrt{s}=13 TeV$ data set (b), the ALICE $\sqrt{s_{NN}}=5.02 TeV |\eta|<0.3$ data set (c), the ALICE $\sqrt{s_{NN}}=5.02 TeV -0.8<\eta<-0.3$ data set (d), and the ALICE $\sqrt{s_{NN}}=5.02 TeV -1.3<\eta<-0.8$ data set. Fit parameters can be found in Table ~\ref{tab:expo}. }
\end{figure}

The results for these small $p_T$ fits can be seen in Figure~\ref{fig:explowpt} and in Table~\ref{tab:expo}, although we caution that the limited amount of data in this region can lead to over-fitting.
 
\ignore{
{{
\begin{table}[ht]
\begin{centering}
\begin{tabular}{|c|c|c|c|c|c|c|}\hline
Dataset &\multicolumn{3}{c|}{$A\,exp(-x/B)$} & \multicolumn{3}{c|}{$A\,exp(-x^2/B)$} \\\hline
 & A/10 (GeV$^-2$) & B (GeV) & $\chi^2$/NDF & A/10 (GeV$^-2$) & B (GeV$^2$) & $\chi^2$/NDF  \\\hline
 \ignore{ATLAS: $\sqrt{s}=13$ TeV}\cite{Aad:2016mok} &  0.110$\pm$0.00306  &  0.268$\pm$0.00636  & 0.00381 & 0.0482$\pm$0.00489 & 0.429$\pm$0.0300 &  0.0350 \\\hline
 \ignore{ATLAS: $\sqrt{s}=8$ TeV} \cite{Aad:2016xww} &  0.107$\pm$0.00292  & 0.261 $\pm$0.00579  & 0.00312 & 0.0442$\pm$0.00486 & 0.417$\pm$0.0283 &  0.0308 \\\hline

\ignore{ALICE} \cite{1405}, black \ignore{ $|\eta|<0.3$} &  0.173$\pm$0.000271  &  0.205$\pm$ 0.00136 & 2.35 & 0.147$\pm$0.000967 & 0.114$\pm$0.00688 & 314  \\\hline
\ignore{ALICE} \cite{1405}, red \ignore{ $-0.8<\eta<-0.3$} &  0.179$\pm$ 0.000425 & 0.194 $\pm$0.00193  & 4.48 & 0.152$\pm$0.00105 & 0.106$\pm$0.00681 & 517  \\\hline
\ignore{ALICE} \cite{1405}, blue\ignore{ $-1.3<\eta<-0.8$} &  0.182$\pm$0.000335  & 0.193 $\pm$0.00151  & 4.31 & 0.155$\pm$0.00101 & 0.104$\pm$0.00631 &  612 \\\hline

\end{tabular}
\caption{Fits of Eq.(~\ref{eq:flatlowpt}) and Eq.(~\ref{eq:by2}) to small $p_T<1$ GeV data. In the last three entries, the colors correspond to the color scheme used in e.g. Figure \ref{fig:1405plot}.}
\label{tab:expo}
\end{centering}
\end{table} }}
}

{{
\begin{table}[ht]
\begin{centering}
\begin{tabular}{|c|c|c|c|}\hline
Dataset &\multicolumn{3}{c|}{$A\,exp(-x/B)$} \\\hline
& A/10 (GeV$^-2$) & B (GeV) & $\chi^2$/NDF \\\hline
ATLAS: $\sqrt{s}=13$ TeV \cite{Aad:2016mok} &  0.110$\pm$0.00306  &  0.268$\pm$0.00636  & 0.00381 \\\hline
ATLAS: $\sqrt{s}=8$ TeV \cite{Aad:2016xww} &  0.107$\pm$0.00292  & 0.261 $\pm$0.00579  & 0.00312 \\\hline
ALICE  $|\eta|<0.3$ \cite{1405}&  0.173$\pm$0.000271  &  0.205$\pm$ 0.00136 & 2.35 \\\hline
ALICE $-0.8<\eta<-0.3$ \cite{1405} &  0.179$\pm$ 0.000425 & 0.194 $\pm$0.00193  & 4.48 \\\hline
ALICE $-1.3<\eta<-0.8$ \cite{1405} &  0.182$\pm$0.000335  & 0.193 $\pm$0.00151  & 4.31 \\\hline
Dataset& \multicolumn{3}{c|}{$A\,exp(-x^2/B)$} \\\hline
& A/10 (GeV$^-2$) & B (GeV$^2$) & $\chi^2$/NDF \\\hline  
ATLAS: $\sqrt{s}=13$ TeV \cite{Aad:2016mok}& 0.0482$\pm$0.00489 & 0.429$\pm$0.0300 &  0.0350 \\\hline
ATLAS: $\sqrt{s}=8$ TeV \cite{Aad:2016xww}& 0.0442$\pm$0.00486 & 0.417$\pm$0.0283 &  0.0308 \\\hline
ALICE  $|\eta|<0.3$ \cite{1405} & 0.147$\pm$0.000967 & 0.114$\pm$0.00688 & 314  \\\hline
ALICE $-0.8<\eta<-0.3$ \cite{1405} & 0.152$\pm$0.00105 & 0.106$\pm$0.00681 & 517  \\\hline
ALICE $-1.3<\eta<-0.8$ \cite{1405}& 0.155$\pm$0.00101 & 0.104$\pm$0.00631 &  612 \\\hline

\end{tabular}
\caption{Fits of Eq.(~\ref{eq:flatlowpt}) and Eq.(~\ref{eq:by2}) to small $p_T<1$ GeV data. In the last three entries, the colors correspond to the color scheme used in e.g. Figure \ref{fig:1405plot}.}
\label{tab:expo}
\end{centering}
\end{table} }}

\end{document}